\documentclass[journal]{new-aiaa}

\usepackage{graphicx}

\usepackage{amssymb}
\usepackage{color}
\usepackage{amsmath}
\usepackage[tight,footnotesize]{subfigure}
\usepackage{float}
\usepackage{multirow}
\usepackage{booktabs}
\usepackage{subeqnarray}
\usepackage{epstopdf}
\usepackage{wrapfig}
\usepackage{soul}
\usepackage{siunitx}
\usepackage{enumitem}
\usepackage{longtable,tabularx}
\usepackage{algorithm}
\usepackage[noend]{algorithmic}
\usepackage{subfigure}
\usepackage{subfigmat}

\usepackage{setspace}

\usepackage{tikz}
\usetikzlibrary{shapes,arrows,positioning,calc}
\tikzset{
	block/.style = {draw, fill=white, rectangle, minimum height=3em, minimum width=4em},
	alertblock/.style = {draw, color = red, 
		fill=white, rectangle, minimum height=3em, minimum width=4em},
	tmp/.style  = {coordinate}, 
	sum/.style= {draw, fill=white, circle, node distance=1cm},
	input/.style = {coordinate},
	output/.style= {coordinate},
	pinstyle/.style = {pin edge={to-,thin,black}}
}

\usepackage{amsthm}

\newtheorem{thm}{Theorem}
\newtheorem{prop}{Proposition}

\newtheorem{prob}{Problem}

\newtheorem{rmk}{Remark}

\newcommand{\real}[1][]{\mathbb{R}^{#1}}                                
\newcommand{\nat}[1][]{\mathbb{N}^{#1}}                                 

\newcommand{\defeq}{:=}                                                 
\newcommand{\msub}[1]{_\mathrm{#1}}                                     
\newcommand{\msup}[1]{^\mathrm{#1}}                                     

\newcommand{\eye}[1]{\boldsymbol{I}_{#1}}                                          

\renewcommand{\leq}{\leqslant}                                          
\renewcommand{\geq}{\geqslant}                                          

\newcommand{\intset}[1]{\{#1\}}

\newcommand{\entropy}{H}

\newcommand{\mydef}[1]{{\textit{#1}}}

\newcommand{\revnew}[1]{\textcolor{red}{#1}}

\newcommand{\ppl}{ath-planning}
\newcommand{\mpl}{otion-planning}

\newcommand{\astar}{A$^*$}

\newcommand{\matlab}{MATLAB\textsuperscript{\textregistered}}

\newcommand{\figpath}{fig}

\newcommand{\eqnnt}[1]{\hyperref[#1]{(\ref*{#1})}}
\newcommand{\eqnsnt}[2]{\hyperref[#1]{(\ref*{#1})}
	and~\hyperref[#2]{(\ref*{#2})}}
\newcommand{\eqnsernt}[2]{\hyperref[#1]{(\ref*{#1})}--\hyperref[#2]{(\ref*{#2})}}

\newcommand{\eqn}[1]{\hyperref[#1]{Eqn.~(\ref*{#1})}}
\newcommand{\eqns}[2]{\hyperref[#1]{Eqns.~(\ref*{#1})} and~\hyperref[#2]{(\ref*{#2})}}
\newcommand{\eqnser}[2]{\hyperref[#1]{Eqns.~(\ref*{#1})}--\hyperref[#2]{(\ref*{#2})}}
\newcommand{\eqnf}[1]{\hyperref[#1]{Equation~(\ref*{#1})}}
\newcommand{\eqnfs}[2]{\hyperref[#1]{Equations~(\ref*{#1})} and~\hyperref[#2]{(\ref*{#2})}}

\newcommand{\scn}[1]{\hyperref[#1]{\S\ref*{#1}}}
\newcommand{\scns}[2]{\hyperref[#1]{\S\ref*{#1}} and~\hyperref[#2]{\ref*{#2}}}
\newcommand{\scnser}[2]{\hyperref[#1]{\S\ref*{#1}}--\hyperref[#2]{\ref*{#2}}}

\newcommand{\fig}[1]{\hyperref[#1]{Fig.~\ref*{#1}}}
\newcommand{\figs}[2]{\hyperref[#1]{Figs.~\ref*{#1}} and~\hyperref[#2]{\ref*{#2}}}
\newcommand{\figser}[2]{\hyperref[#1]{Figs.~\ref*{#1}}--\hyperref[#2]{\ref*{#2}}}
\newcommand{\figf}[1]{\hyperref[#1]{Figure~\ref*{#1}}}
\newcommand{\figfs}[2]{\hyperref[#1]{Figures~\ref*{#1}} and~\hyperref[#2]{\ref*{#2}}}
\newcommand{\figfser}[2]{\hyperref[#1]{Figures~\ref*{#1}}--\hyperref[#2]{\ref*{#2}}}

\newcommand{\tbl}[1]{\hyperref[#1]{Table~\ref*{#1}}}
\newcommand{\tbls}[2]{\hyperref[#1]{Tables~\ref*{#1}} and~\hyperref[#2]{\ref*{#2}}}
\newcommand{\tblser}[2]{\hyperref[#1]{Tables~\ref*{#1}}--\hyperref[#2]{\ref*{#2}}}

\newcommand{\apx}[1]{\hyperref[#1]{Appendix~\ref*{#1}}}

\newcommand{\chp}[1]{\hyperref[#1]{Chapter~\ref*{#1}}}
\newcommand{\chps}[2]{\hyperref[#1]{Chapters~\ref*{#1}} and~\hyperref[#2]{(\ref*{#2})}}
\newcommand{\chpser}[2]{\hyperref[#1]{Chapters~\ref*{#1}}--\hyperref[#2]{(\ref*{#2})}}

\newcommand{\prb}[1]{\hyperref[#1]{Problem~\ref*{#1}}}
\newcommand{\prp}[1]{\hyperref[#1]{Prop.~\ref*{#1}}}
\newcommand{\prpf}[1]{\hyperref[#1]{Proposition~\ref*{#1}}}

\newcommand{\corref}[1]{\hyperref[#1]{Cor.~\ref*{#1}}}
\newcommand{\algoref}[1]{\hyperref[#1]{Algorithm~\ref*{#1}}}
\newcommand{\asmref}[1]{\hyperref[#1]{Assumption~\ref*{#1}}}

\newcommand{\thmref}[1]{\hyperref[#1]{Theorem~\ref*{#1}}}
\newcommand{\thmsref}[2]{\hyperref[#1]{Theorems~\ref*{#1}} and~\hyperref[#2]{\ref*{#2}}}
\newcommand{\thmserref}[2]{\hyperref[#1]{Theorems~\ref*{#1}}--\hyperref[#2]{\ref*{#2}}}

\newcommand{\lemref}[1]{\hyperref[#1]{Lemma~\ref*{#1}}}

\newcommand{\algline}[1]{\hyperref[#1]{Line~\ref*{#1}}}
\newcommand{\alglines}[2]{\hyperref[#1]{Lines~\ref*{#1}} and~\hyperref[#2]{\ref*{#2}}}
\newcommand{\alglineser}[2]{\hyperref[#1]{Lines~\ref*{#1}}--\hyperref[#2]{\ref*{#2}}}

\def\wsp{\mathcal{W}}

\def\threat{c}

\def\edges{E}

\def\vpath{\boldsymbol{\pi}}
\newcommand{\pathIncidence}[1]{\boldsymbol{v}_{\vpath}}

\def\nSensor{N\msub{S}}

\newcommand{\termThreshold}{\varepsilon}

\renewcommand{\vec}[1]{\boldsymbol{#1}}

\def\path{\boldsymbol{\rho}}

\def\wsp{\mathcal{W}}

\let\classAND\AND
\let\AND\relax
\usepackage{algorithmic}

\let\AND\classAND
	
\newcommand{\xState}{\vec{x}}
\newcommand{\qConfig}{\vec{q}}
\newcommand{\qgreedy}{\vec{q}\msub{gr}}
\newcommand{\zMeas}{\vec{z}}
\newcommand{\QProcCovar}{Q}
\newcommand{\PEECovar}{P}
\newcommand{\RMeasCovar}{R}
\newcommand{\gridPath}{\vec{v}}

\newcommand{\nGridPt}{N\msub{G}}
\newcommand{\nParameter}{N\msub{P}}
\newcommand{\paramVec}{\boldsymbol{\Theta}}
\newcommand{\HMeas}{C}
\newcommand{\UKFGain}{L}

\def\timeStep{\Delta t}

\title{Optimal Coupled Sensor Placement and 
	Path-Planning in Unknown Time -Varying Environments} %

\author{Prakash Poudel\footnote{Graduate Research Assistant, 
		Aerospace Engineering Department. AIAA Student Member.
		Email: \texttt{ppoudel@wpi.edu}} 
		and Raghvendra V. Cowlagi\footnote{Associate Professor, 
		Aerospace Engineering Department. AIAA Senior Member.
		Corresponding author. Email: \texttt{rvcowlagi@wpi.edu}}}
\affil{Worcester Polytechnic Institute, 100 Institute Rd., Worcester MA, USA 01609.}
\begin{document}
	
\maketitle
\begin{abstract}  
We address path-planning for a mobile agent to navigate in an unknown environment with
minimum exposure to a spatially and temporally varying threat field. The threat
field is estimated using pointwise noisy measurements from a mobile sensor
network. For this problem, we present a new information gain measure for
optimal sensor placement that quantifies reduction in uncertainty in the path
cost rather than the environment state. This measure, which we call the
context-relevant mutual information (CRMI), couples the sensor placement and
path-planning problem. We propose an iterative coupled sensor configuration and
path-planning (CSCP) algorithm. At each iteration, this algorithm places
sensors to maximize CRMI, updates the threat estimate using new measurements,
and recalculates the path with minimum expected exposure to the threat. The
iterations converge when the path cost variance, which is an indicator of risk,
reduces below a desired threshold. We show that CRMI is submodular, and
therefore, greedy optimization provides near-optimal sensor placements while
maintaining computational efficiency of the CSCP algorithm. Distance-based
sensor reconfiguration costs are introduced in a modified CRMI measure, which
we also show to be submodular. Through numerical simulations, we demonstrate
that the principal advantage of this algorithm is that near-optimal
low-variance paths are achieved using far fewer sensor measurements as compared
to a standard sensor placement method.
\let\thefootnote\relax\footnotetext{
	Preliminary results from this work were presented at 
	the 2025 AIAA SciTech Forum and Exposition, 
	January 6 - 10, Orlando, FL, USA. AIAA paper number 2025-2069.
}
\end{abstract}

\def\figpath{.}

\onehalfspacing
\section*{Nomenclature}

\begin{longtable}{p{0.08\columnwidth} p{0.4\columnwidth} 
		| p{0.08\columnwidth} p{0.35\columnwidth}}
		\toprule
		\textbf{Symbol} & \textbf{Meaning} &
		\textbf{Symbol} & \textbf{Meaning} \\
		\midrule
		
		CSCP & {Coupled sensor configuration and p\ppl} &
		CRMI & {Context-relevant mutual information} \\
				
		$\nParameter$ & Threat state dimension &
		$\nSensor$ & Number of sensors \\
		
		$\mathcal{W}$ & Compact 2D workspace & 
		$\nGridPt$ & Number of grid points \\
		
		$\gridPath$ & Path in grid topological graph & $\delta$ & Grid spacing \\
		
		$\xState$ & Cartesian position coordinates & 
		$\threat(\xState,t)$ & Threat intensity \\
		
		$\boldsymbol\Phi(\xState), $ & Spatial basis function & 
		$\overline{\xState}, a_n$ & Constants used in spatial basis functions \\
		
		$\paramVec(t), \widehat{\paramVec}(t)$ & True and estimated threat states &
		$J(\gridPath),\widehat{J}(\gridPath)$ & True and estimated path cost \\
		 
		$\zMeas$ & Measurement & $\HMeas$ & Measurement model \\
		
		$\boldsymbol\omega, \QProcCovar $ & Process noise and error covariance & 
		$\boldsymbol\eta, \RMeasCovar $ & Measurement noise and error covariance \\
		
		$\qConfig$ & Sensor configuration & $\termThreshold $ & CSCP termination threshold \\ 
		
		\multicolumn{4}{l}{$\PEECovar_{\paramVec\paramVec}, \PEECovar_{\paramVec\zMeas}, 
		\PEECovar_{\zMeas\zMeas}, \PEECovar_{J J}, \PEECovar_{J \zMeas}$ \qquad
		 Various estimation error covariances and cross-covariances} \\

		$\mathcal{R}_n\msup{sup}$ & Region of significant support & 
		$\mathcal{K}$ & Path-relevant set \\ 
		
		$I(\paramVec;\zMeas(\qConfig))$ & Standard mutual information (SMI) &
		$I({J};\zMeas(\qConfig))$ & CRMI \\
		
		$I\msup{mod}(\qConfig)$ & Modified CRMI &
		$\entropy$ & Entropy \\
		
		%
		
		\bottomrule
\end{longtable}

\renewcommand{\revnew}[1]{#1}
\renewcommand{\hl}[1]{#1}

\doublespacing
\section{Introduction}
We consider scenarios where a mobile agent navigating in an unknown environment
can leverage measurements collected by a network of spatially distributed
sensors. The unknown environment may include various adverse attributes, which
we abstractly represent by a spatiotemporally-varying scalar field and refer to
as the \mydef{threat field}. The threat field represents unfavorable areas such
as those associated with various natural or artificial phenomena, such as
wildfires, harmful gases in the atmosphere, or the perceived risk of adversarial
attack.

\revnew{
\figf{fig:Flood_map} provides a motivating example of a time-varying
flood map based on data collected during Hurricane Harvey at spatially
distributed gauge stations over 15-minute time intervals~\cite{harvey-data}.
Initially, there is low water discharge with no signs of flooding
in \fig{fig:Flood_map}(a). Significant flooding is visible in
\figf{fig:Flood_map}(c) at $t = 1500$ minutes, which subsequently
recedes. The changing flood levels may be considered as a
spatiotemporally-varying ``threat'' to which exposure of, say, an
emergency first-response vehicle, should be minimized.
}

We address the problem of p\ppl\ with minimum threat exposure in such an
environment. Because the environment is unknown, an important task is to place
the sensors in appropriate locations, which is called the \mydef{sensor
	placement} problem, or more generally, the sensor \mydef{configuration} problem.
If we had at our disposal an abundance of sensors and computational resources to
process large amounts of sensor data, then the placement / configuration problem
would be trivial. We would simply place sensors to ensure maximum area coverage.

\def\thisfigwidth{0.46\columnwidth}
\begin{figure}[t]
	\centering
	\subfigure[$t=500$ min.]{
		\includegraphics[width=\thisfigwidth]{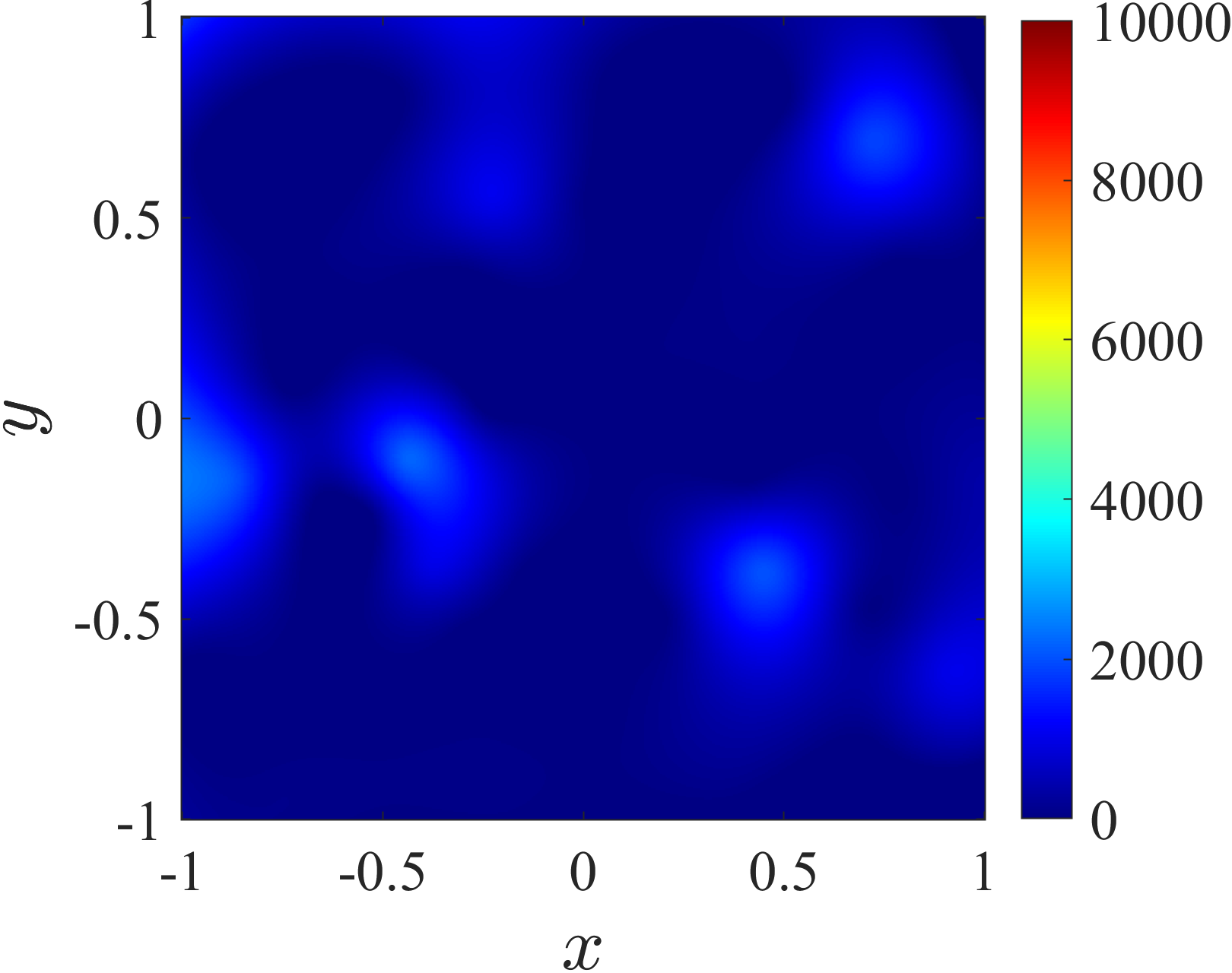}
	}
	\subfigure[$t=1000$ min.]{
		\includegraphics[width=\thisfigwidth]{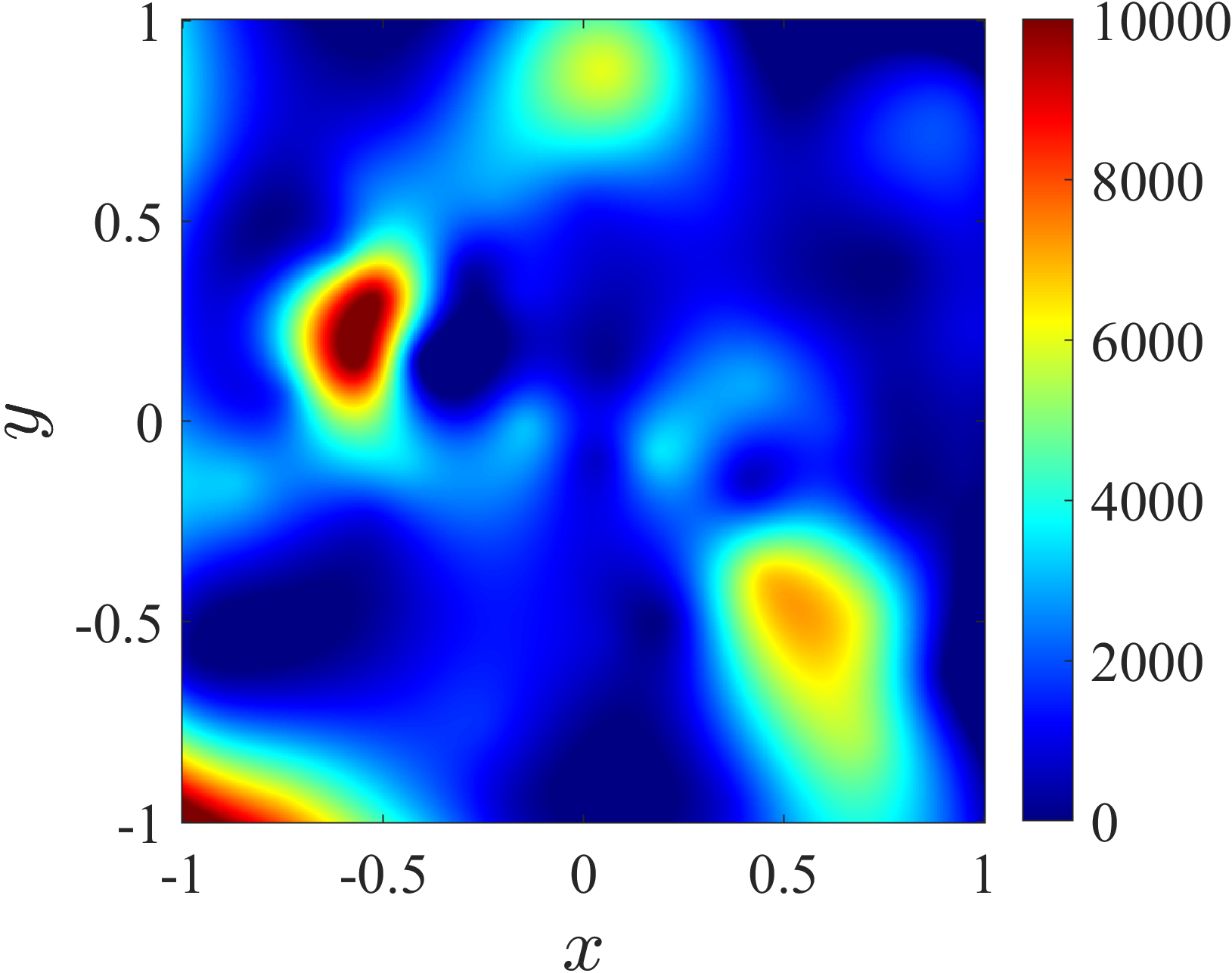}
	}
	\subfigure[$t=1500$ min.]{
		\includegraphics[width=\thisfigwidth]{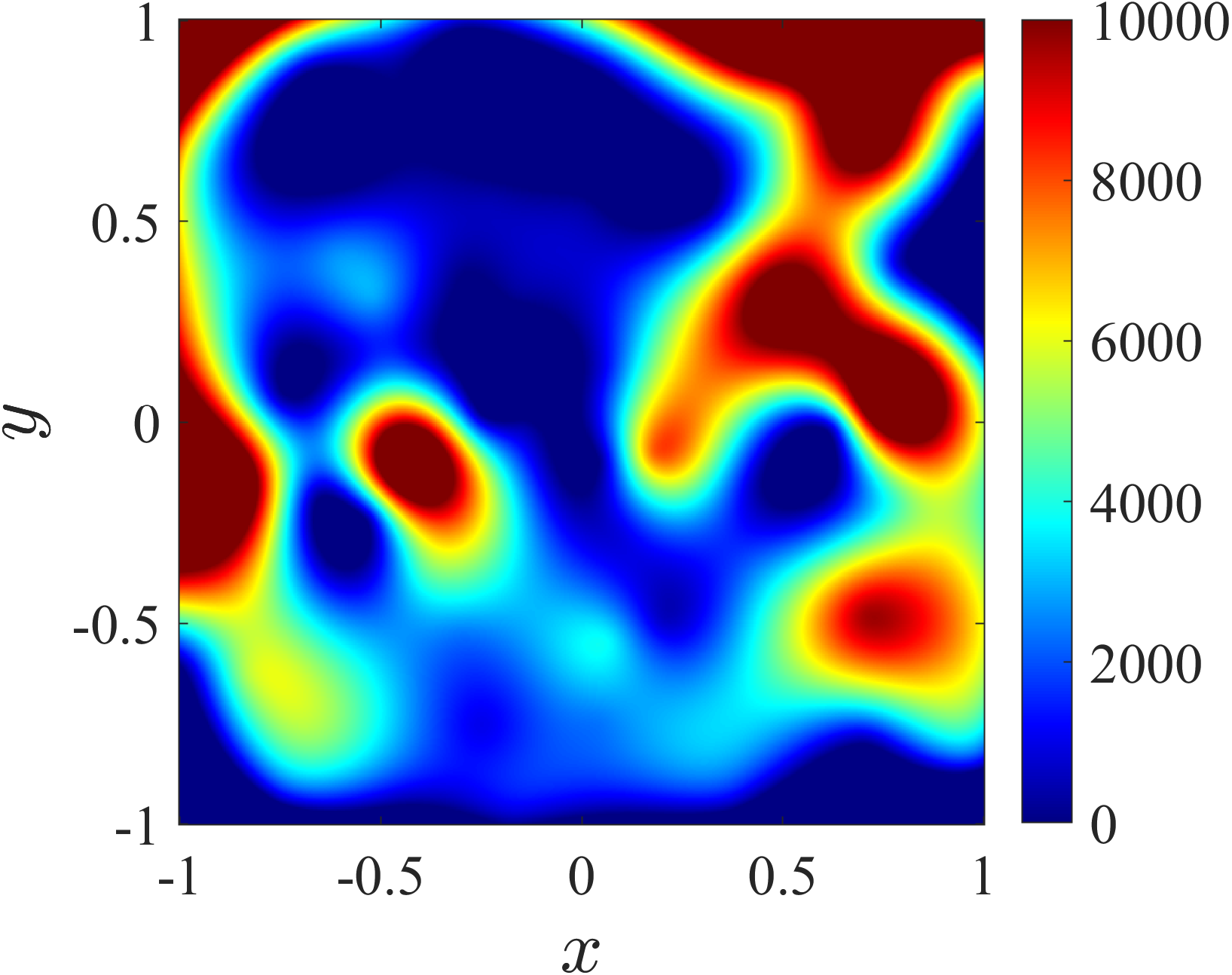}
	}
	\subfigure[$t=2000$ min.]{
		\includegraphics[width=\thisfigwidth]{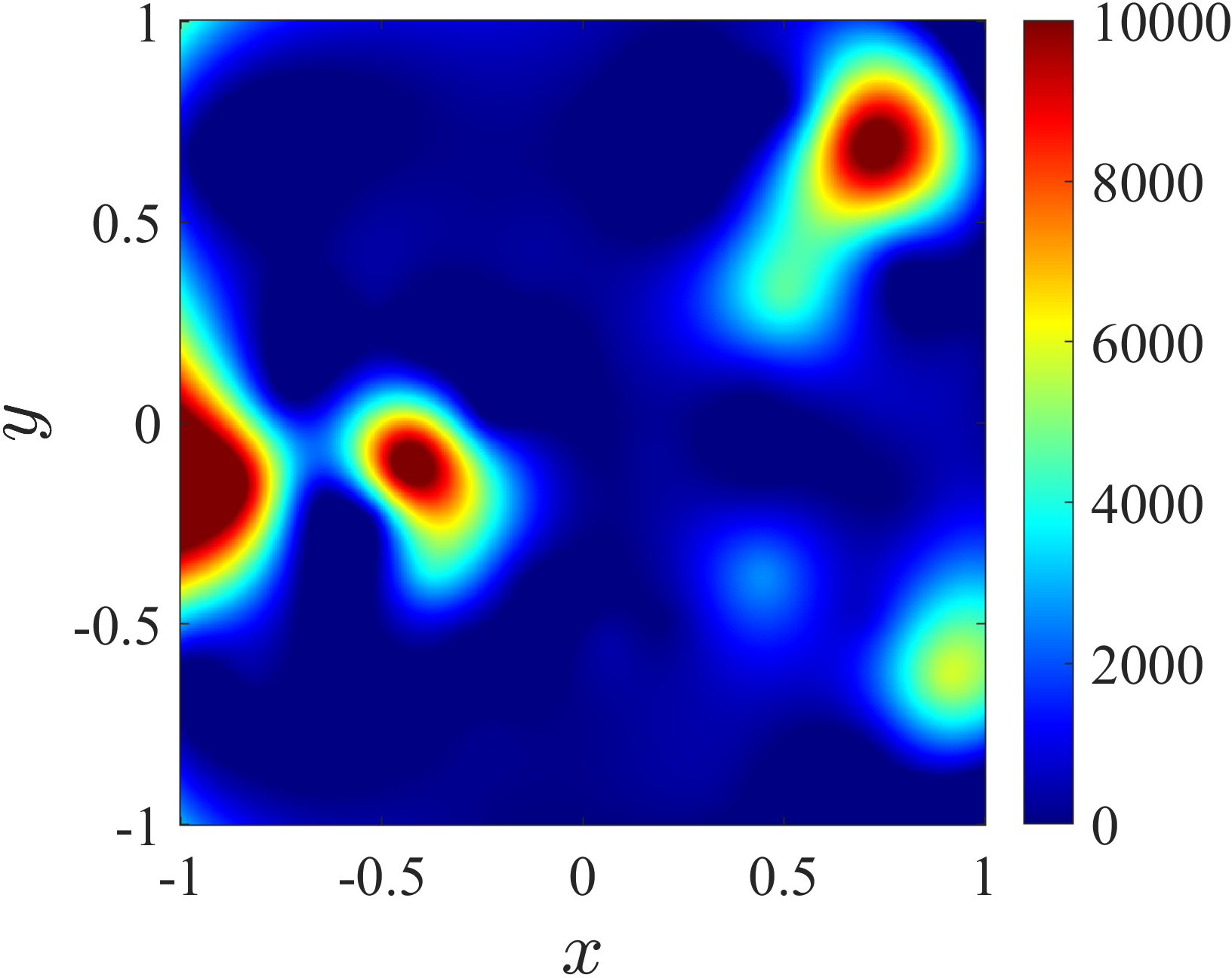}
	}
	\caption{Visualization of the flood map at different times (minutes).}
	\label{fig:Flood_map}
\end{figure}

In practical applications, however, sensor networks may be constrained by size
as well as energy usage. \revnew{In the flood evolution example, a sensor
	network of unmanned aerial vehicles (UAVs) may be depoyed for real-time
	surveillance of flooding. } Due to cost and battery limitations, it may not be
possible to achieve full area coverage quickly enough to inform the actions of a
ground robot to safely navigate the environment. This situation exemplifies the
broader problem of p\ppl\ with a \emph{minimal} number of sensor measurements,
and in turn, highlights the need for optimal sensor configuration \emph{in the
	context} of p\ppl. This problem lies at the intersection of several research
areas including estimation, p\ppl, and sensor placement, which we briefly review
next. We note that the problem of interest here is quite different from the
simultaneous localization and mapping (SLAM) problem, where the p\ppl\ and
sensing is coupled due to the assumption of fully onboard sensing. By contrast,
we consider a distributed sensor network separate from the mobile agent.

Of these areas, perhaps estimation is the most mature \cite{Lewis2017optimal}.
The literature on estimation involves different techniques including Kalman
filter, maximum likelihood estimator \cite{Lewis2017optimal}, and Bayesian
filter \cite{thrun2006}. Application of the extended Kalman filter (EKF), the
unscented Kalman filter (UKF) \cite{Julier2004}, or the particle filter
\cite{doucet2009tutorial} is common for nonlinear dynamical systems.
\revnew{Several data-driven estimation techniques have been employed for
spatiotemporal modeling of hazardous regions, particularly in robotics,
environmental monitoring, and surveillance. These include a supervised learning
approach using Gaussian process regression for threat field estimation
\cite{Laurent2023}, a statistical generalized additive model for landslide
hazard estimation \cite{FANG2024} and  and a combined approach using Bayesian
inference and random forest for the spatial prediction of wildfires
\cite{cisneros2023combined}.}

Path- and m\mpl\ are similarly mature areas of research. Generally, p\ppl\ under
uncertainty involves finding paths that minimize the expected cost. Classical
approaches to path-planning include cell decomposition, probabilistic roadmaps,
and artificial potential field techniques \cite{LaValle2006,Patle2019}.
Dijkstra's algorithm, $\textrm{A}^{\ast}$, and their variants are
branch-and-bound optimization algorithms that leverage heuristics to effectively
steer the path search towards the goal. While classical path-planning methods
are powerful, they are inherently limited by the accuracy of the environment’s
available information. An accurate representation of the environment is
difficult if the environment’s states or dynamics are unknown. Modern approaches
to path planning leverage advanced methodologies such as adaptive informative
path-planning~\cite{popovic2024learning}, coverage
path-planning~\cite{chen2021clustering} and informed sampling-based path
planning~\cite{chintam2024informed}. More recently, learning based techniques,
particularly deep reinforcement learning
\cite{Ruckin2022,qin2023deep,wen2024drl,xue2024bidirectional} and fuzzy logic
\cite{kamil2022} are reported for addressing environmental uncertainty.
\revnew{In risk-aware path planning, the objective is not only to find
a feasible or optimal path but also to minimize exposure to uncertain or
hazardous regions \cite{Cai2024}. Recent risk-aware path planning techniques
integrate reinforcement learning with failure mode and effect analysis to
ensure safe and complete coverage in hazardous environments
\cite{riskaware2025}.}

Different sensor placement approaches have been employed depending on the type
of application and parameters that need to be measured. Greedy approaches based
on information-based metrics are presented in \cite{Ranieri2014,Kreucher2003,Soderlund2019}. Machine learning-based sensor placement techniques are reported for
efficient sensing with a minimal number of sensors and measurements as possible
\cite{Kasper2015,Wang2020}. Information theory-based sensor placement techniques
utilize performance metrics such as the Fisher information matrix (FIM)
\cite{Kangsheng2006}, entropy \cite{Wang2004}, Kullback-Leibler (KL) divergence
\citep{Blasch2010}, mutual information \cite{Krause2008}, and frame potential
\cite{Ranieri2014} to maximize the amount of valuable information gathered
from the surrounding environment. Similarly, \cite{robbiano2021} utilizes two
information measures, one associated with mutual information based on objection
detection, and another with mutual information based on classification of the
detected objects. With all these performance metrics, the intention is to
maximally reduce some quantification of the uncertainty. More recently, a sensing and path-planning method based on reinforcement learning has been reported, which evaluates performance using a technique called proximal policy optimization (PPO) \cite{hoffmann2020}. In robotics and aerospace applications, \emph{active sensing} plays a
crucial role in linking perception and action, enabling systems to gather
meaningful information that supports intelligent decision-making. Some examples
of active sensing include information-driven or cooperative active
sensing~\cite{jang2020multi, park2020cooperative,la2014cooperative},
uncertainty-aware active sensing~\cite{macdonald2019active}, and active sensing
using machine learning~\cite{li2021attention,wu2020achieving}.

The problem of minimizing sensor reconfiguration costs, such as the distance
traveled by mobile sensors, is commonly studied in the mobile sensor network
literature, but less so in the context of sensor placement. Reconfiguration
becomes an important issue when multiple iterations of sensor configuration and
estimation are conducted, which in turn may be necessary when the number of
sensors is small. Some examples include consideration of reconfiguration cost of
the sensor network topology~\cite{leong2014network}, or the total energy
consumption of the sensor
network~\cite{ramachandran2015measuring,grichi2017new}.

In this paper we consider the problem of optimal sensor placement coupled with
path-planning in an unknown dynamic environment. Specifically, we are interested in
sensor placement to collect information of most relevance to the p\ppl\ problem.
The objective is to find a near-optimal path with high confidence, i.e., low variance
in the path cost, with a minimal number of sensor measurements. We aim to compare such
a coupled sensor configuration and p\ppl\ (CSCP) method against decoupled methods, where
sensor configuration is achieved by optimizing a metric that does not consider the p\ppl\ problem in any way.
This is a relatively new research problem. Prior works by the second author and
co-workers address this problem for static (i.e., time-invariant) environments.
A heuristic task-driven sensor placement approach called the interactive
planning and sensing (IPAS) for static environments is reported in
\cite{Cooper2019}. The IPAS method outperforms several decoupled sensor
placement methods in terms of the total number of measurements needed to achieve
near-optimal paths. Sensor configuration for location and field-of-view is
reported in \cite{Laurent2023}, also for static fields. Sensor placement for
multi-agent p\ppl\ based on entropy reduction is presented in \cite{fang2021}.

The novelty of this work is that we consider a time-varying threat field and
provide a new sensor placement method. Specifically, we develop the so-called
context-relevant mutual information (CRMI), which quantifies the amount of
information in configuration-dependent sensor measurements in the context of
reducing uncertainty in path cost, rather than the environment state estimation
error (as a decoupled method typically does). We develop an iterative algorithm
for CSCP. At each iteration, a threat estimate is first computed using sensor
measurements. Next, a p\ppl\ algorithm finds a path of minimum expected threat.
Next, optimal sensor placements are computed to maximize the path-dependent
CRMI, and the iterations repeat. We compare this CSCP-CRMI method to a decoupled
method that finds optimal sensor placement by maximizing the ``standard'' mutual
information (SMI). The metric of comparison is based on the number of
measurements needed to achieve a path cost with variance no greater than a
user-specified threshold. The results of this comparison show that the CSCP-CRMI
method significantly outperforms the decoupled method. We further extend the
CSCP method to incorporate sensor reconfiguration costs into the cost function.
The objective is to collectively maximize the path-dependent CRMI and minimize
sensor movement. A comparison is performed between the results of the CSCP
methods, one ignoring and the other incorporating the sensor reconfiguration
cost.

Preliminary results from this work were recently discussed in conference papers
\cite{poudel2024coupled,poudel2025reconfiguration}. In this paper we further
extend the conference paper works by introducing an approximation algorithm based on
the greedy placement of sensors. Additionally, we show the submodularity
property of both the CRMI and the modified CRMI, and demonstrate that the greedy
placement of sensors guarantees a near-optimal solution. A proof of convergence
of the CSCP algorithm is also discussed.

The rest of the paper is organized as follows. In~\scn{sec-problem}, we
introduce the elements of the problem. In \scn{sec-cscp}, we present and analyze
the new CRMI measure and the CSCP-CRMI algorithm, followed by a discussion on
reconfiguration costs and greedy optimization for near-optimal sensor placement.
We present numerical simulation results in \scn{sec-results} and conclude the
paper in \scn{sec-conclusions}.

\section{Problem Formulation}
\label{sec-problem}

Let $\real, \nat$ denote the sets of real and natural numbers, respectively. We
denote by $\intset{N}$ the set \{1, 2, \ldots, $N$\}, and by $\eye{N}$ the
identity matrix of size $N$, for any $N\in\nat.$

Consider a closed square region denoted by $\mathcal{W}\subset\real[2]$ and
referred as the \textit{workspace}, within which the mobile agent (called the
\mydef{actor}) and a network of mobile sensors operate. In this workspace,
consider a  grid consisting of $\nGridPt$ uniformly spaced points. The
coordinates of these points in a prespecified Cartesian coordinate axis system
are denoted by $\xState_{i} = (x_i,y_i)$, for each $i\in{\nGridPt}$. The distance between
the adjacent grid points is denoted by $\delta$. The mobile agent traverses grid
points according to the ``4-way adjacency rule", such that the adjacent points
are top, down, left, and right. Furthermore, we assume a constant speed such
that the actor's transitions to adjacent grid points occurs in a constant time
step $\timeStep.$

We formulate the path planning problem for the actor as a graph search problem
on a graph, $\mathcal{G} = (V,E)$ with $V = \intset{\nGridPt}$ such that each
vertex in $V$ is uniquely associated with a grid point. The set of edges $E$ in
this graph consists of pairs of grid points that are geometrically adjacent to
each other.

A \mydef{threat field}, denoted as
$c:\mathcal{W}\times\real_{\geq0}\rightarrow\real_{>0}$, is a time-varying
scalar field that takes strictly positive values, indicating regions with higher
intensity that are potentially hazardous and unfavorable. \hl{A path between two
prespecified initial and goal vertices, $i_{s}, i_{g}\in V$, is defined as a
finite sequence $\gridPath = \{v_{0},v_{1},\ldots,v_{L}\}$ of successively
adjacent vertices. This sequence starts at the initial vertex $v_{0} = i_{s}$
and ends at the goal vertex $v_{L} = i_{g}$, where $L\in\mathbb{N}$ represents
the number of vertices in the sequence.} The edge transition costs, which account
for the expenses incurred when an actor moves between vertices in a graph, are
determined by a scalar function $g:E\rightarrow\real_{>0}$. This function
assigns a value to each edge in the graph, representing the associated cost or
effort required for traversal and is defined as, 
\begin{equation} g((i,j),t) =
	\threat(\xState_{j},t),\thickspace \textrm{for} \quad 
	i,j\in\intset{\nGridPt},\quad (i,j)\in \edges.
\end{equation} 
The cost $J(\gridPath)\in\real_{>0}$ indicates the total threat exposure for an
actor on its traversal along a path $\gridPath$ and is defined as the sum of
edge transition costs, $J(\gridPath) =\delta\sum_{\ell=1}^{L}g((v_{\ell-1},v_{\ell}),\ell\timeStep)$. The main
problem of interest is to find a path~$\gridPath^{*}$ with minimum cost. Because
the threat field is unknown and time-varying, its estimation is essential. A
network of $\nSensor$ sensors, where $\nSensor \ll \nGridPt,$ can be used to
measure the intensity of threat. \hl{These sensor measurements are denoted
$\zMeas(\xState,t;\qConfig) = \{z_{1}(\xState,t;q_1),
z_{2}(\xState,t;q_2),\ldots, z_{N_{s}}(\xState,t;q_{N_{s}})\}.$} Sensors are
placed at distinct grid points, and the set of these grid points is called the
\textit{sensor configuration}, $\qConfig =
\{q_{1},q_{2},\ldots,q_{N_{s}}\}\subset\intset{\nGridPt}$.

The threat field is considered to be a stochastic quantity with a predictive
model involving uncertainty. Specifically, we consider a finite parametrization
$c(\xState,t) \defeq 1 + \sum_{n=1}^{\nParameter}\theta_{n}(t)\phi_{n}(\xState)
= 1 + \boldsymbol\Phi^{\intercal}(\xState)\paramVec(t)$, with
$\boldsymbol\Phi(\xState)\defeq[\phi_{1}(\xState) \ldots
\phi_{{\nParameter}}(\xState)]^{\intercal}$, and $\phi_{n}(\xState) \defeq \exp(
-(\xState-\overline{\xState}_{n})^{\intercal} (\xState-\overline{\xState}_{n}) /
2a_{n})$ representing the basis functions for each $n\in\intset{\nParameter}$.
\hl{Here, $\nParameter$ denotes the number of parameters} \revnew{(or bases),}
\hl{
representing the} \mydef{threat state}, \hl{and $a_n$ and $\overline{\xState}_n$ are
constants. The locations of the basis functions $\boldsymbol\Phi$ are fixed and
remain unchanged throughout the entire sensing and path planning process.
Although the actor has prior knowledge of the functional forms of these
functions, the threat parameter $\paramVec(t) \defeq
[\theta_{1}(t)\ldots\theta_{{\nParameter}}(t)]^{\intercal}$ is unknown and must
be estimated.} The values of the constants $a_{n}\in\real_{>0}$ and
$\overline{\xState}_{n}\in \mathcal{W}$ are prespecified and chosen in such a
manner that the combined interiors of the significant support regions cover
$\wsp.$ The parameter $a_{n}$ is chosen to minimize the overlap between the
basis functions, ensuring better distinction and independence among them
\cite{Cooper2019}.
\revnew{In general, the basis functions should be chosen to 
approximate threat field data, e.g., the flood map data in
\fig{fig:Flood_map}.}

%

The temporal evolution of the threat is modeled by $\dot{\paramVec}(t) =
A_{c}\paramVec(t) + \boldsymbol\omega(t)$, where $\boldsymbol\omega(t) \sim
\mathcal{N}(0, \QProcCovar_{c})$ is white process noise with $\QProcCovar_{c}
\defeq \sigma_{P}^2 \eye{\nParameter}$. Such a model may be available either from
an underlying physical model of the threat, or it may be derived from data, or a
combination of both. As an illustrative example, the solution to a heat
diffusion equation, $\frac{\partial{c}}{\partial{t}} =
\alpha(\frac{\partial^{2}{c}}{\partial{x^{2}}} +
\frac{\partial^{2}{c}}{\partial{y^{2}}})$ can be approximated by $c(\xState,t) =
1 + \boldsymbol\Phi^{\intercal}(\xState)\paramVec(t)$ such that $\paramVec(t)$
satisfies $\dot{\paramVec}(t) =
\alpha\frac{\boldsymbol\Phi^{\intercal}}{|\boldsymbol\Phi|^{2}} \nabla^{2}
\boldsymbol\Phi\paramVec(t)$, where $A_{c} \defeq \alpha
\frac{\boldsymbol\Phi^{\intercal}}{|\boldsymbol\Phi|^{2}}
\nabla^{2}\boldsymbol\Phi$.

\revnew{
We restrict the scope of this paper to linear threat field 
dynamics for the purpose of establishing a proof of convergence of the
proposed method. However, the method itself is not limited
to linear dynamics. We implement the Unscented Kalman 
Filter in our current threat field estimator in anticipation of
this future extension. Our current CSCP implementation will 
work as-is for nonlinear threat dynamics, but the CRMI calculations
will be approximations. Other works in the literature, e.g., \cite{Adurthi2020}
have studied MI calculations for nonlinear systems, which we can
easily adapt to CRMI in the future.
}

Discretization in time of this model is easily accomplished by the series
expansion $A \defeq \eye{\nParameter} + A_{c}\timeStep + \frac{(A_{c})^{2}
(\timeStep)^{2}}{2!} + \ldots $ and $\QProcCovar \defeq \QProcCovar_{c}\timeStep
+ \frac{(A_{c} \QProcCovar_{c}  + \QProcCovar_{c}A_{c}^{\intercal})
(\timeStep)^{2}}{2!} + \ldots$ terminated at a desired 
order~\citep{Lewis2017optimal}. The discretized system dynamics are:
\begin{align}
	\paramVec_{k} &= A \paramVec_{k-1} + \boldsymbol\omega_{k-1},
\end{align}
\noindent
where $\boldsymbol\omega_{k-1} \sim \mathcal{N}(0, \QProcCovar)$ for each $k \in
\nat.$

The measurements obtained from each sensor are modeled as
\begin{align}
	\zMeas_{k} &\defeq c(\xState_{{q}_{k}},t) + \boldsymbol\eta_{k} = 
	\HMeas_{k}(\qConfig)\paramVec_{k} + \boldsymbol\eta_{k}, 
	\label{eq-sensor-model} \\
	\mbox{where \quad} \HMeas_{k}(\qConfig) &\defeq 
	\left[\boldsymbol\Phi(\xState_{\qConfig_{k,1}}) 
	\quad \boldsymbol\Phi(\xState_{\qConfig_{k,2}}) 
	\quad \ldots \quad
	\boldsymbol\Phi(\xState_{\qConfig_{k,\nSensor}})\right]^{\intercal},
	\label{eq-sensor-Hmatrix}
\end{align}
and $\boldsymbol\eta_{k} \sim 
\mathcal{N}(0, \RMeasCovar)$ is zero mean measurement noise 
with covariance $\RMeasCovar \succ 0$.

We generate stochastic estimates of the threat state with mean value 
$\widehat{\paramVec}_k$ and estimation error covariance $\PEECovar.$
For any path, $\gridPath = 
\{v_{0},v_{1},\ldots,v_{L}\}$ in $\mathcal{G}$, the cost of the path is 

\begin{align}
	J(\gridPath) \defeq \delta{\sum_{\ell=1}^{L}}\threat(\xState_{\gridPath_{\ell}},t) = \delta \left(L + 
	{\sum_{\ell=1}^{L}} \boldsymbol\Phi^{\intercal}(\xState_{\gridPath_{\ell}})\paramVec_{\ell}\right).
	\label{eq-path-cost-defn}
\end{align}
Here, $\paramVec_\ell$ represent the threats at different iterations (or time
steps) equivalent to the iterations required for an actor to follow the
path~$\gridPath$. The cost $J$ becomes a random variable with distribution
dependent on $\paramVec.$ \revnew{If $\paramVec$ is Gaussian, then $J$ is also
Gaussian because it is linearly dependent on $\paramVec.$} Let
$\mathcal{V}\subset\mathcal{W}$ be the set of grid point locations in the
workspace associated with each $v_\ell$ in the path $\gridPath$. \hl {Note that while $\theta_{n}(t)$ and $\phi_{n}$ represent the total number of threat parameters and the basis functions involved in the threat field generation, $\theta_{m}(t)$ and $\phi_{m}$ denote the subset of threat parameters and associated basis functions that lie within the path and are relevant for computing the path cost $J(\gridPath)$.} The set $\mathcal{K} \defeq \left\{m \in
\intset{\nParameter} :\mathcal{V}\cap \mathcal{R}_m\msup{sup} \neq \emptyset
\right\}$ is defined as the path-relevant set of threat states for any path
$\gridPath$. In other words, $\mathcal{K} $ consists of the set of indices $m$,
representing the threat parameters $\theta_m$ or their corresponding basis
functions $\phi_m$, such that the path lies within the total region of
significant support $\mathcal{R}_m\msup{sup}$ defined by $\phi_m$. 


\hl{An important characteristic associated with the convergence of the path-planning
algorithm is the} \mydef{\hl{risk}} \hl{of the path denoted by $\rho(\gridPath)$.}
\revnew{For a Gaussian $J$, the risk of the path $\gridPath$ is defined as
$\rho(\gridPath) \defeq   \widehat{J}(\gridPath) +
\sqrt{\textrm{Var}[{J}(\gridPath)]}$ \citep{rockafellar2000optimization}.} Since
$\nSensor \ll \nGridPt$, it is not possible to obtain good estimates with only
one set of measurements, and it is required to take measurements repeatedly over
a finite number of iterations.

\begin{prob}
	For a prespecified termination threshold, $\termThreshold > 0$ and
	some finite iterations $k = 0, 1,\ldots, M$, find a sequence of sensor
	configurations $\qConfig^{*}_{k}$ and a path $\gridPath^{*}$ with
	minimum expected cost $\widehat{J}(\gridPath^{*})$ and such that
	$\textrm{Var}[{J}(\gridPath^{*})]\leq\termThreshold.$
	\label{problem_pp}
\end{prob}
Alternatively, one may consider a requirement that the path risk 
$\rho(\gridPath^{*})$ be lower than a prespecified threshold.

\section{Coupled Sensing and Planning}
\label{sec-cscp}
\mydef{Coupled sensor configuration and path-planning} (CSCP) is our proposed
iterative approach to solve Problem~1. At each iteration, a sensor configuration
is determined, and measurements of the threat field are collected. The optimal
sensor configuration is found by maximizing an information measure that we call
\mydef{context-relevant mutual information} (CRMI). These sensor measurements
are used to update the threat field estimates in an estimator. \revnew{For
	future compatibility with nonlinear threat dynamics, an unscented Kalman filter
	(UKF) is used for the estimating the parameters $\paramVec.$ The reader
	interested is referred to Appendix~B for a brief description of the UKF.} Next,
the path plan is modified based on the new threat field estimate, and this
process continues until the path cost variance is reduced below a prespecified
threshold~$\termThreshold$.

\revnew{In this paper, we assume that the actor is a \emph{planning agent} that does not
move before the CSCP algorithm is complete. In other words, we may think of this
implementation of CSCP as occurring in a virtual world for planning the future
movements of the actor and sensors. The reason for this restriction in scope is
to be able to provide a proof of convergence of the CSCP method as a baseline
for future applications.
In our recent work \cite{PoudelDesRochesCowlagi-2025ACC}, we implement a version
of CSCP where the actor moves simultaneously with the CSCP iterations, and the
sensors need a finite non-zero duration to relocate. The simulations in
\cite{PoudelDesRochesCowlagi-2025ACC} demonstrate a successful implementation
and provide sensor resource benefits similar to those described in the present
manuscript, pending a formal proof of convergence.
}

In what follows, we provide details of this
iterative process, analysis, and an illustrative example.

\subsection{Context-Relevant Mutual Information (CRMI)}
For any time step $k$, the \mydef{mutual information (MI)} between the state
$\paramVec_{k}$ and measurement $\zMeas_{k}$ random variables is defined as
\citep{Thomas1991} \[I(\paramVec_{k};\zMeas_{k})\defeq 
\int\int\textit{p}(\paramVec_{k},\zMeas_{k}) \log \left(\frac
{\textit{p}(\paramVec_{k},\zMeas_{k})}
{\textit{p}(\paramVec_{k})\textit{p}(\zMeas_{k})}\right)\,d\paramVec_{k}\,d\zMeas_{k},\] where 
$\textit{p}(\paramVec_{k})$, $\textit{p}(\zMeas_{k})$ and 
$\textit{p}(\paramVec_{k},\zMeas_{k})$ represent the probability density functions (PDFs) of 
state, measurement and a joint PDF of state and measurement, respectively. The joint PDF 
$\textit{p}(\paramVec_{k},\zMeas_{k})$ is a multivariate normal distribution with mean 
$(\widehat{\paramVec}_{k|k-1}, \widehat{\zMeas}_{k})$ and covariance
\begin{equation*}
\begin{bmatrix} {\PEECovar}_{\paramVec\paramVec_{k|k-1}} & \PEECovar_{{\paramVec \zMeas}_{k|k-1}} 
\\ 
\PEECovar^{\intercal}_{{\paramVec \zMeas}_{k|k-1}} & \PEECovar_{{\zMeas\zMeas}_{k|k-1}} 
\end{bmatrix}.
\end{equation*}
Here, $\PEECovar_{\paramVec\paramVec_{k|k-1}}$ is obtained from the UKF
algorithm. The covariance of the measurement random vector
$\PEECovar_{{\zMeas\zMeas}_{k|k-1}}$  and cross covariance between the state and
measurement random vectors $\PEECovar_{{\paramVec \zMeas}_{k|k-1}}$ depend on
the sensor configuration $\qConfig$. At each grid point, these covariances are
determined as { 
\begin{align}
	\PEECovar_{{\zMeas\zMeas}_{k|k-1}}  
	&\defeq \mathbb{E}\left[\left(\vec{z}_k - \widehat{\vec{z}}_k\right)\left(\vec{z}_k - \widehat{\vec{z}}_k\right)^{\intercal}\right] \nonumber \\
	&= \mathbb{E}\left[\left(\HMeas_k(\qConfig)\left(\paramVec_k - \widehat{\paramVec}_{k|k-1}\right) + \left(\boldsymbol\eta_k - \widehat{\boldsymbol\eta}_k\right)\right)\left(\HMeas_k(\qConfig)\left(\paramVec_k - \widehat{\paramVec}_{k|k-1}\right) + \left(\boldsymbol\eta_k - \widehat{\boldsymbol\eta}_k\right)\right)^{\intercal}\right] \nonumber \\
	&= \HMeas_k(\qConfig)\mathbb{E}\left[\left(\paramVec_k - \widehat{\paramVec}_{k|k-1}\right)\left(\paramVec_k - \widehat{\paramVec}_{k|k-1}\right)^{\intercal}\right]\HMeas_k^{\intercal}(\qConfig) + \mathbb{E}\left[\left(\boldsymbol\eta_k - \widehat{\boldsymbol\eta}_k\right)\left(\boldsymbol\eta_k - \widehat{\boldsymbol\eta}_k\right)^{\intercal}\right] \nonumber \\
	&= \HMeas_{k}(\qConfig)\PEECovar_{\paramVec\paramVec_{k|k-1}} \HMeas_{k}^{\intercal}(\qConfig) + \RMeasCovar_{k}, \\
	\PEECovar_{{\paramVec\zMeas}_{k|k-1}}  
	&\defeq \mathbb{E}\left[\left(\paramVec_k - \widehat{\paramVec}_{k|k-1}\right)\left(\vec{z}_k - \widehat{\vec{z}}_k\right)^{\intercal}\right] \nonumber \\
	&\defeq \mathbb{E}\left[\left(\paramVec_k - \widehat{\paramVec}_{k|k-1}\right)\left(\HMeas_k(\qConfig)\left(\paramVec_k - \widehat{\paramVec}_{k|k-1}\right) + \left(\boldsymbol\eta_k - \widehat{\boldsymbol\eta}_k\right)\right)^{\intercal}\right] \nonumber \\
	&= \mathbb{E}\left[\left(\paramVec_k - \widehat{\paramVec}_{k|k-1}\right)\left(\paramVec_k - \widehat{\paramVec}_{k|k-1}\right)^{\intercal}\right]\HMeas_k^{\intercal}(\qConfig) + \mathbb{E}\left[\left(\paramVec_k - \widehat{\paramVec}_{k|k-1}\right)\left(\boldsymbol\eta_k - \widehat{\boldsymbol\eta}_k\right)^{\intercal}\right] \nonumber \\
	&= \PEECovar_{\paramVec\paramVec_{k|k-1}} \HMeas_{k}^{\intercal}(\qConfig).
\end{align}}
The mutual information between the state $\paramVec_{k}$
and measurement $\zMeas_{k}(\qConfig)$ is then written as \citep{Adurthi2020}:
 \begin{align}
I(\paramVec_{k};\zMeas_{k}(\qConfig)) &= 
\frac{1}{2}\log\left(\frac{|\PEECovar_{\paramVec\paramVec_{k|k-1}}|}{|\PEECovar_{\paramVec\paramVec_{k|k-1}}
- \PEECovar_{{\paramVec \zMeas}_{k|k-1}} \PEECovar^{-1}_{{\zMeas\zMeas}_{k|k-1}} 
\PEECovar^{\intercal}_{{\paramVec \zMeas}_{k|k-1}}|}\right).
\label{eq-mi}
\end{align}

The MI $I(\paramVec_{k};\zMeas_{k}(\qConfig))$ depends on the sensor
configuration $\qConfig,$ and in that sense it quantifies ``informativeness'' of
a configuration~$\qConfig.$ A canonical method of finding the optimal sensor
configuration is to maximize the MI ($I$) over~$\qConfig.$ Note, however, that
this MI $I(\paramVec_{k};\zMeas_{k}(\qConfig))$ is entirely decoupled from the
path-planning problem, in that it is no way contextualized by the optimal path
to be found. We introduce a new information measure called the context-relevant
mutual information (CRMI), which couples the sensor configuration (placement)
and path-planning problems.

We define CRMI as the mutual information between the path cost and the
measurements. CRMI may be thought of as a formalization of the vague notion
\emph{``sensors `near' the planned path are more informative than those farther
	away.''} To this end, consider that, for any path $\gridPath$, the expected
\revnew{(mean)} cost is $\widehat{J}(\gridPath) \defeq \delta\left(L +
\sum_{\ell=1}^{L} \boldsymbol\Phi^{\intercal}(\xState_{\gridPath_{\ell}})
\widehat{\paramVec}_\ell\right).$ The joint PDF $\textit{p}(J_{k},\zMeas_{k})$
between the path cost and measurement variables is
\begin{align*}
	\textit{p}(J_{k},\zMeas_{k}) &= \mathcal{N}\left(\begin{bmatrix} J_{k}\\ \zMeas_{k} 
	\end{bmatrix} : 
	\begin{bmatrix} \widehat{J}_{k|k-1}\\ \widehat{\zMeas}_{k} \end{bmatrix}
	, \begin{bmatrix} {\PEECovar}_{J J_{k|k-1}} & \PEECovar_{{J \zMeas}_{k|k-1}} \\ 
	\PEECovar^{\intercal}_{{J \zMeas}_{k|k-1}} & \PEECovar_{{\zMeas\zMeas}_{k|k-1}} 
	\end{bmatrix}\right).
\end{align*}

\revnew{
The variance of the path cost is
\begin{align*}
	{\PEECovar}_{J J_{k|k-1}} &\defeq
	\mathbb{E}\left[\left(J(\gridPath) - \widehat{J}(\gridPath)\right)^{2}\right] = 
	\mathbb{E}\left[\left(\delta\sum_{\ell=1}^{L} 
	\boldsymbol\Phi^{\intercal}(\xState_{\gridPath_{\ell}})
	\left(\paramVec_\ell - \widehat{\paramVec}_\ell\right) \right)^{2}\right].
\end{align*}
Using the formula for square of sums, $ \left(\sum_{i}a_{i}\right)^{2} = \sum_{i}a_{i}^{2} + 2\sum_{i<j}a_{i}a_{j}$
 \begin{align}
 	\PEECovar_{J J_{k|k-1}} &= \delta^{2}\left(\mathbb{E}\left[\sum_{\ell=1}^{L}\left( 
 	\boldsymbol\Phi^{\intercal}(\xState_{\gridPath_{\ell}})\left(\boldsymbol\Theta_{\ell} - 
 	\widehat{\paramVec}_{\ell}\right) \right)^{2} + 2\sum_{\substack{\ell<m \\ 
 			\ell,m\in\intset{L}}}^{L}\left(\boldsymbol\Phi^{\intercal}(\xState_{\gridPath_{\ell}})
 	\left(\paramVec_{\ell} - 
 	\widehat{\paramVec}_{\ell}\right)\boldsymbol\Phi^{\intercal}(\vec{x}_{\gridPath_{m}})
 	\left(\paramVec_{m} - 
 	\widehat{\paramVec}_{m}\right)
 	\right) \right]\right) \nonumber \\
 	&= \delta^{2}\left(\mathbb{E}\left[\sum_{\ell=1}^{L}\left( 
 	\boldsymbol\Phi^{\intercal}(\xState_{\gridPath_{\ell}})\left(\paramVec_{\ell} - 
 	\widehat{\paramVec}_{\ell}\right) \right)^{2} + 2\sum_{\substack{\ell<m \\ 
 			\ell,m\in\intset{L}}}^{L}\left(\boldsymbol\Phi^{\intercal}(\xState_{\gridPath_{\ell}})
 	\left(\boldsymbol\Theta_{\ell} - \widehat{\paramVec}_{\ell}\right)\left(\paramVec_{m} - \widehat{\paramVec}_{m}
 	\right)^{\intercal}\boldsymbol\Phi(\xState_{\gridPath_{m}})\right) \right]\right). \nonumber 
\end{align} 
}
{ 
The first term $\mathbb{E}\left[\sum_{\ell=1}^{L}\left( 
\boldsymbol\Phi^{\intercal}(\xState_{\gridPath_{\ell}})\left(\boldsymbol\Theta_{\ell} - 
\widehat{\boldsymbol\Theta}_{\ell}\right) \right)^{2}\right]$ can be further expressed as:
\begin{align}
	\PEECovar_{J J_{k|k-1}} 
	&= \sum_{\ell=1}^{L} 
	\mathbb{E}\left[
	\boldsymbol\Phi^{\intercal}(\xState_{\gridPath_{\ell}})
	\left(\boldsymbol\Theta_{\ell} - \widehat{\paramVec}_{\ell}\right)
	\left(\boldsymbol\Theta_{\ell} - \widehat{\paramVec}_{\ell}\right)^{\intercal}
	\boldsymbol\Phi(\xState_{\gridPath_{\ell}})
	\right] \nonumber \\
	&= \sum_{\ell=1}^{L} 
	\boldsymbol\Phi^{\intercal}(\xState_{\gridPath_{\ell}})
	\mathbb{E}\left[
	\left(\boldsymbol\Theta_{\ell} - \widehat{\paramVec}_{\ell}\right)
	\left(\boldsymbol\Theta_{\ell} - \widehat{\paramVec}_{\ell}\right)^{\intercal}
	\right]
	\boldsymbol\Phi(\xState_{\gridPath_{\ell}}) = \sum_{\ell=1}^{L} 
	\boldsymbol\Phi^{\intercal}(\xState_{\gridPath_{\ell}})
	P_{k_{\ell}}
	\boldsymbol\Phi(\xState_{\gridPath_{\ell}}). \nonumber
\end{align}
}
Therefore, \begin{align}
	\PEECovar_{J J_{k|k-1}} 
	& = \delta^{2}\sum_{\ell=1}^{L}\left(\boldsymbol\Phi^{\intercal}(\xState_{\gridPath_{\ell}})
         \PEECovar_{k_{\ell}}\boldsymbol\Phi(\xState_{\gridPath_{\ell}}) \right) + 
          2\delta^{2}\sum_{\substack{\ell<m \\ 
		\ell,m\in\intset{L}}}^{L}\left(\boldsymbol\Phi^{\intercal}(\xState_{\gridPath_{\ell}})
         \PEECovar_{k_{\ell m}}\boldsymbol\Phi(\xState_{\gridPath_{m}})\right).
         \label{eq-covar1}
\end{align}

The calculation of $\PEECovar_{J J_{k|k-1}}$ requires the determination of $\boldsymbol\Phi$ 
and the error covariance $\PEECovar$ for every grid point $\gridPath_{l}$ lying on the path. 
$\PEECovar_{k_{\ell}}$ and $\PEECovar_{k_{\ell m}}$ are determined by propagating the UKF prediction 
steps for iterations equivalent to the path length. 
The covariance of the measurement and the cross covariance between the path cost and the 
measurement random vector are formulated as,
{ 
\begin{align}
	\PEECovar_{{J \zMeas}_{k|k-1}} 
	&= \mathbb{E}\left[\left(J(\gridPath) - \widehat{J}(\gridPath)\right)\left(\vec{z}_k - \widehat{\vec{z}}_k\right)^{\intercal}\right] \nonumber \\
	&= \mathbb{E}\left[\left(\delta\sum_{\ell=1}^{L} 
	\boldsymbol\Phi^{\intercal}(\xState_{\gridPath_{\ell}})\left(\paramVec_{\ell} - 
	\widehat{\paramVec}_{\ell}\right) \right)\left(\HMeas_k(\qConfig)\left(\paramVec_k - \widehat{\paramVec}_{k}\right) + \left(\boldsymbol\eta_k - \widehat{\boldsymbol\eta}_k\right)\right)^{\intercal}\right] \nonumber \\
	&= \delta\mathbb{E}\left[\left(\sum_{\ell=1}^{L} 
	\boldsymbol\Phi^{\intercal}(\xState_{\gridPath_{\ell}})\left(\paramVec_{\ell} - 
	\widehat{\paramVec}_{\ell} \right)\left(\paramVec_k - \widehat{\paramVec}_{k}\right)^{\intercal}
	\HMeas_k^{\intercal}(\qConfig)\right) \right] = \delta\sum_{\ell=1}^{L}
	\left(\boldsymbol\Phi^{\intercal}(\xState_{\gridPath_{\ell}})\PEECovar_{k_{\ell}} 
	\right)\HMeas_{k}^{\intercal}(\qConfig). \label{eq-covar2}\\
	\PEECovar_{{\zMeas\zMeas}_{k|k-1}} &= \mathbb{E}\left[\left(\zMeas_k 
	-\widehat{\zMeas_k}\right)\left(\zMeas_k -\widehat{\zMeas_k}\right)^{\intercal}\right] 
	=  
	\HMeas_{k}(\qConfig)\PEECovar_{\paramVec\paramVec_{k|k-1}}\HMeas_{k}^{\intercal}(\qConfig) + \RMeasCovar_{k}. \label{eq-covar3}
\end{align}
}

 Finally, the CRMI is calculated as
 \begin{align}
 I({J}_{k};\zMeas_{k}(\qConfig))=
 \frac{1}{2}\log\left(\frac{|\PEECovar_{JJ_{k|k-1}}|}{|\PEECovar_{JJ_{k|k-1}} - 
 \PEECovar_{{J \zMeas}_{k|k-1}} \PEECovar^{-1}_{{\zMeas\zMeas}_{k|k-1}} \PEECovar^{\intercal}_{{J 
 \zMeas}_{k|k-1}}|}\right).
	\label{eq-crmi}
\end{align}

This definition of the CRMI is the critical step in the proposed coupled sensor placement
and path planning (CSCP) algorithm, described next.
Similar to MI, CRMI is the difference between entropy and conditional entropy given 
sensor measurements, namely:
\begin{align}
	I({J};\zMeas(\qConfig)) = \entropy(J) - \entropy(J \mid \zMeas(\qConfig)),
	\label{eqn-entropy-gain}
\end{align}
where $\entropy(J)$ and $\entropy(J \mid \zMeas(\qConfig))$
denote the entropy of $J$ and its entropy conditional given 
sensor measurements~$\zMeas(\qConfig).$

\subsection{CSCP Algorithm}
The CSCP algorithm described in \algoref{alg-cscp}
initializes with $\widehat{\paramVec}_{0} = \mathbf{0}$ and $\PEECovar_{0}=\chi\eye{\nParameter}$,
where $\chi$ is a large arbitrary number. The initial sensor placement $\qConfig_0$ is arbitrary. At the initial iteration, an optimal path $\gridPath^{*}_{0}$ of minimum expected cost 
$\mathbb{E}\left[J(\gridPath^{*}_{0})\right]$ is calculated. \revnew{This threat estimate 
initialization implies that until a measurement is taken, the algorithm assumes all threat states 
to be zero. As a result, the CSCP planner is ``optimistic'' in that it plans paths through regions 
associated with threat states that were not previously measured or estimated. The CRMI-based sensor 
placement then ensures sensors are placed close to this path, thereby ensuring that the sensors 
explore the entire workspace.} 

The description in \algoref{alg-cscp} is quite general, and its various steps
can be implemented using different methods of the user's choice. At each
iteration $k$, the algorithm calculates the variance
$\textrm{Var}[{J}(\gridPath^{*}_{k})]$ of the cost of the path
$\gridPath^{*}_{k}$ per~\eqnnt{eq-covar1}. The algorithm terminates whenever the
variance of the path cost reduces below a prespecified threshold $\termThreshold
> 0.$ The method of computation of the optimal path $\gridPath^{*}_{k}$ is the
user's choice: for most practical applications, Dijkstra's algorithm (our choice
for implementation) or \astar\ algorithm will suffice.

The optimal sensor configuration in \algline{line-config} can be calculated by
optimizing some measure of information gain. In a \emph{decoupled} approach, we
may optimize the standard MI in \eqnnt{eq-mi}. In the proposed CSCP method, we
optimize the CRMI in \eqnnt{eq-crmi}. The method of optimization is left to the
user, and is not the focus of this paper. For a small number of grid points, we
can determine $\qConfig^{*}_{k}$ by mere enumeration. 
\revnew{In prior works, e.g.,~\cite{Laurent2023}, we have found success in 
implementing evolutionary global optimization methods for sensor
configuration, albeit using a different reward function. It is possible
to apply such methods for CRMI maximization as well.}

With an optimal sensor placement, a new set of measurements is recorded, which
are then used to update the state estimate for $\paramVec.$ Yet again, the
specific method of estimation is the user's choice. We choose the
linearization-free UKF method for future applications to nonlinear threat
dynamics, briefly described in the Appendix. %
This iterative process continues until the termination criteria
$\textrm{Var}[(\widehat{J}(\gridPath)]\leq\termThreshold$ is satisfied.

\begin{algorithm}	
	\begin{algorithmic}[1]
		\STATE Set $k=0, \widehat{\paramVec}_{0} = \mathbf{0},$ and 
		$\PEECovar_{0}=\chi\eye{\nParameter}$;
		\STATE Initialize sensor placement $\qConfig_{0}\subset\intset{\nGridPt}$;
		\STATE Find $\gridPath^{*}_{0}= \arg\min (\widehat{J}_{0}(\gridPath) ) $;
		\WHILE {$\textrm{Var}[({J}(\gridPath^{*}_{k})]>\termThreshold$}
		
		\STATE  Determine $I({J}_{k};\zMeas_{k}(\qConfig))$ per \eqnnt{eq-crmi};
		\label{line-CRMI}
		\STATE  Find optimal sensor configuration\\ $\qConfig^{*}_{k} \defeq 
		\arg\max_{\qConfig} I({J}_{k};\zMeas_{k}(\qConfig))$;
		\label{line-config}
		
		\STATE  Obtain new sensor measurements $\zMeas_k(\qConfig^*_k)$;
		\STATE Update $\widehat{\paramVec}_{k}, \PEECovar_{k}$ ;
		\STATE Find $\gridPath^{*}_{k} \defeq \arg\min(\widehat{J}_{k}(\gridPath))$;
		\label{line-dijkstra}
		\STATE Increment iteration counter $k=k+1$.
		\ENDWHILE
	\end{algorithmic}
	\caption{Coupled Sensor Configuration and Planning (CSCP)}
	\label{alg-cscp}
\end{algorithm}

\subsection{CRMI Optimization}

\label{sec-approx_alg}
Finding the optimal sensor configuration by maximizing CRMI is challenging
because the number of feasible sensor configurations suffers combinatorial 
explosion with increasing number of sensors $\nSensor.$
To resolve this issue, we execute greedy optimization of one sensor at a time,
which enormously reduces the computation time. When the objective function 
is submodular, the sub-optimality due to greedy optimization remains bounded.
Therefore, we consider submodularity of the proposed CRMI.

%
%
A brief definition of submodularity is as follows.
Consider a finite set $\Omega$ and a real-valued function $f: 2^\Omega \rightarrow \real.$
For any two subsets $\mathcal{X}, \mathcal{Y} \subseteq \Omega$ such that 
$\mathcal{X} \subseteq \mathcal{Y},$ the function $f$ is said to be
submodular if
\begin{align}
	f(\mathcal{X} \cup\left\{x\right\}) - f(\mathcal{X}) 
	\geq f(\mathcal{Y} \cup x) - f(\mathcal{Y}).
	\label{eq-submodular}
\end{align}
for each $x \in \Omega \backslash \mathcal{Y}.$ The inequality~\eqnnt{eq-submodular}
expresses the property of diminishing returns, i.e., the increase in $f$ due to 
the introduction of a new element in a set diminishes with the size of that set.
In the context of sensor placement, this means that if a information gain measure
is submodular, then the information gain due to the placement of a new sensor 
diminishes with the number of sensors already placed.

\begin{prop}
	\label{prop-submodular}
	The CRMI $I({J};\zMeas(\qConfig))$ is submodular.
\end{prop}
\begin{proof}
 \revnew{Refer to Appendix A.}
\end{proof}

%

\begin{algorithm}
	\begin{algorithmic}[1]
		\STATE Set $\qgreedy = \emptyset$;
		\FOR{$i = 1$ to $\nSensor$}
		\STATE Set $\qConfig \defeq \intset{\nGridPt} \backslash \qgreedy$;
		\STATE Calculate $q_i^{*} \defeq \arg\max_{q \in \qConfig} I(J;\zMeas(\qConfig))$;
		\STATE $\qgreedy = \qgreedy \cup q_i^{*}$.
		\ENDFOR
	\end{algorithmic}
	\caption{Greedy CRMI Optimization}
	\label{alg-greedy}
\end{algorithm}

\subsection{Greedy Sensor Placement}
\label{ssec-greedy} 
We implement a greedy
sensor placement algorithm, as shown in \algoref{alg-greedy}, in which the
sensor locations are chosen in sequence such that the choice of next sensor
maximizes the CRMI. This approach aims to select a set of sensors that
collectively provides the maximum information relevant to the path planning. As
described in \algoref{alg-greedy}, greedy optimization initializes the empty
configuration $\qgreedy=\emptyset$ and iterations are carried out until
$\nSensor$ sensors are selected. At each iteration, the greedy optimal
configuration is the scalar $q^* \in \intset{\nGridPt} \backslash \qgreedy$
that maximizes CRMI $I(J;\zMeas(q)).$ The configuration $\qgreedy$ is then
updated to include $q^*.$
 
\begin{thm}\citep{nemhauser1978analysis}
	\label{them-greedy}
	The greedy placement algorithm for any monotone submodular function 
	provides a performance guarantee of $(1-\frac{1}{e})$ times the optimal
	value.
\end{thm}

We denote the maximum $I({J};\zMeas(\qConfig^{\ast}))$ as an optimal value, and 
$I(J;\zMeas(\qgreedy))$ as 
the approximate mutual information value. For any $k=\nSensor$ sensor elements chosen by the 
greedy 
algorithm, it follows from 	\thmref{them-greedy} that
\begin{align*}
	I(J;\zMeas(\qgreedy)) &\geq
	\left(1-\left(1-\frac{1}{k}\right)^{k}\right)I(J;\zMeas(\qConfig^{\ast}))
	\geq \left(1-\frac{1}{e}\right)I(J;\zMeas(\qConfig^{\ast})). 
\end{align*}


\subsection{Sensor Reconfiguration Cost}
\label{ssec-src}
\begin{figure}
	\centering
	\includegraphics[width=0.33\columnwidth]{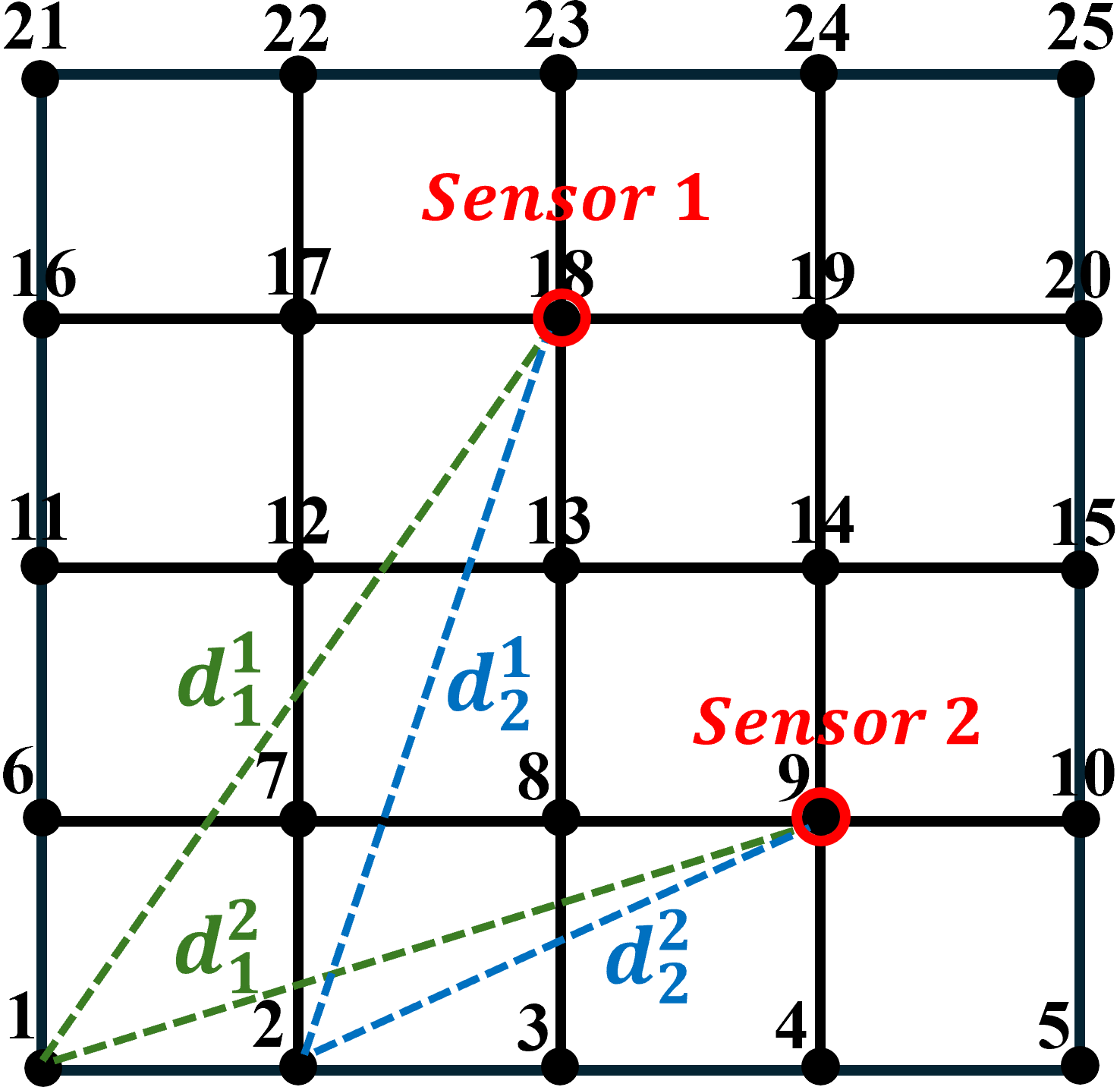}
	\caption{Illustration of sensor reconfiguration cost.}
	\label{fig:Sensor_reconfiguration_cost}
\end{figure}

Sensor reconfiguration cost refers to the cost of moving sensors from one location
to another in the grid space. We consider a sensor reconfiguration cost based on
the Euclidean distance between the new and previous sensor locations, as illustrated
in~\fig{fig:Sensor_reconfiguration_cost}. Here, $d_{i}^{j}$ is the Euclidean 
distance between the $i\msup{th}$ grid point and the location of the $j\msup{th}$ sensor. 

Informally, at the $k\msup{th}$ iteration of the CSCP algorithm,
the objective is to find new sensor locations $\qConfig_{k+1}^*$ that maximize
the CRMI while minimizing the cost of sensor reconfiguration from the current 
configuration $\qConfig^*_k = \{q_1^{k*}, \ldots, q_{\nSensor}^{k*}\}.$
To this end we define
\begin{align}
	I\msup{mod}(\qConfig) \defeq I({J_k};\vec{z}_k(\vec{q})) 
	+ \alpha_1 - \alpha_2 \min_{j,\ell \in \intset{\nSensor} \times \intset{\nSensor}}\| q_\ell - 
	q_{j}^{k*} \|, 
	\label{eq-sensorcost}
\end{align}
where $\alpha_1, \alpha_2$ are constants. We choose $\alpha_1$ to be a relatively large value,
e.g., proportional to the size of the overall workspace, whereas $\alpha_2$ is chosen
based on the user's preference for reducing the reconfiguration cost.

\begin{prop}
	\label{cor-submodularity}
	The modified CRMI $I\msup{mod}(\qConfig)$ is submodular.
\end{prop}
\begin{proof}
\revnew{Refer to Appendix A.}
\end{proof}

\subsection{Convergence of the Proposed CSCP Algorithm}

In this section, we show that the proposed CSCP algorithm converges
in a finite number of iterations. To this end, first consider the
following result.


\begin{prop}
	\label{prop-cscp_convergence}
	If $A$ is Schur, then the CSCP algorithm terminates 
	in a finite number of iterations.
\end{prop}
\begin{proof}
		If all modes of $A$ are stable, then the pair $(A, \HMeas_{k}(\qConfig))$ is uniformly 
		detectable for any $\qConfig.$ Furthermore, because $Q = \sigma_{P}^2 \eye{\nParameter},$
		the pair $(A,Q)$ is controllable. By~\cite{anderson1981detectability}, it follows that the
		estimation error is exponentially stable. Consequently, for any $\termThreshold > 0$ there 
		is a finite iteration number $M \in \nat$ such that $\mathrm{tr}(\PEECovar_k) \leq 
		\termThreshold,$ after which the CSCP algorithm terminates.
\end{proof}

\begin{rmk}
	\label{rmk-cscp_convergence}
	Whereas \prpf{prop-cscp_convergence} is sufficient, it is not necessary for the 
	convergence of the CSCP algorithm. A less restrictive criterion for convergence, 
	which does not assume $A$ to be Schur, is discussed below.
\end{rmk}

At the $k\msup{th}$ iteration of the CSCP method, the predicted error covariance
is defined as $\PEECovar_{k|k-1} \defeq A \PEECovar_{k-1|k-1} A^{\intercal} +
\QProcCovar_{k-1}$. Here, the term $A \PEECovar_{k-1|k-1} A^{\intercal}$ grows
the uncertainty from the previous step's posterior covariance based on the $A$,
and $\QProcCovar_{k-1}$ accounts for the uncertainty introduced due to process
noise. The measurement update is defined as $\PEECovar_{k|k}\defeq
\PEECovar_{k|k-1} - \UKFGain_k\HMeas_{k}\PEECovar_{k|k-1}$. The uncertainty
growth in the system is $\{\PEECovar_{k|k-1} - \PEECovar_{k-1|k-1} = A
\PEECovar_{k-1|k-1} A^{\intercal} - \PEECovar_{k-1|k-1} + \QProcCovar_{k-1}\}$.
Similarly, the reduction in the uncertainty of the system after the measurement
update is $\{\PEECovar_{k|k-1} - \PEECovar_{k|k} =
\UKFGain_k\HMeas_{k}\PEECovar_{k|k-1}\}$. The convergence of the CSCP method is
guaranteed if the following criterion is satisfied for every time steps.
\begin{align*}
	\operatorname{tr} (\UKFGain_k\HMeas_{k}\PEECovar_{k|k-1}) \geq \operatorname{tr} (A 
	\PEECovar_{k-1|k-1} A^{\intercal} - \PEECovar_{k-1|k-1} + \QProcCovar_{k-1}).
\end{align*}
Note that this criterion cannot be verified \emph{a priori} as it depends on the 
estimation error covariances computed during the algorithm's execution.

\section{Results and Discussion}
\label{sec-results}

In this section, we first provide an illustrative example of the proposed CSCP-CRMI
method. Second, we compare the proposed method against a decoupled method that
finds optimal sensor placement using the standard (path-independent) MI; for
brevity we call this decoupled method CSCP-SMI. Third, we study the effects of
varying numbers of sensors, threat parameters, and grid points on the CSCP-CRMI
method. Fourth, we perform comparative study between the CSCP-CRMI and greedy
sensor placements, and observe the equivalency as well as differences between
the two approaches. Finally, we conduct a comparative study between the CSCP
method with and without the sensor reconfiguration cost. \revnew{All simulations are
performed within a square workspace $\mathcal{W} = [-1, 1] \times [-1, 1]$ using
non-dimensional units in a Cartesian coordinate axes system.}

\revnew{A \matlab-based implementation of the CSCP method used for
producing these results is available at this repository:
\url{https://github.com/prakashpoudel2014/CSCP\_time\_varying.}
}
\subsection{Illustrative Example}

\def\thisfigwidth{0.46\columnwidth}
\begin{figure}[t]
	\centering
	\begin{subfigmatrix}{2}
		\subfigure[$k=1$]{\includegraphics[width=\thisfigwidth]{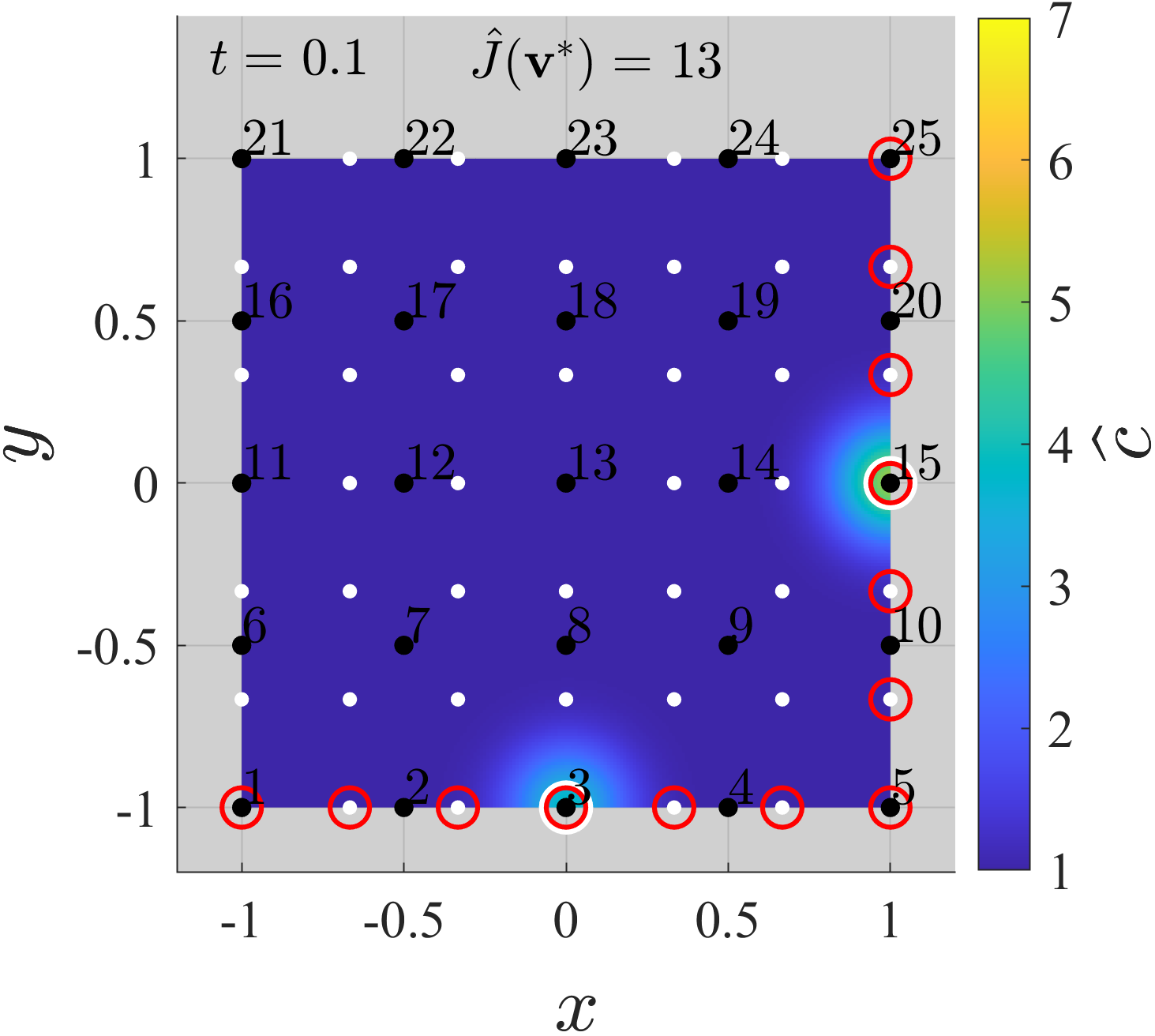}}
		\subfigure[$k=5$]{\includegraphics[width=\thisfigwidth]{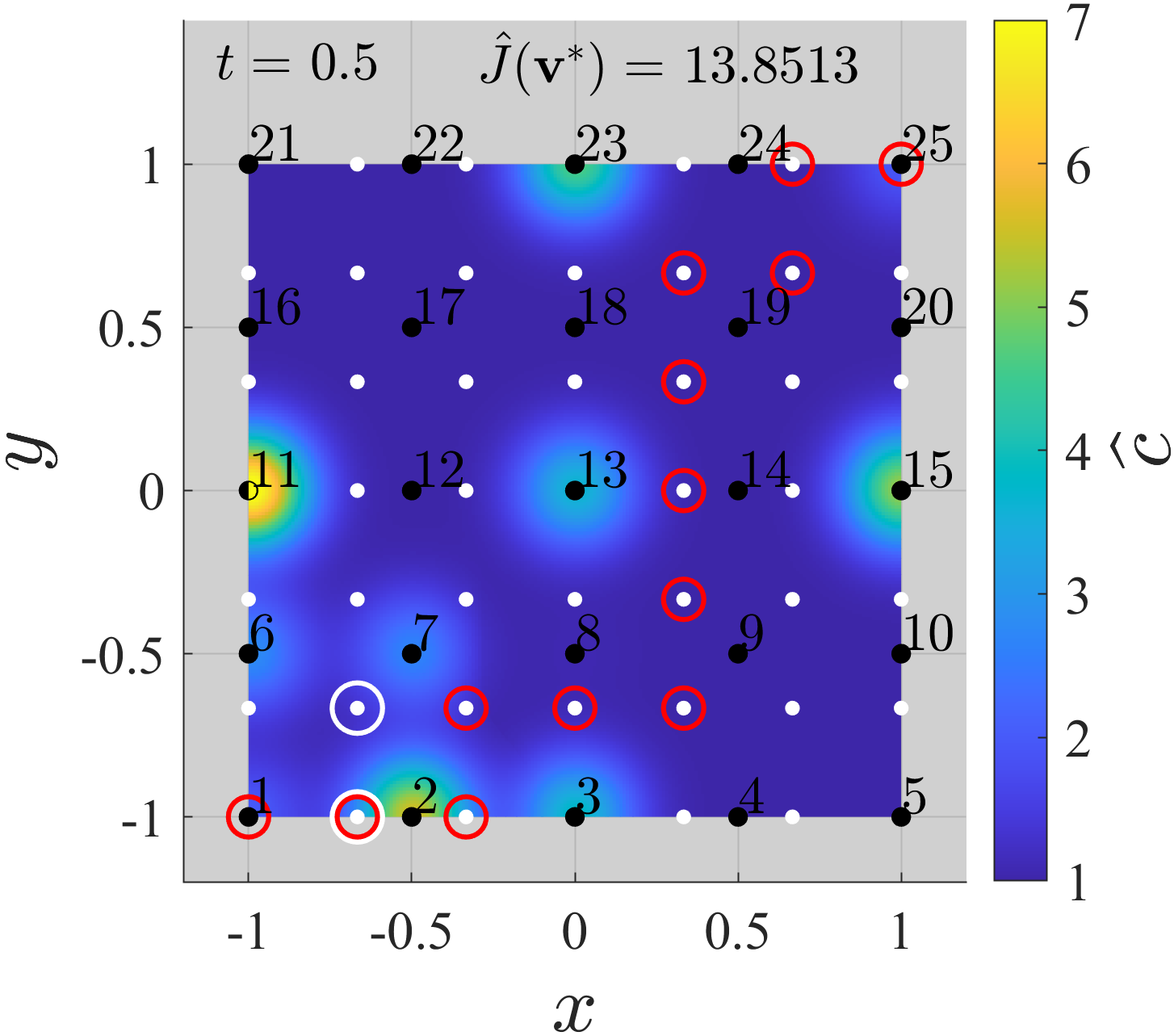}}
		\subfigure[$k=11$]{\includegraphics[width=\thisfigwidth]{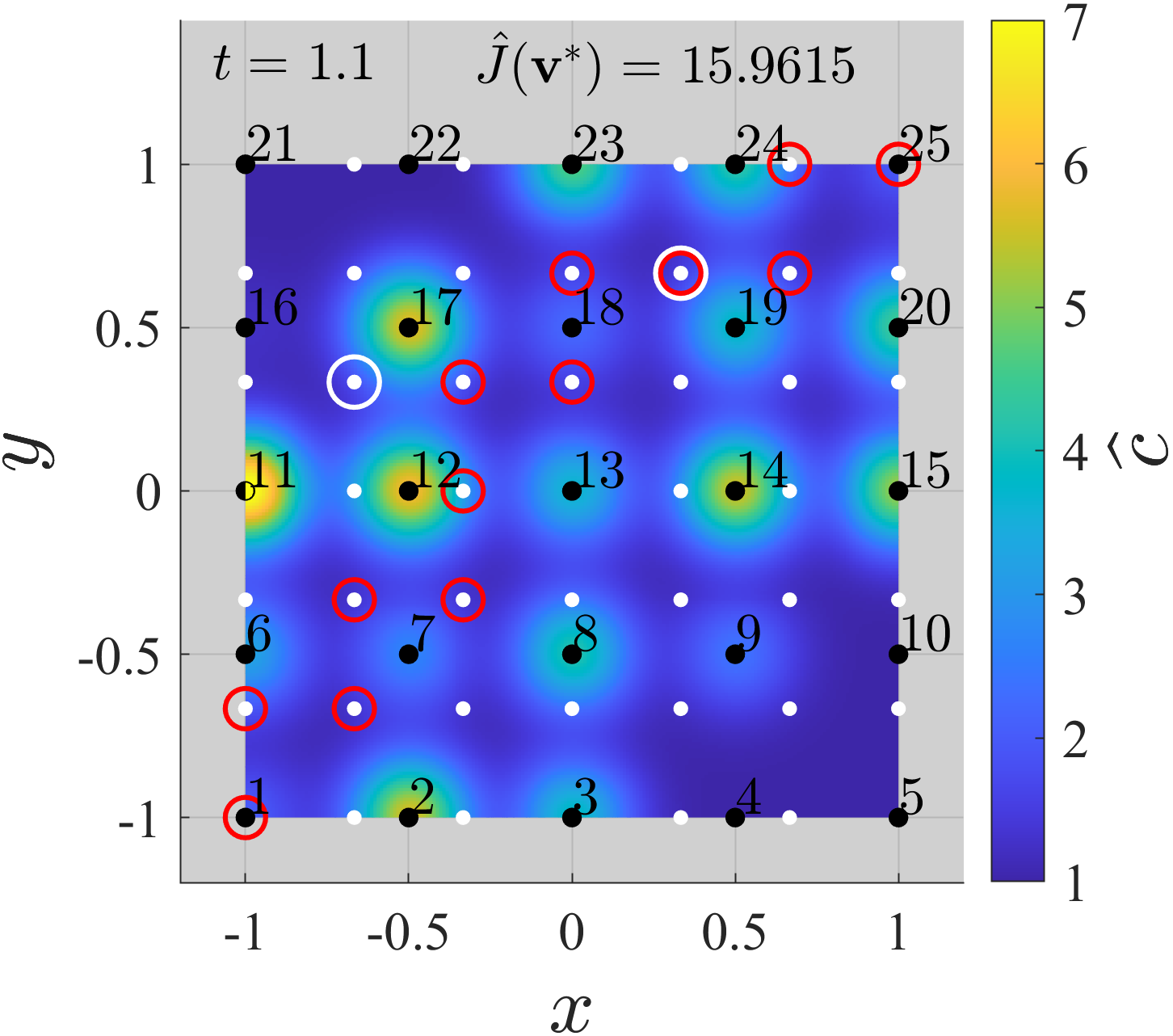}}
		\subfigure[$k=15$]{\includegraphics[width=\thisfigwidth]{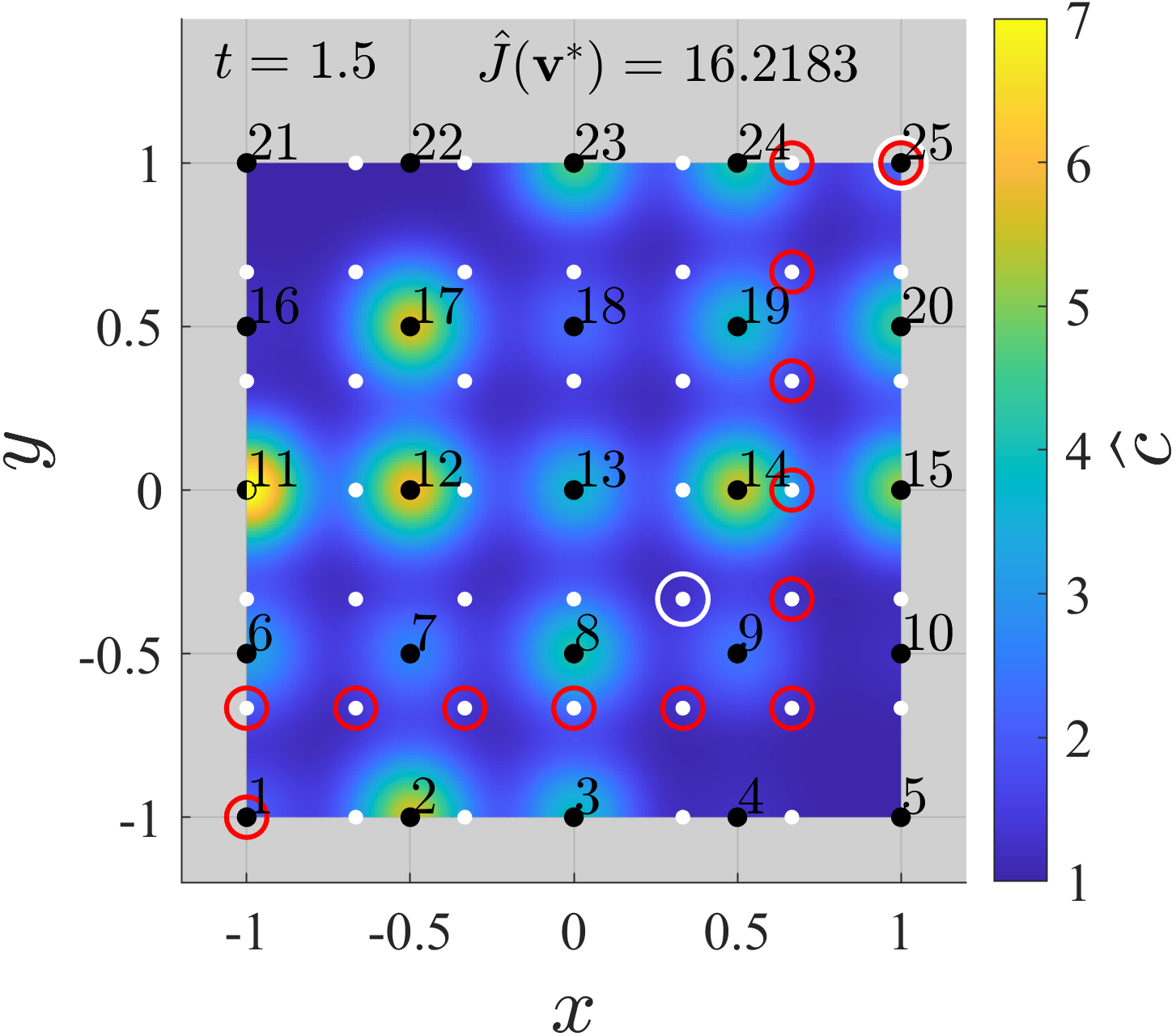}}	
	\end{subfigmatrix}
	\caption{Visualization of CSCP-CRMI process for $\nParameter = 25$ and $\nGridPt = 49$.}
	\label{fig:Fig.CSCP_visualization}
\end{figure}

\begin{figure}
	\centering
	\begin{subfigmatrix}{2}
		\subfigure[Error ($k=1$)]{\includegraphics[width=\thisfigwidth]{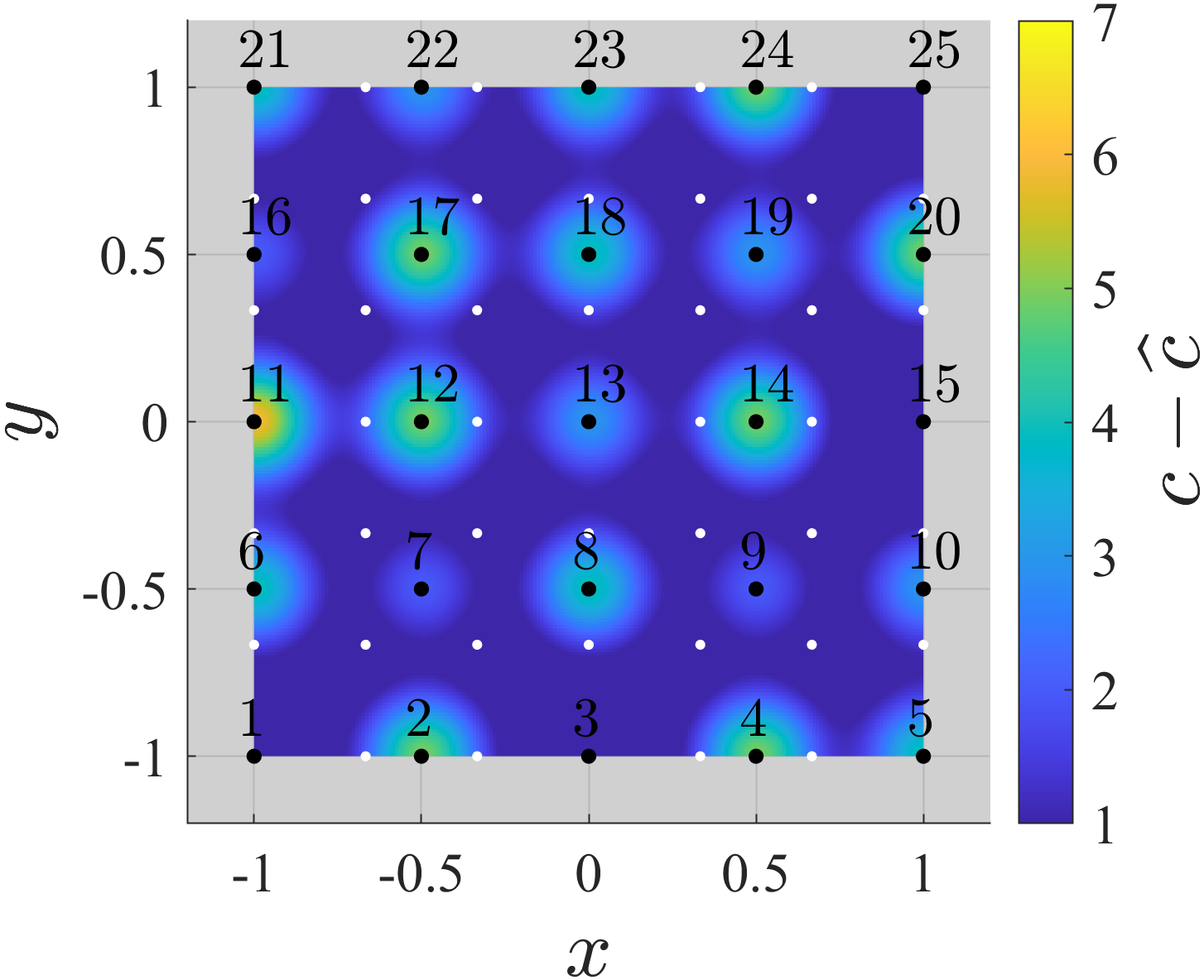}}
		\subfigure[Error ($k=5$)]{\includegraphics[width=\thisfigwidth]{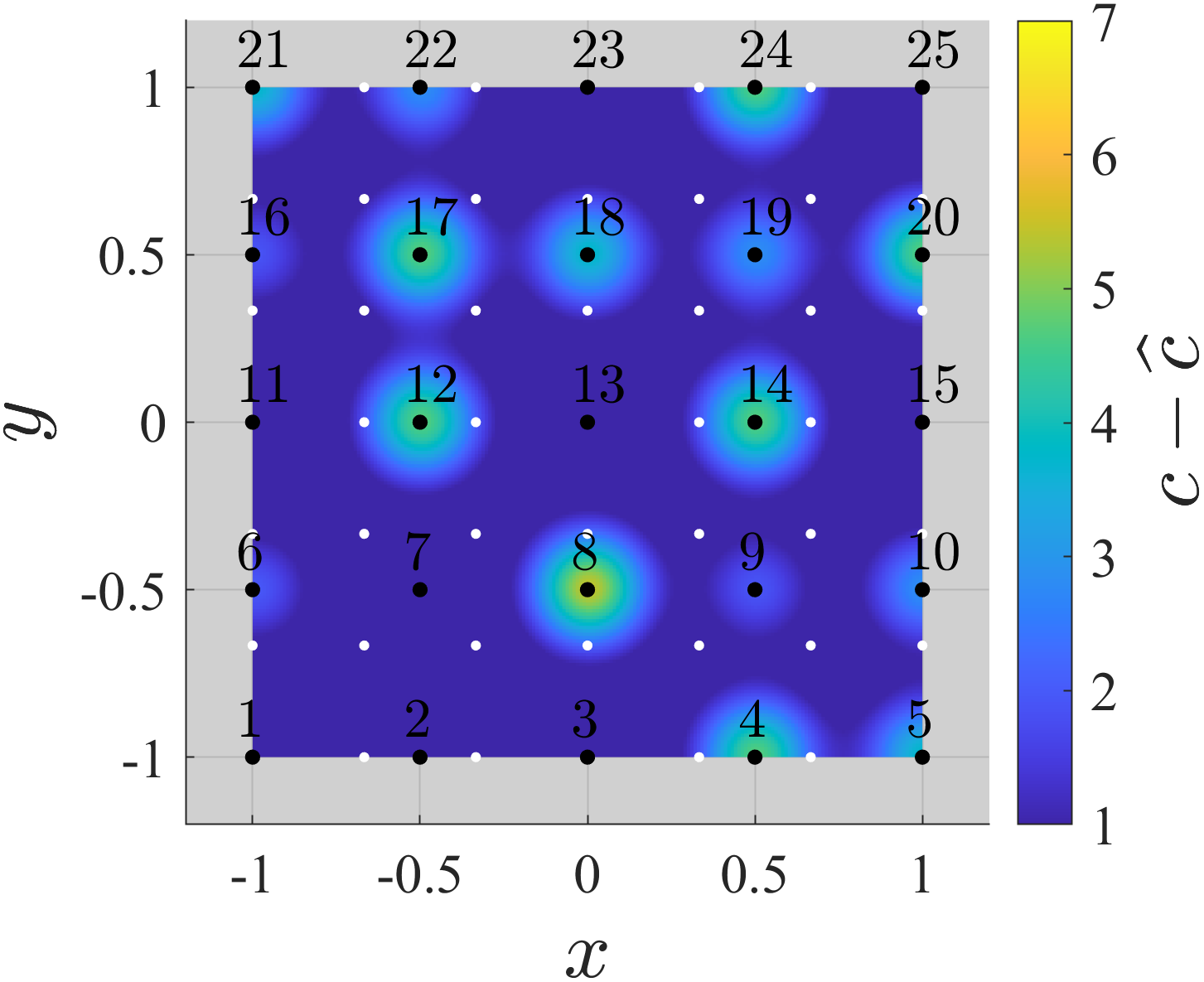}}
		\subfigure[Error ($k=11$)]{\includegraphics[width=\thisfigwidth]{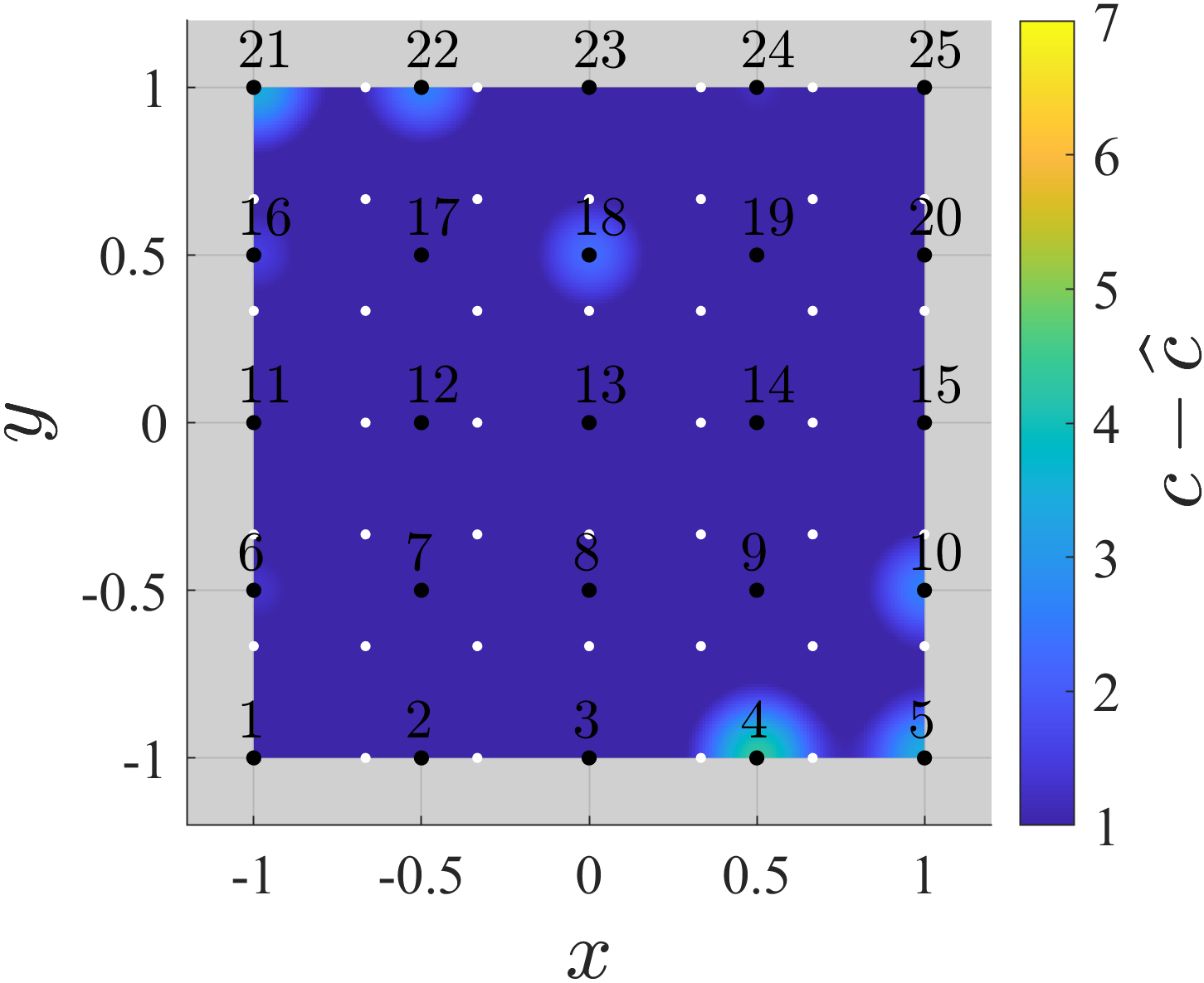}}
		\subfigure[Error ($k=15$)]{\includegraphics[width=\thisfigwidth]{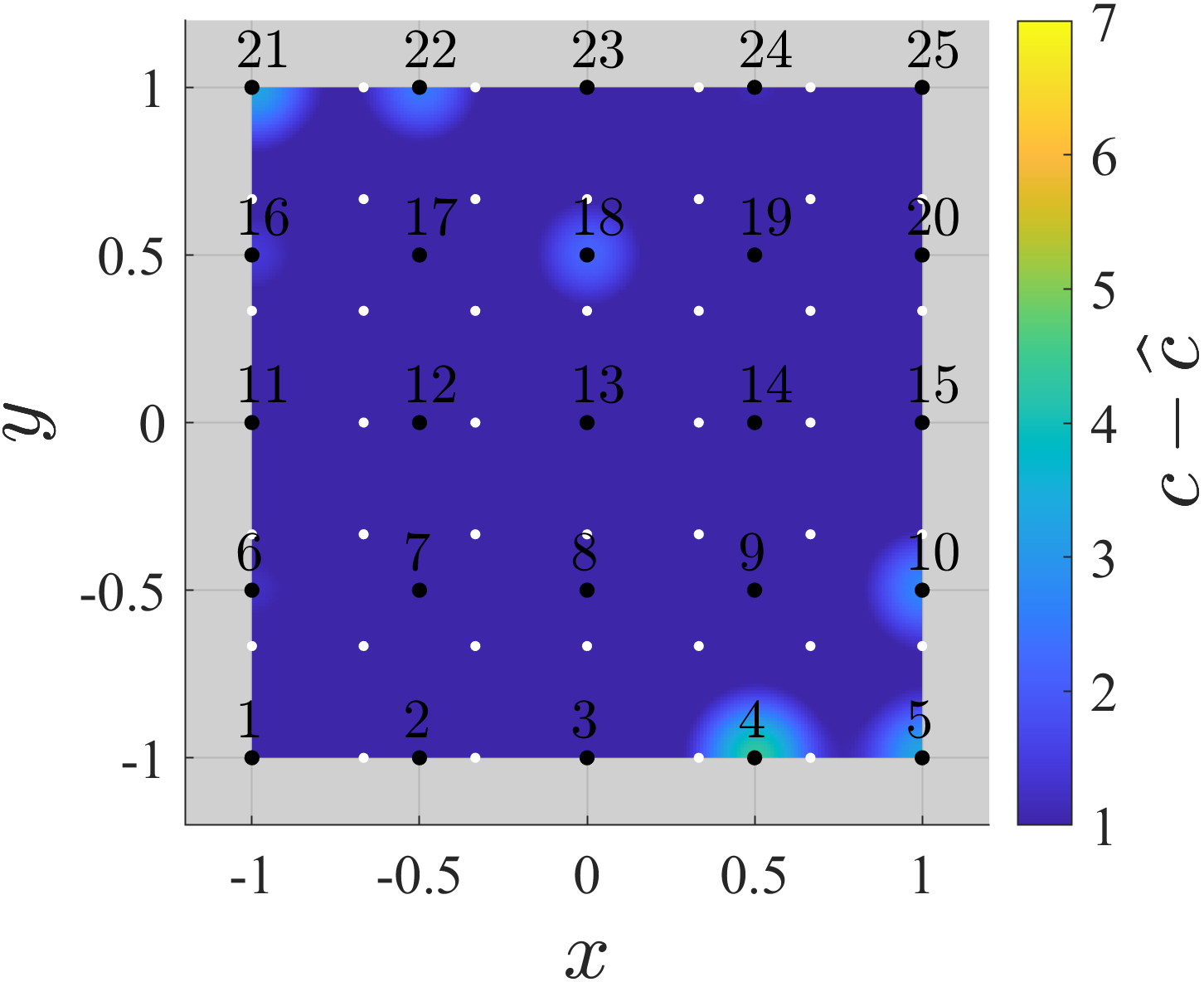}}
	\end{subfigmatrix}
	\caption{Error between the mean estimate and ground truth during CSCP-CRMI, for $\nParameter = 25$ and $\nGridPt = 49$.}
	\label{fig:Fig.CSCP_visualization_error}
\end{figure}

\begin{figure}
	\centering
	\begin{subfigmatrix}{2}
		\subfigure[Path cost mean.]
		{\label{fig:path_cost1}
			\includegraphics[width= 0.47\columnwidth]{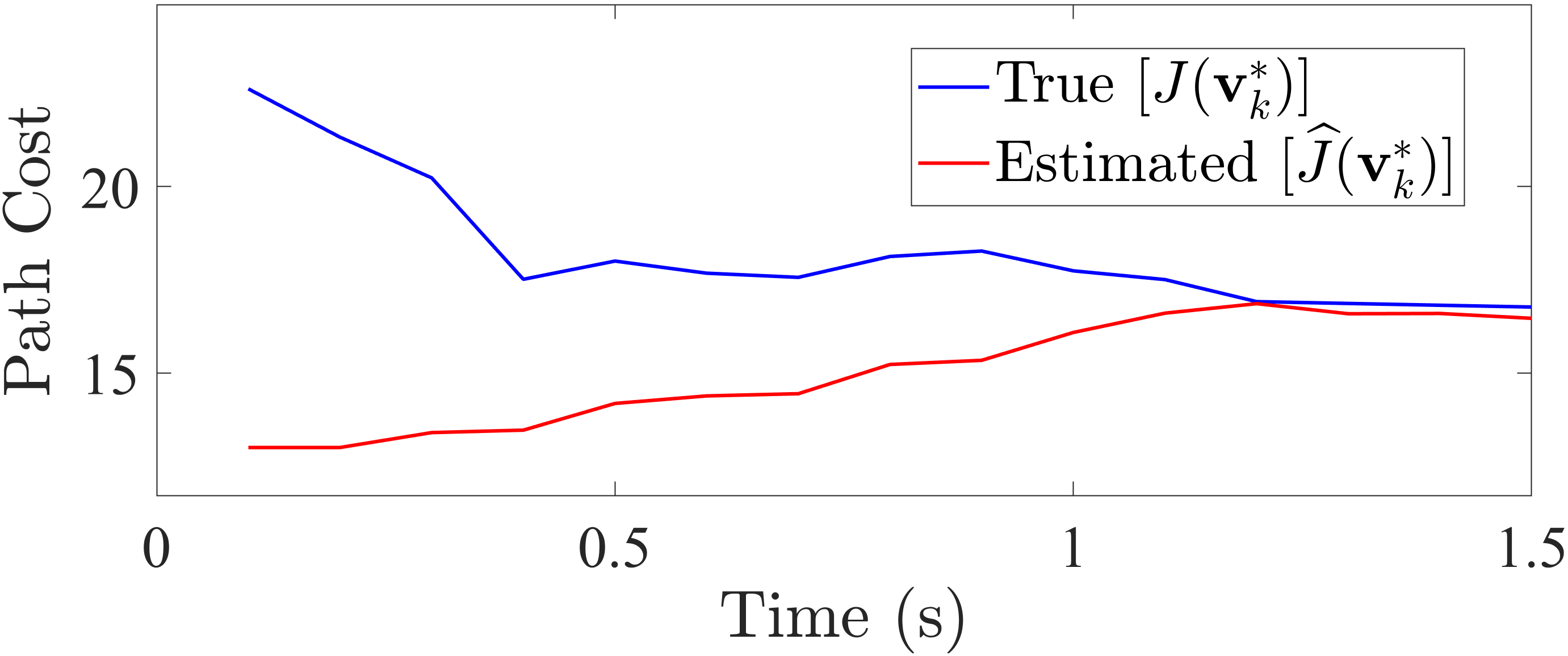}}
		\hspace{0.01\columnwidth}
		\subfigure[Path cost variance.]{\label{fig:Path_cost_var1}\includegraphics[width= 
			0.47\columnwidth]{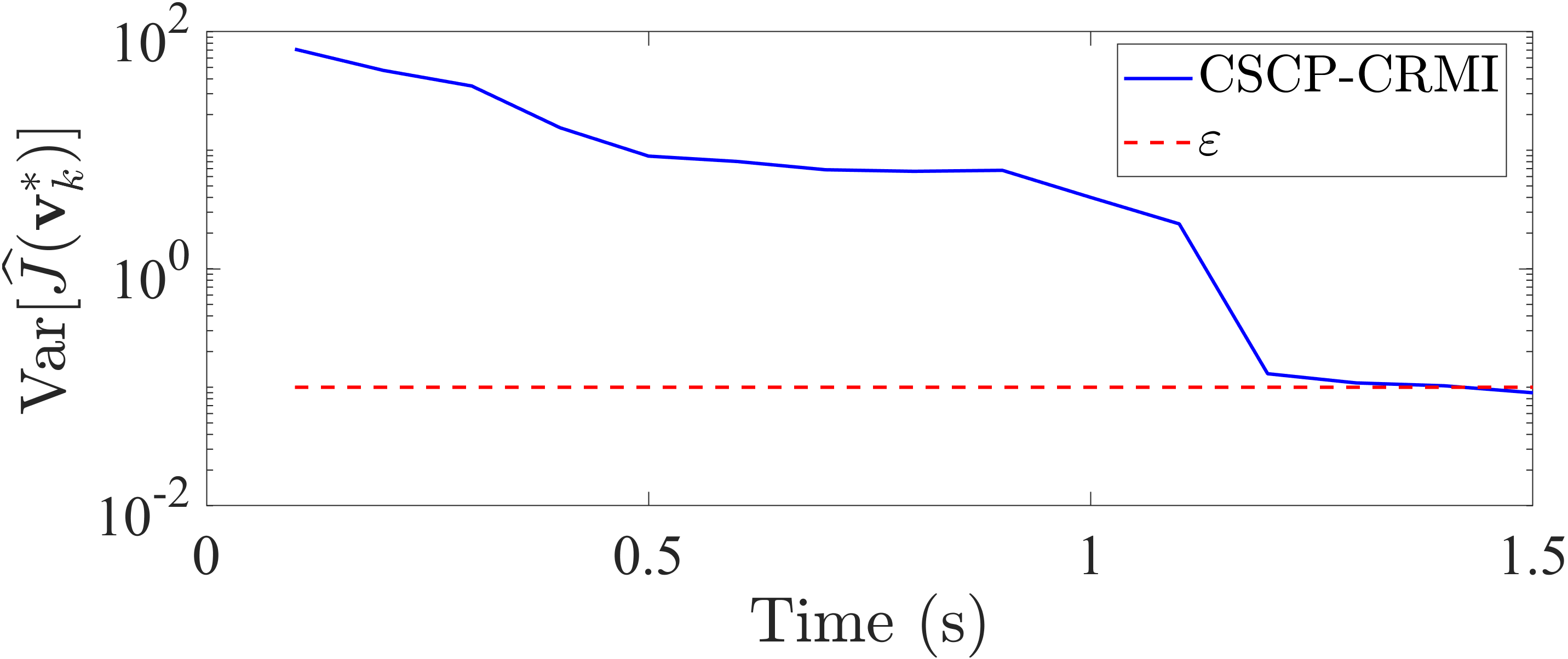}}		
	\end{subfigmatrix}
	\caption{Convergence of CSCP-CRMI algorithm.}
	\label{fig-path-var-comp2}
\end{figure}

The implementation of CSCP-CRMI algorithm on an illustrative example is shown in
\fig{fig:Fig.CSCP_visualization}. The number of threat parameters, grid points,
and sensors are $\nParameter=25$, $\nGridPt=49$, and $\nSensor=2$, respectively.
The threat parameters $\nParameter$, indicated by the black dots and numbered
from 1 to 25, are uniformly spaced in the workspace. The white dots represent
the grid points. The initial and the goal points are represented by the bottom
left and the top right grid points in the map. The evolution of the threat field
estimate $\widehat{c}$ for different time steps, namely $k=1, 5, 11$, and $15$
is shown by a colormap. The path $\gridPath^{*}_{k}$ of minimum estimated cost
is indicated by red circles, and the sensor placement $\qConfig_{k}$ is shown by
white circles, as illustrated in the \fig{fig:Fig.CSCP_visualization}(a)-(d).
For a specified threshold $\termThreshold = 0.1$, the algorithm terminates at
$k=15$ iterations, and the optimal path $\gridPath^{*}$ is achieved. \revnew {\fig{fig:Fig.CSCP_visualization_error}(a)-(d) show the absolute error between the true threat field $\threat$ and the estimated field $\widehat{\threat}$, computed as
$| \threat - \widehat{\threat} |$, for the respective iterations $k =1,5,11,15$. These plots illustrate how the estimation accuracy improves as the algorithm progresses.}

The comparison between the true and estimated path cost is illustrated in
\fig{fig:path_cost1}. Upon termination, the true and estimated path costs are
nearly identical, with $J(\gridPath^{*}_{k}) = 16.46$ and
$\widehat{J}(\gridPath^{*}_{k}) = 16.77$, respectively.
\figf{fig:Path_cost_var1} shows the convergence of the proposed CSCP-CRMI
algorithm. The path cost variance
$\textrm{Var}[(\widehat{J}(\gridPath^{*}_{k})]$ decreases with time, and the
algorithm terminates in 15 iterations as the path cost variance decreases below
0.1.

\begin{figure}
	\centering
	\begin{subfigmatrix}{2}
	\subfigure[ CSCP-CRMI method ]{
		\label{fig:daig_err_cov_CRMI}
		\includegraphics[width=0.45\columnwidth]{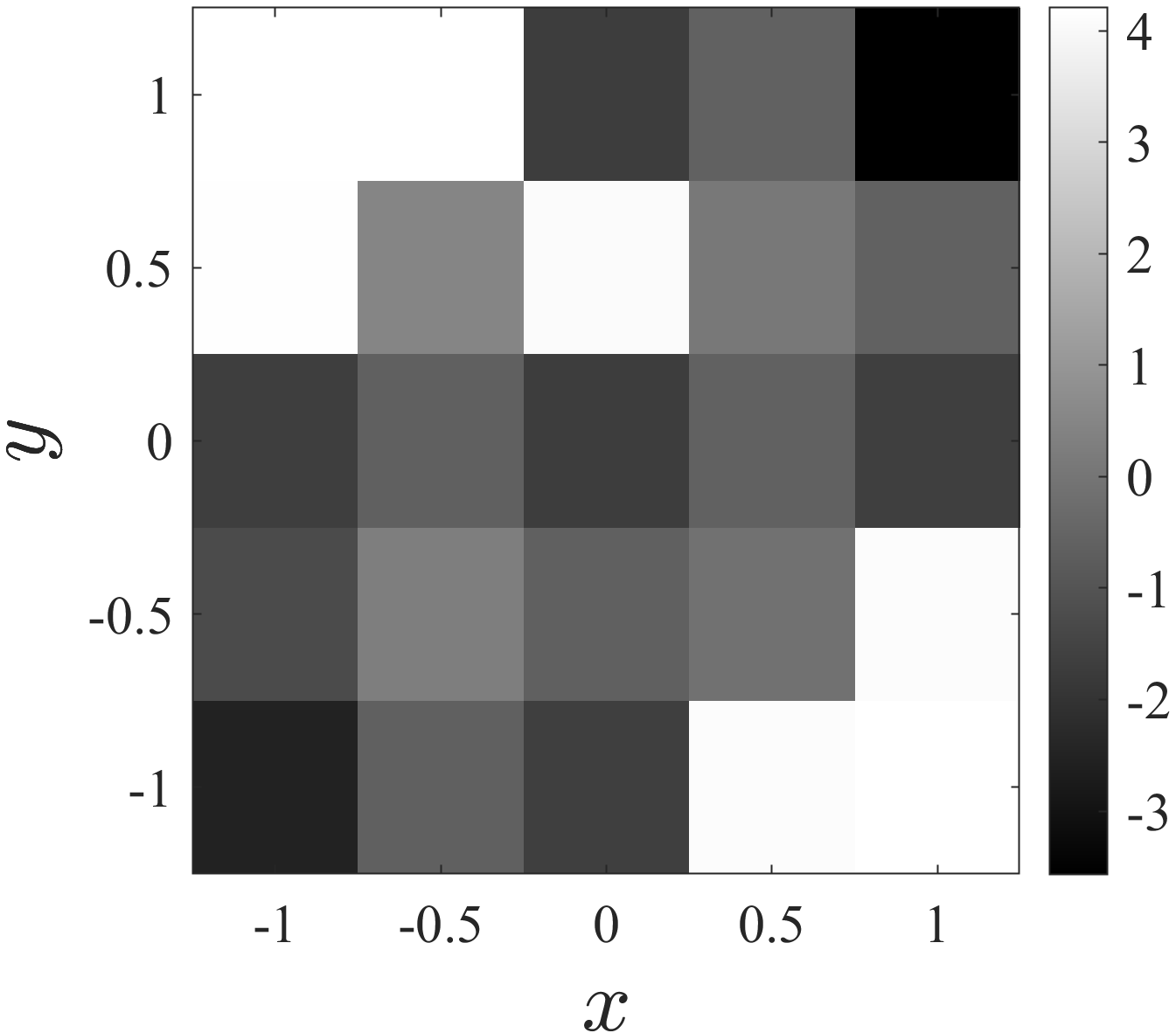}}
	\subfigure[ CSCP-SMI method ]{
		\label{fig:daig_err_cov_SMI}
		\includegraphics[width=0.45\columnwidth]{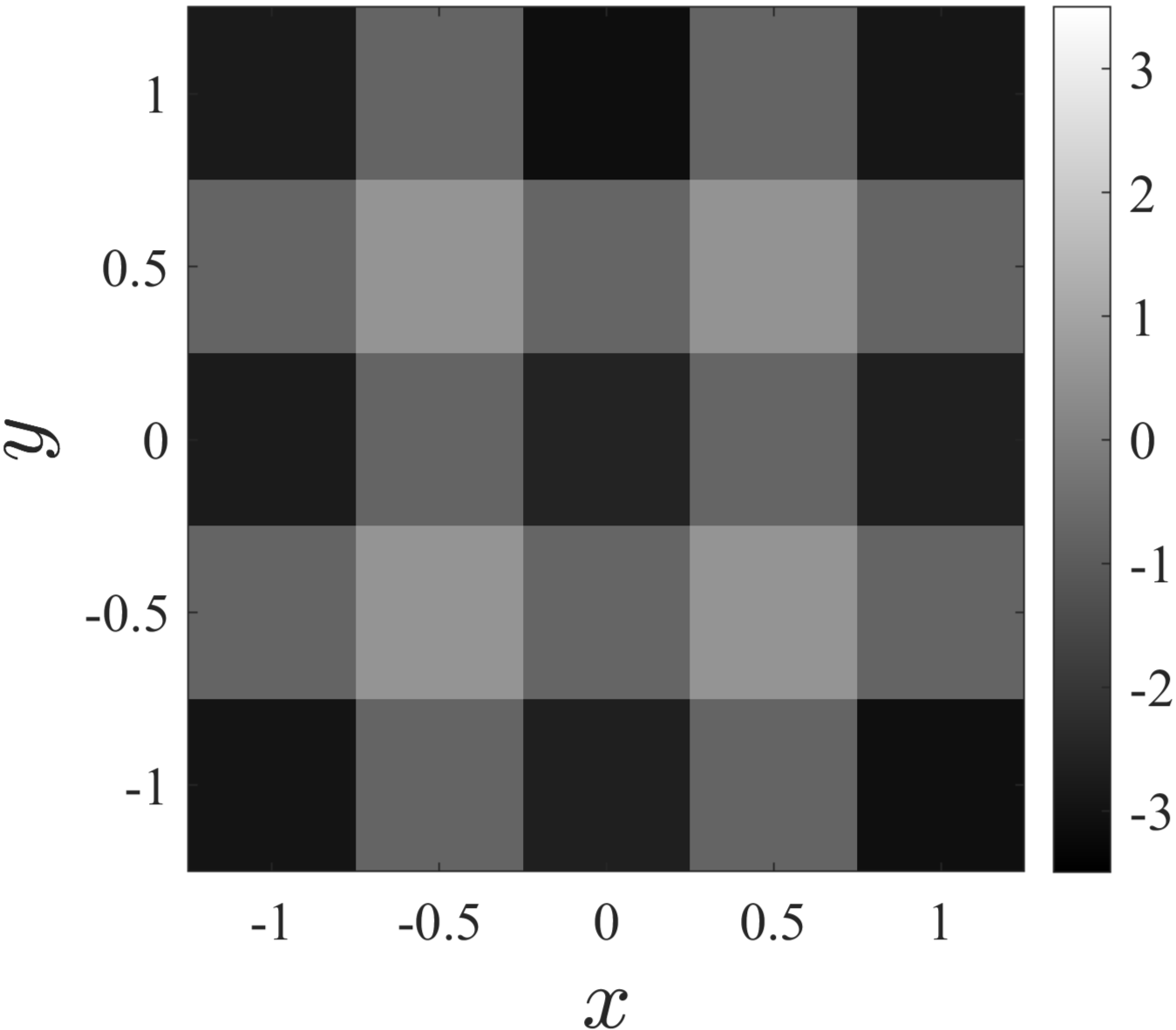}}
	\end{subfigmatrix}
	\caption{Diagonal log-values of error covariance at final iteration.}
	\label{fig:daig_err_cov}
\end{figure}

\subsection{Comparison of CSCP-CRMI and CSCP-SMI}
\label{sec-crmi_smi_comparision}
For comparison, now consider the execution of CSCP-SMI on the same
example as discussed above.

\figf{fig:daig_err_cov} shows the estimation error covariance $\PEECovar$ 
at the final iteration, mapped to the spatial regions of the environment
using the centers of spatial basis functions~$\Phi.$
In a slight departure from convention, \fig{fig:daig_err_cov}
shows the \emph{logarithms} of the diagonal values of $\PEECovar,$ which
explains the negative values despite $\PEECovar$ being a symmetric positive
definite matrix. The reason for this choice of logarithmic values is to 
clearly show the orders of magnitude difference in the estimation error
covariance in different regions of the environment.


The white regions in \fig{fig:daig_err_cov}(a) with high error covariance
represent areas where few, if any, sensors are placed throughout the execution
of CSCP-CRMI. The darker regions represent areas where sensors were placed to
reduce the estimation error covariance values orders of magnitude below those of
the white-colored regions. Compare \fig{fig:daig_err_cov} to the optimal path
found in~\fig{fig:Fig.CSCP_visualization}(d), and we find that the CSCP-CRMI
sensor placement is such that areas around the optimal path generally have lower
estimation error covariance. \revnew{Note that the path does not exclusively
follow regions of minimal covariance. The planning objective is to minimize
cumulative threat exposure, the optimal trajectory occasionally favors areas
with slightly higher estimation uncertainty if the expected threat intensity in
those areas is substantially lower.} Note, crucially, the novelty of this
approach: \emph{the optimal path is at first unknown,} and that the sensor
placement and path-planning are performed iteratively to arrive at these
results.

\begin{figure}[t]
	\centering
	\begin{subfigmatrix}{2}
		\subfigure[ CSCP-CRMI method ]{
			\label{fig:CRMI_Ns_1}
			\includegraphics[width=0.45\columnwidth]{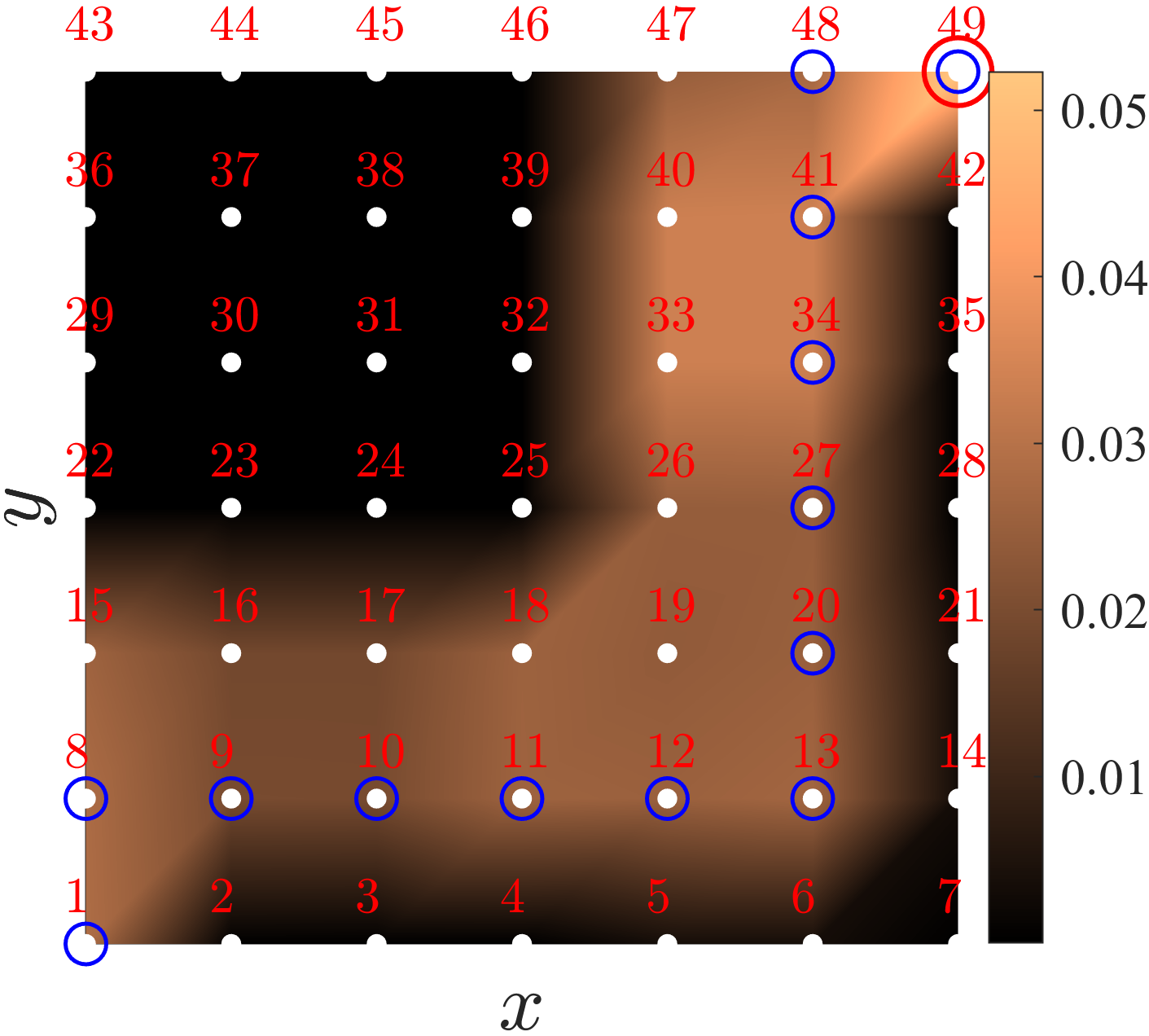}}
		\hspace{\fill}
		\subfigure[ CSCP-SMI method ]{
			\label{fig:SMI_Ns_1}
			\includegraphics[width=0.45\columnwidth]{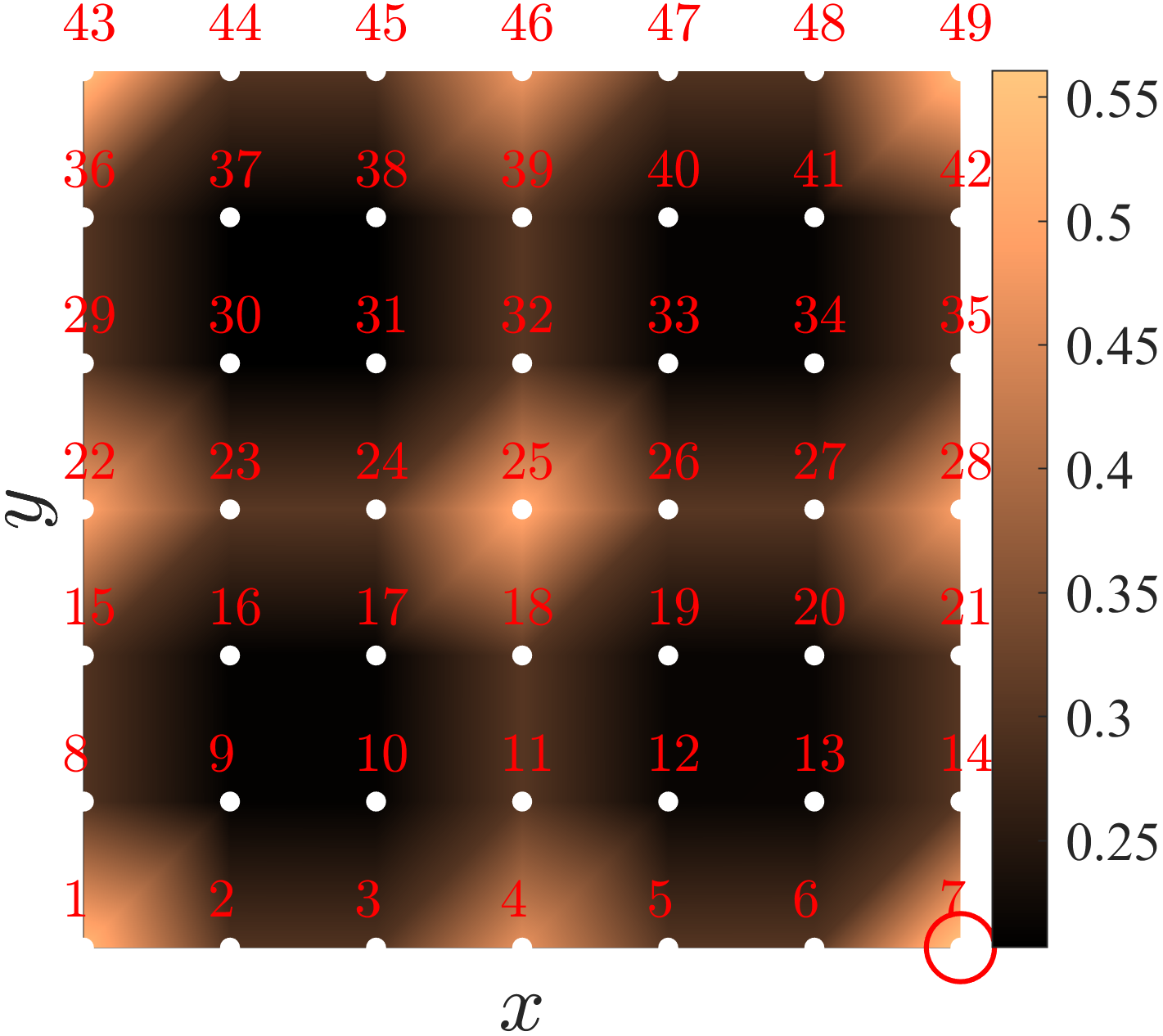}}
	\end{subfigmatrix}
	\caption{Mutual information intensity map.}
	\label{fig:CRMI_map}
\end{figure}

\begin{figure}[t]
	\centering
	\includegraphics[width=0.49\columnwidth]{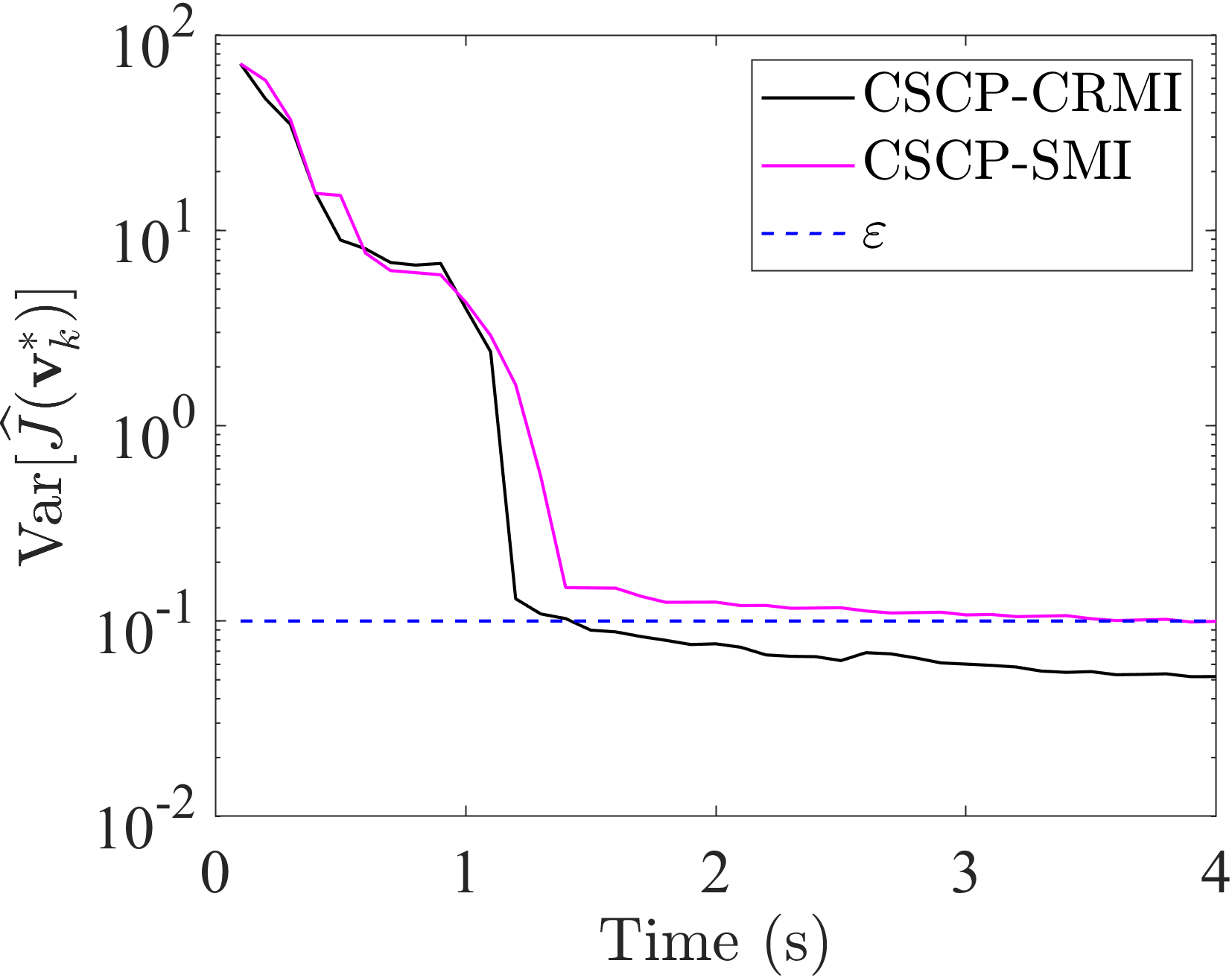}
	\caption{Comparison of path cost variance between CSCP-CRMI and CSCP-SMI 
		with $\nSensor = 2$.}
	\label{fig:path_var_comp}
\end{figure}

\figf{fig:daig_err_cov} (b) shows the estimation error covariance $\PEECovar$ of
CSCP-SMI method. By contrast to \fig{fig:daig_err_cov} (a) for CSCP-CRMI, we
note here spatially uniform covariance values. This means that in CSCP-SMI, the
sensors are placed in such a way that the error covariances in \emph{all}
regions of the environment are low compared to CSCP-CRMI. Whereas this would be
of benefit if we were merely trying to map the threat in the environment, this
uniformly low error covariance is indicative of wasteful sensor placement in the
context of p\ppl. In case of CSCP-CRMI, although there are some regions in that
are not explored, the outcome of the p\ppl\ algorithm is still near-optimal.

An intensity map showing the mutual information values for each grid points from
an illustrative example is shown in \fig{fig:CRMI_map}. Notably, these values
are derived with consideration for only a single sensor. The brown regions
represent the areas with higher CRMI values. It can be observed in
\fig{fig:CRMI_map} (a) that the CRMI regions are more visible around the
vicinity of the path obtained in \fig{fig:Fig.CSCP_visualization}(d). \revnew {The optimal path, indicated by blue circles, is overlaid on the CRMI intensity map for clarity.} Sensor is
placed at the grid point, here number 49, with a maximum CRMI value. Similarly,
\fig{fig:CRMI_map}(b) shows the SMI map representing the mutual information
values between the state and measurement variables. The higher CRMI regions are
observed around the location of the threat parameter, and for this example,
sensor is placed at 7 numbered grid point.

\begin{figure}[t]
	\centering
	\begin{subfigmatrix}{2}
	\subfigure[$\nSensor = 
	3.$]{\includegraphics[width=0.45\columnwidth]{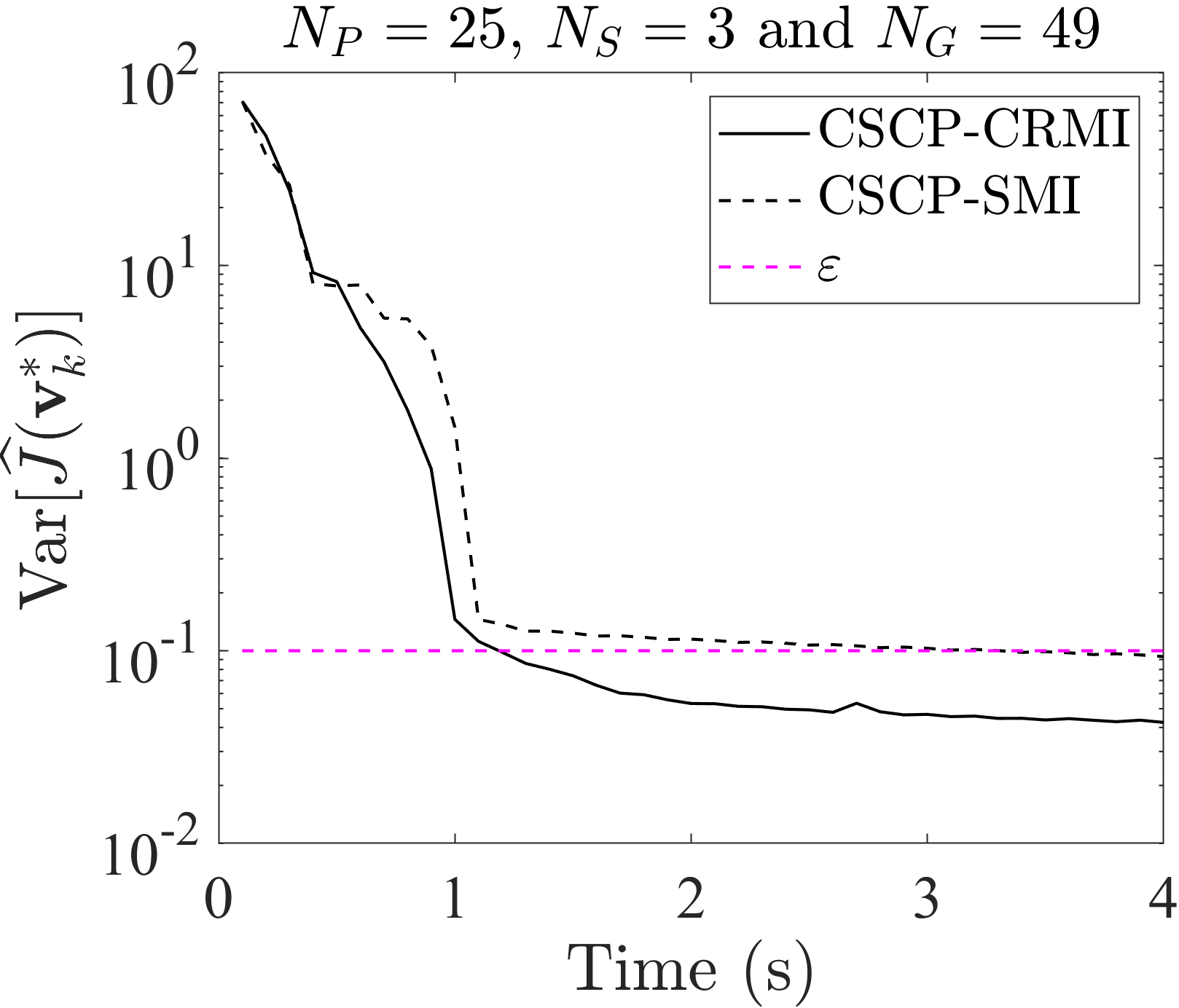}}
	\subfigure[$\nSensor = 
	4.$]{\includegraphics[width=0.45\columnwidth]{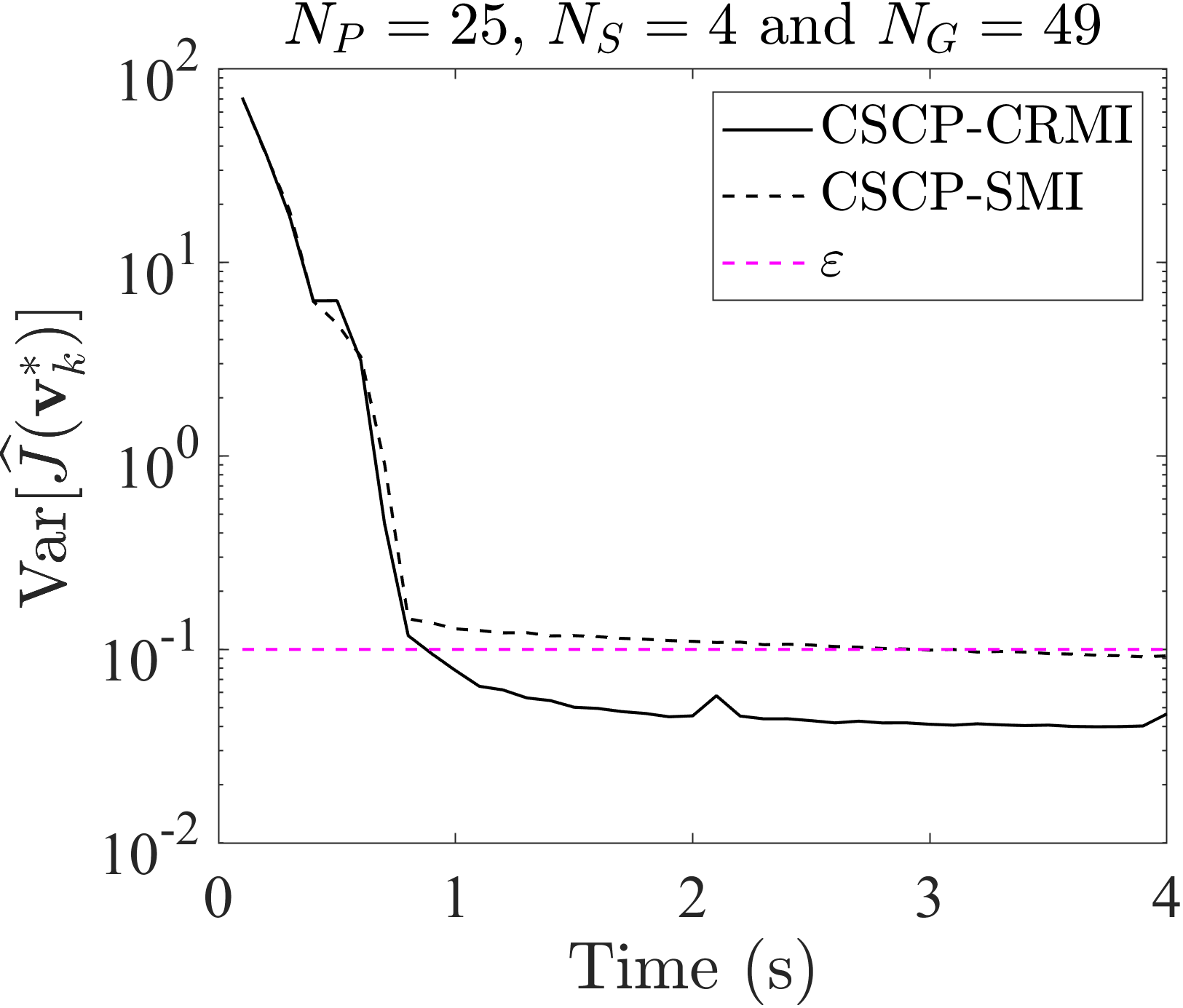}}
	 \end{subfigmatrix}
	\caption{Additional comparison of path cost variance between CSCP-CRMI and CSCP-SMI.}
	\label{fig-path-var-comp_2}
\end{figure}

\begin{figure}[t]
	\centering
	\begin{subfigmatrix}{2}
		\subfigure[ Varying $\nSensor$]{
			\label{fig:pathVar_Ns_varies}
			\includegraphics[width=0.45\columnwidth]{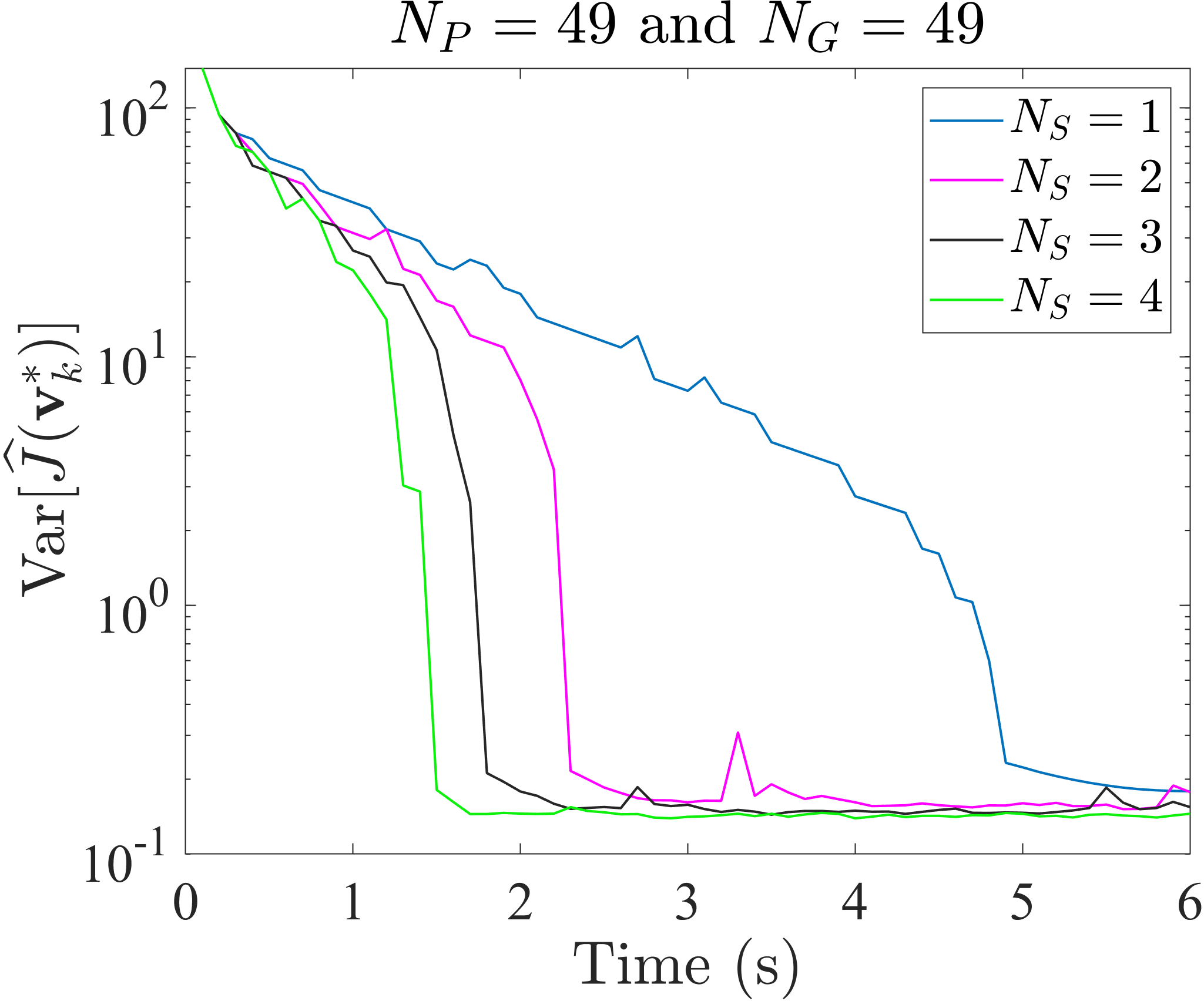}}
		\hspace{\fill}
		\subfigure[Varying $\nParameter$]{
			\label{fig:path_var_Np_varies}
			\includegraphics[width=0.45\columnwidth]{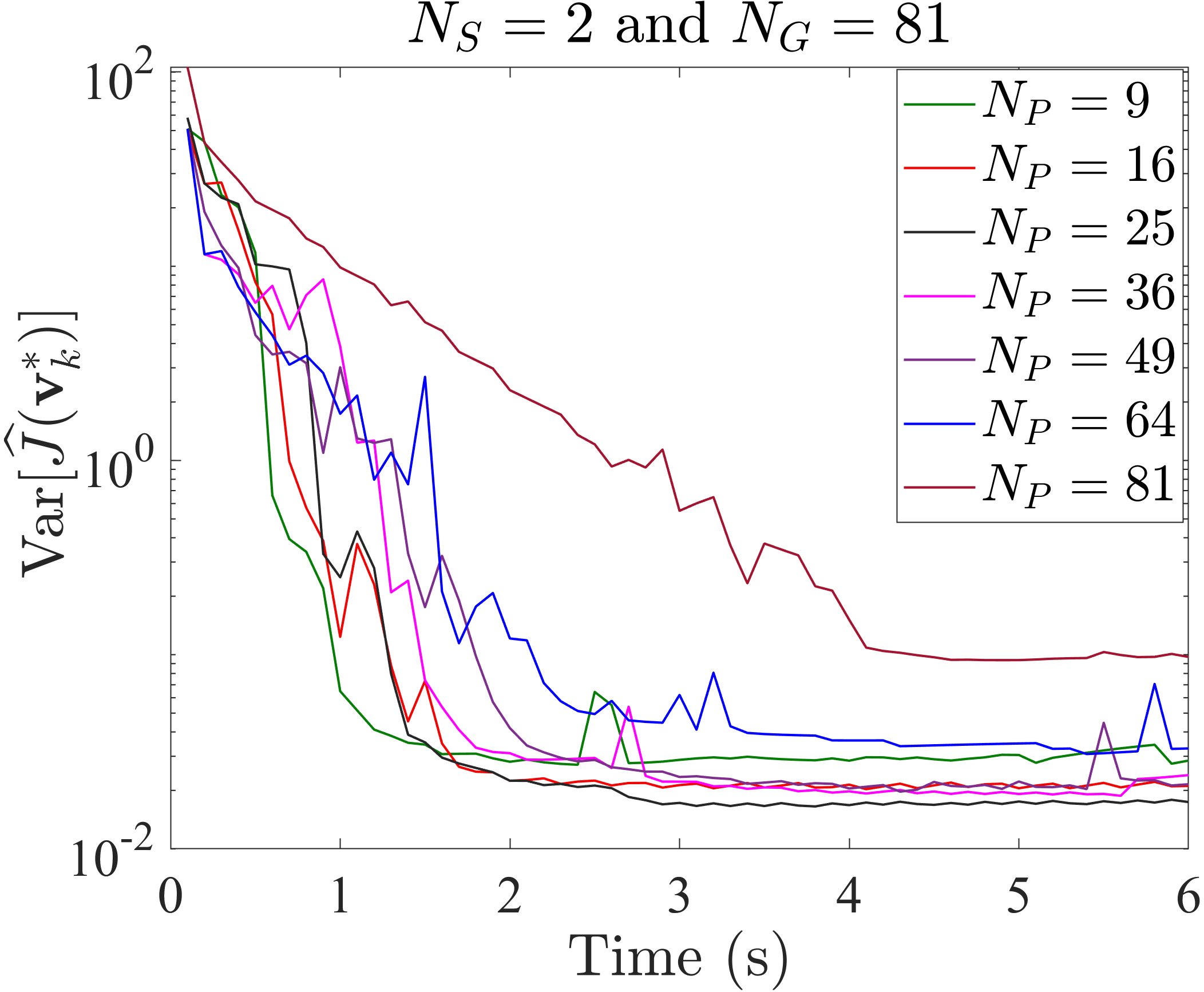}}
	\end{subfigmatrix}
	\caption{Convergence of CSCP-CRMI algorithm for different number of sensors and parameters.}
	\label{fig:pathVar_varying_Ns_Np}
\end{figure}

\figf{fig:path_var_comp} shows a comparison between the path cost variance
$\textrm{Var}[{J}(\gridPath^{*}_{k})]$ of the two methods. \revnew{Note that for
$\termThreshold = 0.1$, the CSCP-SMI algorithm requires 39 iterations to
converge, whereas the CSCP-CRMI algorithm requires only 15 iterations. This
indicates that the number of iterations for the SMI based method is
\emph{160\% larger} than CSCP.}
\figf{fig-path-var-comp_2} shows similarly large differences in convergence
rate for different numbers of sensors.

\figf{fig:pathVar_Ns_varies} shows the variation of path cost variance with
varying number of sensors. For a specified number of threat parameters,
$\nParameter=49$ and the grid points $\nGridPt=49$, better convergence is
achieved with more number of sensors. Using a single sensor will require a
relatively large number of iterations for convergence.

The convergence of the CSCP-CRMI algorithm for different number of threat
parameters $\nParameter$ is shown in \fig{fig:path_var_Np_varies}. For
$\nSensor=2$ and $\nGridPt=81$, CSCP converges faster for fewer threat states 
(e.g., $\nParameter = 9$ or $16$) compared to, say, $\nParameter = 64$ or~$81$.

We also performed comparative analysis for specific number of sensors and
parameters with varying number of grid points. It is observed that the path cost
and the path cost variance remain unchanged for different number of grid points.
This result is a consequence of proper scaling in the path cost formulation,
namely, the scaling of the cost by the grid spacing $\delta.$

\subsection{Analysis of Greedy Optimization}

\begin{figure}[t]
	\centering
	\begin{subfigmatrix}{4}
		\subfigure[ Sensor locations.]{
			\label{fig:CRMI_Greedy Placement Image}
			\centering
			\includegraphics[width=0.45\columnwidth]{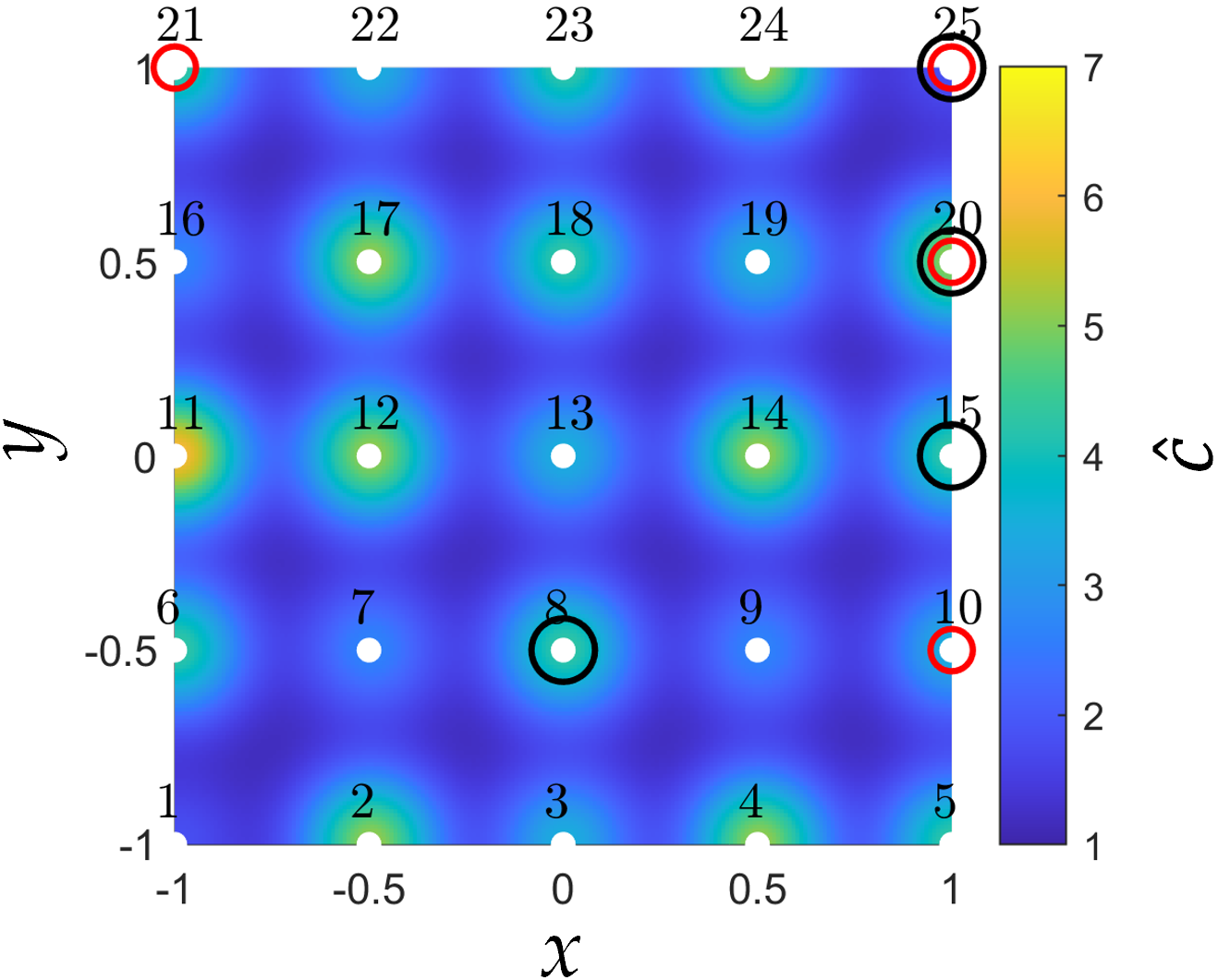}
		}
		\subfigure[Computation time.]{
			\label{fig:computing time}
			\centering
			\includegraphics[width=0.45\columnwidth]{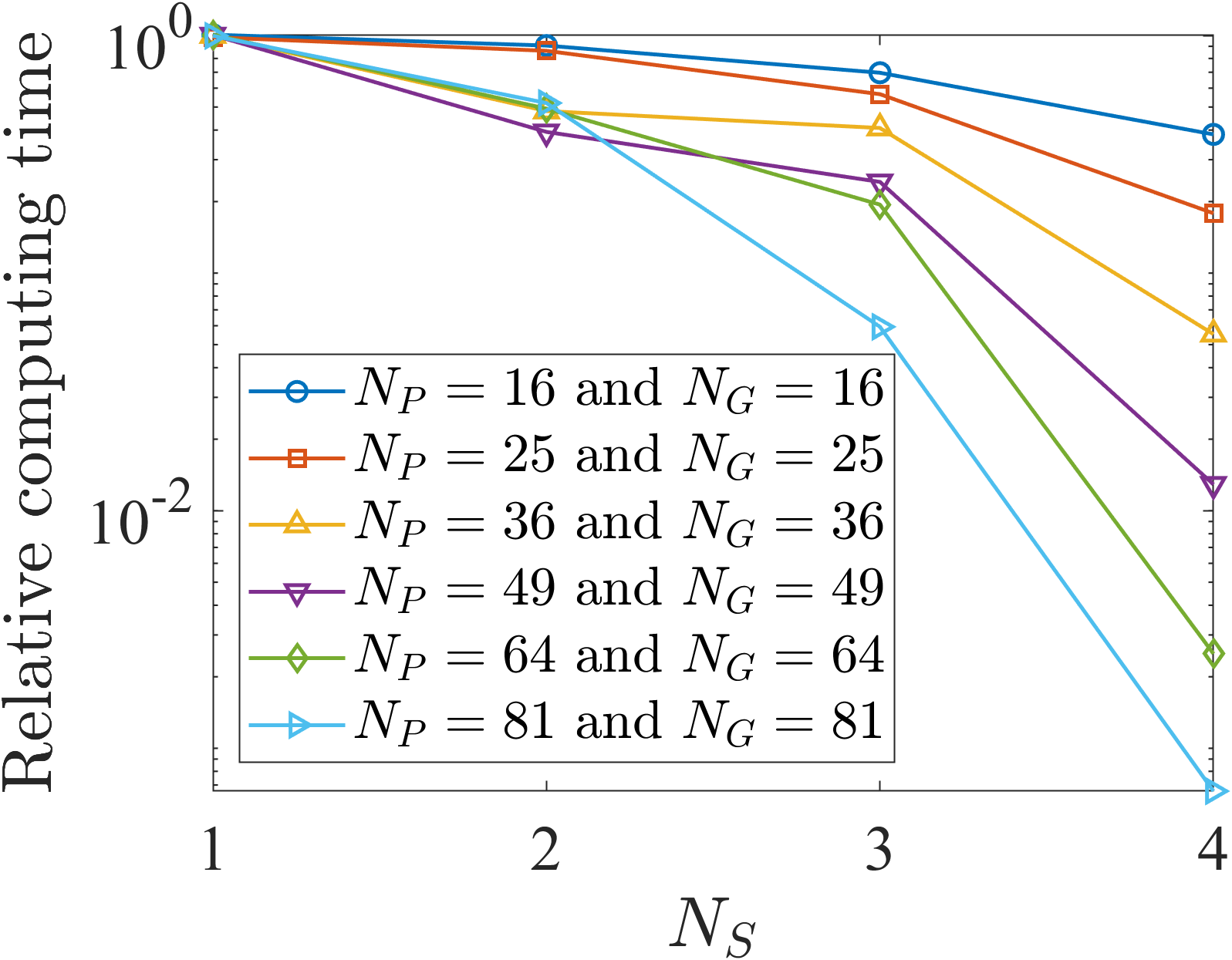}
		}
		\subfigure[Mutual information.]{
			\label{fig:Np_25mod MI plot}
			\centering
			\includegraphics[width=0.45\columnwidth]{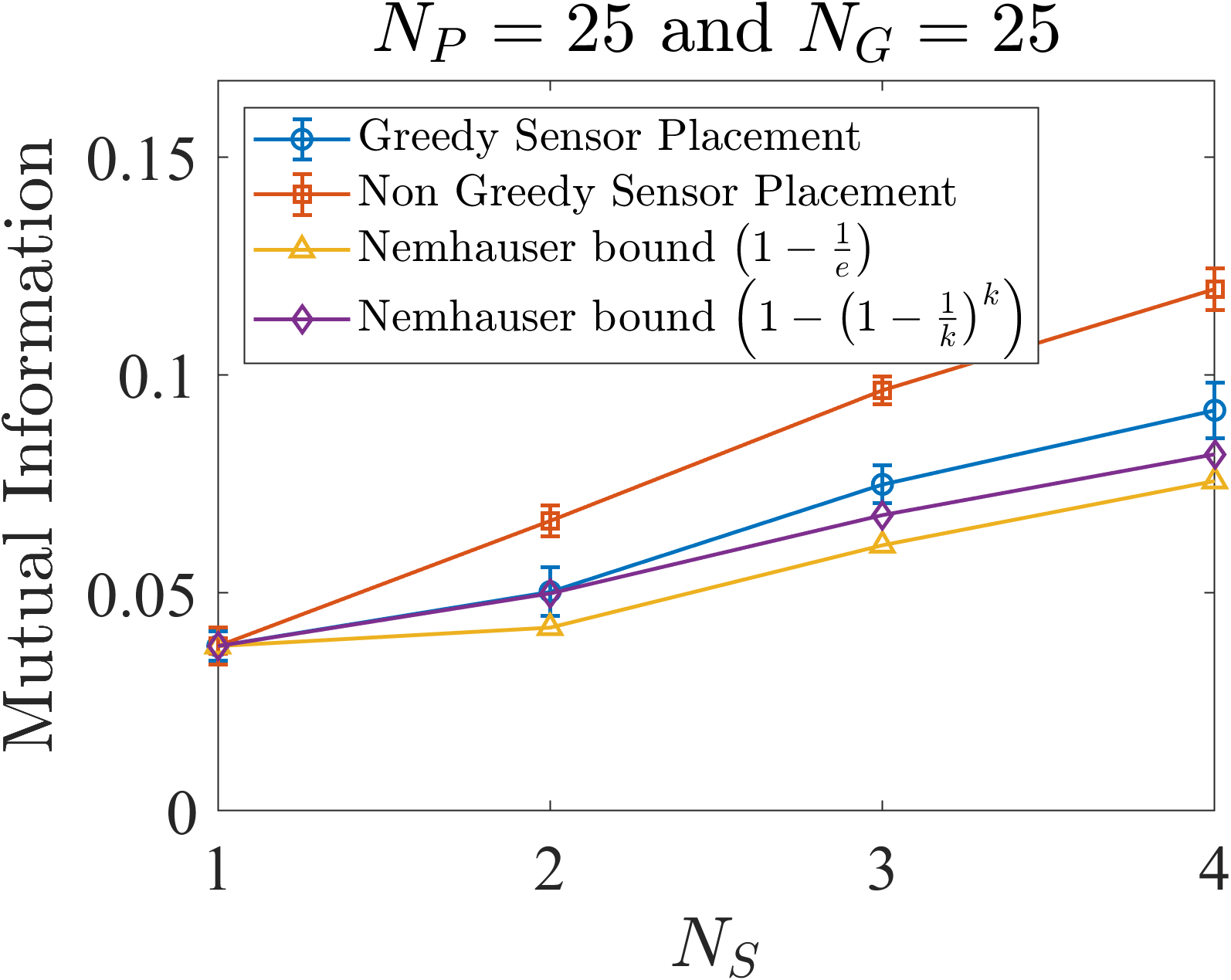}
		}
		\subfigure[Path cost variance.]{
			\label{fig:CRMI Greedy path var plot}
			\centering
			\includegraphics[width=0.45\columnwidth]{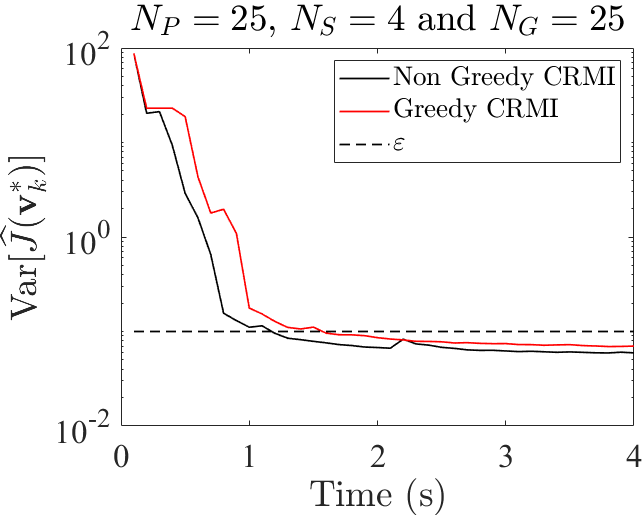}
		}
	\end{subfigmatrix}
	\caption{Comparison between greedy and non greedy CSCP-CRMI.}
	\label{fig:Comparison between CRMI and greedy}
\end{figure}

A comparative example of sensor placements for $\nParameter=25, \nGridPt=25$, and
$\nSensor=4$ using greedy and non-greedy (true optimal) is shown in
\fig{fig:CRMI_Greedy Placement Image}. Red circles in the figure indicate sensor
positions with greedy placement, whereas the black circles indicate true
optimal. Note that the two sensor locations in the grid space are common for
both criteria. This is a specific example where greedy optimization results in
the true optimal sensor configuration. \revnew{For a submodular function, greedy optimization results
in near-optimal configurations.}

\figf{fig:computing time} compares the time required to compute the optimal
sensor configuration using the greedy and non-greedy approaches. The relative
computing time is plotted for different combinations of grid points and number
of sensors. As the number of sensors increases the relative computation time
rapidly decreases, indicating the computational advantages of the greedy method.
A sharp decrease on the relative computation time is observed especially as the
number of grid points increases.

A comparison between the CRMI values for greedy and non-greedy methods is shown
in \fig{fig:Np_25mod MI plot}. The suboptimality bounds described in
\thmref{them-greedy} are also shown. As expected, \fig{fig:Np_25mod MI plot}
indicates that the greedy optimization-based CRMI values lie within these
bounds. Finally, a comparison of the path cost variance is shown in
\fig{fig:CRMI Greedy path var plot}. For this specific example and
$\termThreshold=0.1$, the non-greedy (true optimal) method results in 12 CSCP
iterations for convergence whereas the greedy method requires 16 CSCP
iterations.

\subsection{Analysis of Sensor Reconfiguration Cost}

\begin{figure}[t]
	\centering
	\begin{subfigmatrix}{3}
		\subfigure[Path cost mean.]{\includegraphics[width= 0.45\columnwidth]
			{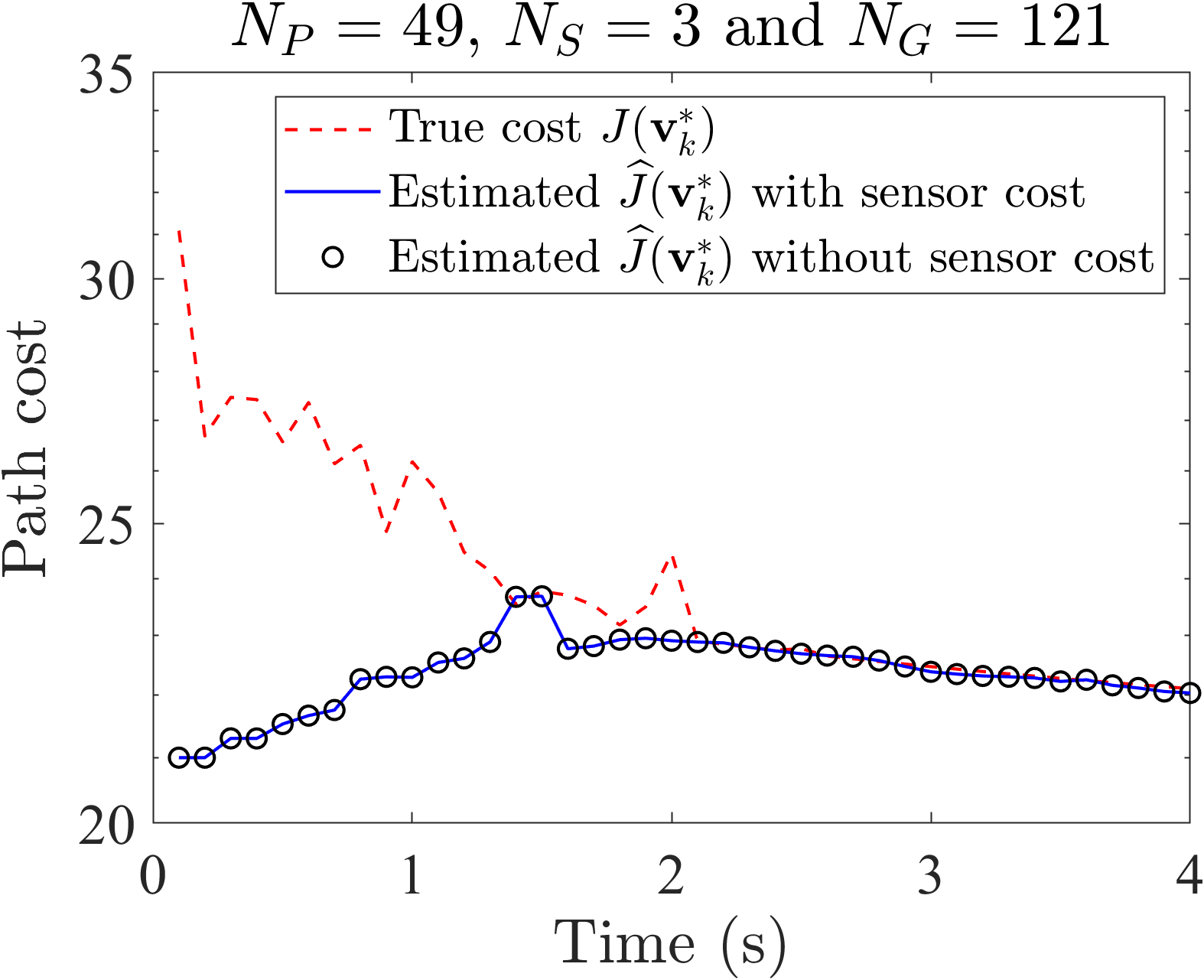}}
		\hspace{0.01\columnwidth}
		\subfigure[Path cost variance.]{\includegraphics[width= 0.45\columnwidth]{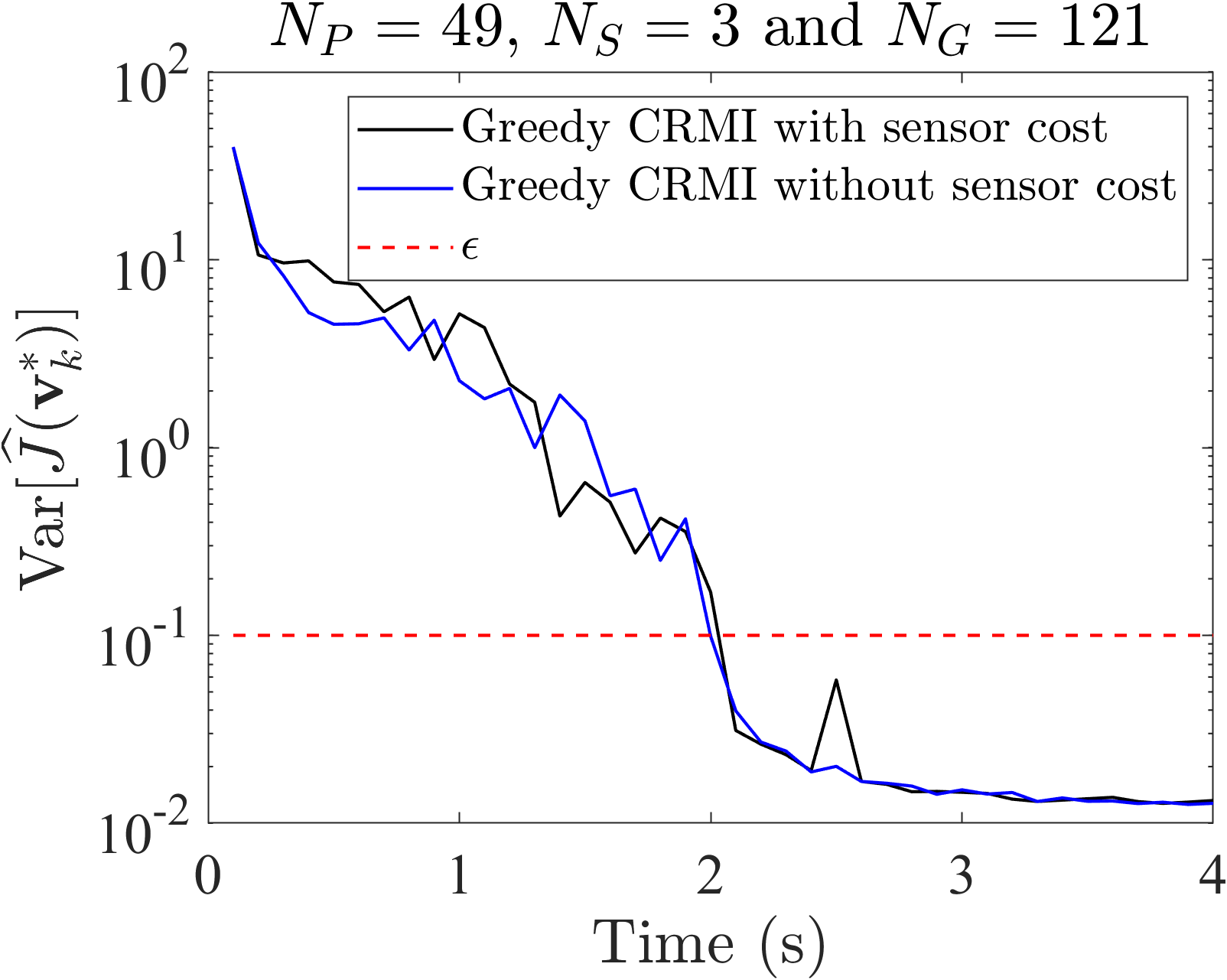}}
		\subfigure[Sum of distances traveled by sensors.]{\includegraphics[width= 
			0.45\columnwidth]{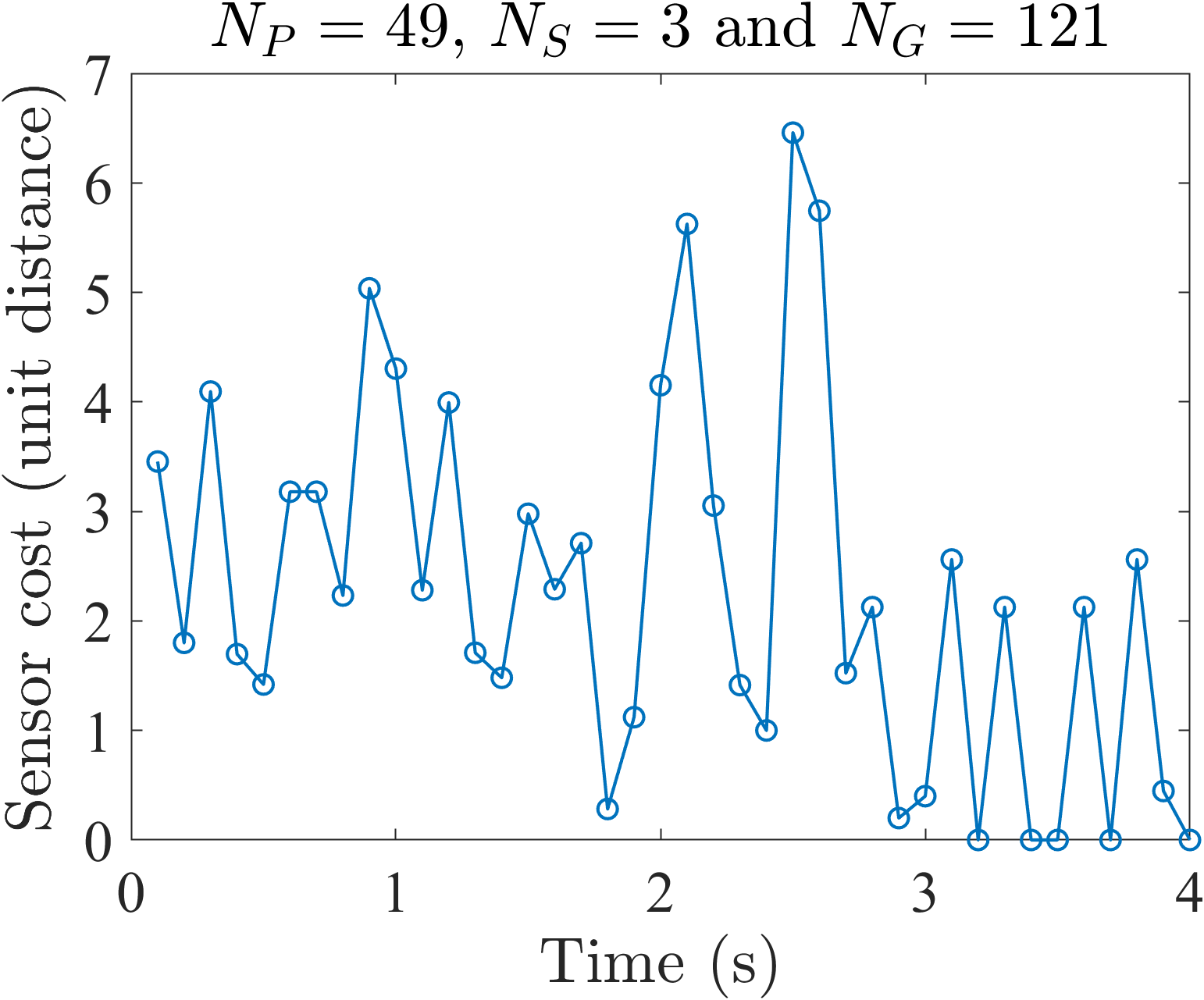}}
	\end{subfigmatrix}
	\caption{Path cost mean, variance over CSCP iterations with and without sensor reconfiguration 
	cost.}
	\label{fig:Fig Cost}
\end{figure}

The number of threat parameters, sensors, and the grid points used for the
analysis are $\nParameter=49$, $\nSensor=3$, and $\nGridPt=121$, respectively.
\figf{fig:Fig Cost}(a) shows the comparison between the true and estimated mean
path costs. Initially, the both estimated mean path costs are much lower than
the true path cost. This is because there is not much information about the
environment, and the estimator relies heavily on the ``optimistic'' prior, which
results in threat estimates with small values. The estimated path costs for both
with and without sensor reconfiguration costs are similar. The CSCP method with
a sensor reconfiguration cost converges in 21 iterations, whereas the method
without considering the sensor reconfiguration cost converges in 20 iterations.
The convergence of both CSCP methods is shown by a path cost variance plot in
\fig{fig:Fig Cost}(b). As the iterative process continues, the path cost
variance $\textrm{Var}[(\widehat{J}(\gridPath^{*}_{k})]$ decreases, and the
algorithm terminates when $\textrm{Var}[(\widehat{J}(\gridPath^{*}_{k})]$ falls
below $\termThreshold = 0.1$. \figf{fig:Fig Cost}(c) shows the sensor
reconfiguration cost values at different time steps. For the computation of
sensor reconfiguration cost, we choose $\alpha_1 = \sqrt{8}$ (diagonal distance
across the workspace) and $\alpha_2 = 0.01.$ For $\nSensor=3$, the sensor cost
is the cumulative sum of the distance traveled by the three sensors at each
iterations.

\begin{figure}[t]
	\centering
	\begin{subfigmatrix}{3}
		\subfigure[Sensor locations at $k=19.$]{\includegraphics[width= 
			0.45\columnwidth]{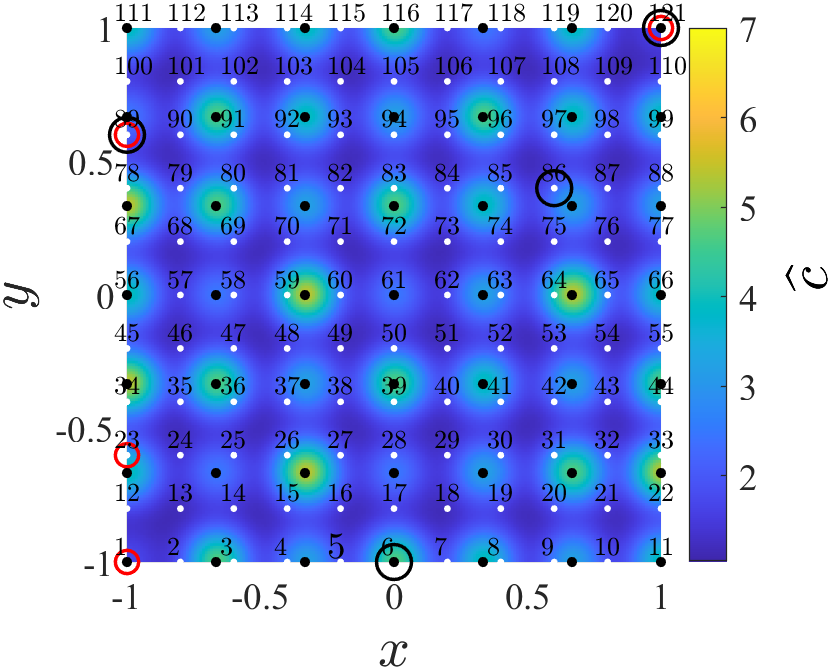}}
		\hspace{0.01\columnwidth}
		\subfigure[Sensor locations at $k=20.$]{\includegraphics[width= 
			0.45\columnwidth]{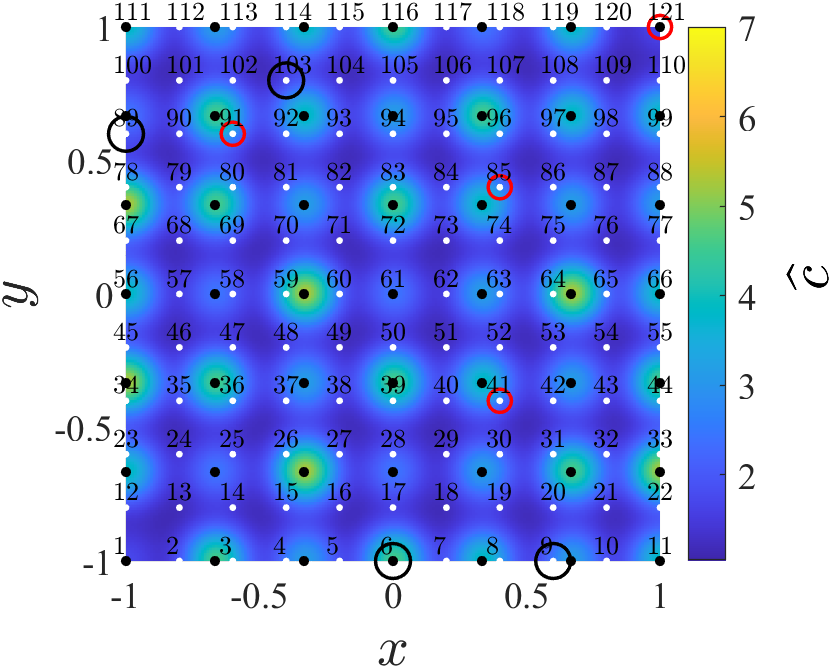}}
		\subfigure[CRMI and modified CRMI.]{\includegraphics[width= 
		0.45\columnwidth]{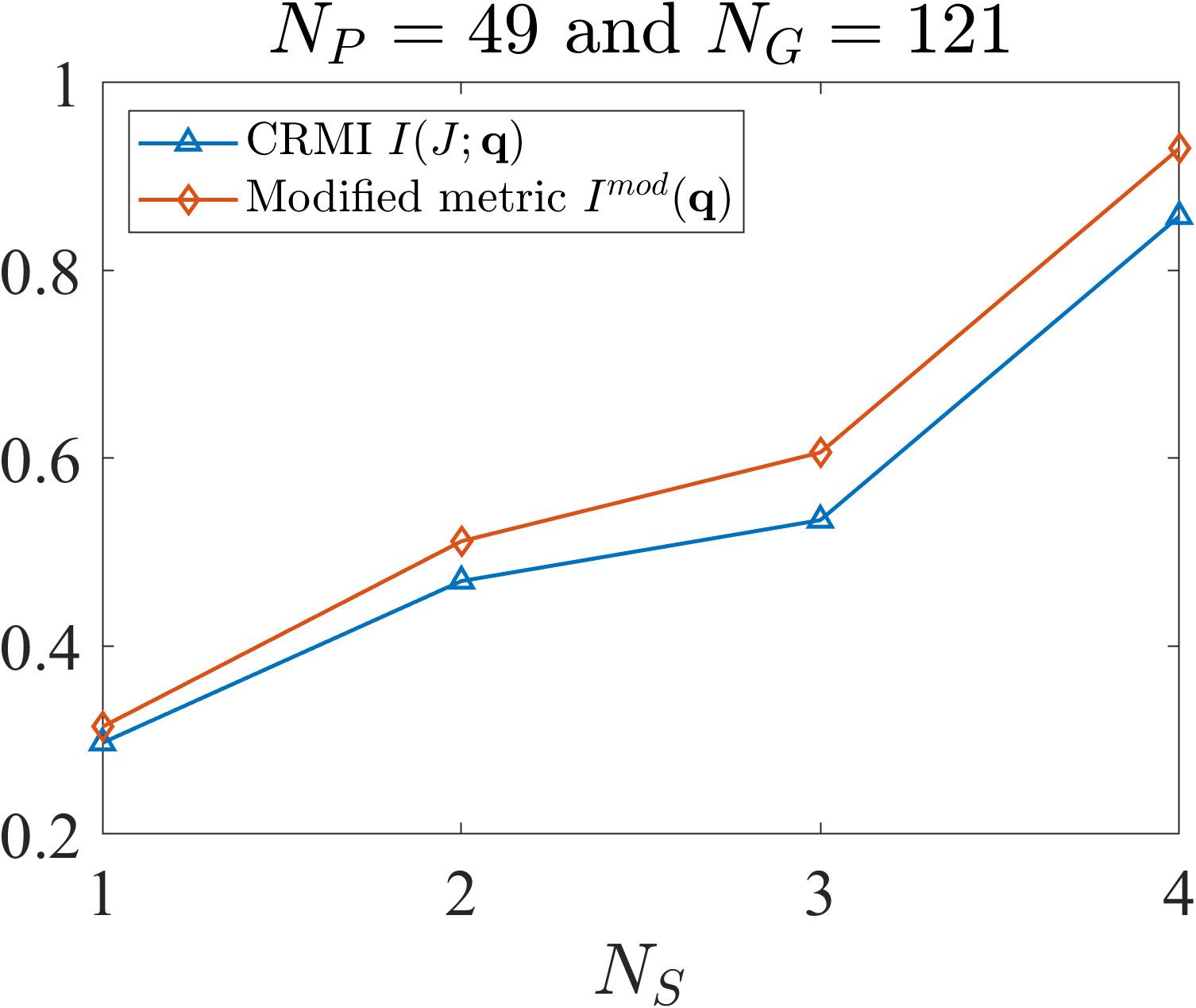}}			
	\end{subfigmatrix}
	\caption{Comparison between the CSCP method with and without 
		sensor reconfiguration cost.}
	\label{fig:Sensor-locations}
\end{figure}

A comparative example of sensor placements for $\nParameter=49, \nGridPt=121$,
and $\nSensor=4$ using CSCP with and without consideration of the sensor
configuration cost is shown in \fig{fig:Sensor-locations}(a) and (b). Red
circles in the figure indicate sensor positions based on maximizing CRMI,
whereas the black circles indicate sensor locations based on maximizing the
modified CRMI. At $k=19$, it can be observed that the two sensor locations in
the grid space are common for methods. As expected, when sensor reconfiguration
cost is considered, the new sensor locations (black circles) at $k=20$ are
relatively closer to the previous locations at $k=19$ as compared to sensor
positions denoted by the red circles across the two iterations. 
\figf{fig:Sensor-locations}(c) shows the values of the CRMI and the modified CRMI 
for different numbers of sensors.

\revnew{Finally, although the basis functions used in the previous results
had uniformly spaced centers and non-significant intersections in
regions of support, the proposed CSCP method is not restricted
by these assumptions. For instance, \figf{fig:CSCP_non_unifrom_field} shows
different CSCP iterations in a threat field constructed using 
overlapping basis functions and non-uniformly spaced centers.}

\def\thisfigwidth{0.46\columnwidth}
\begin{figure}[t]
	\centering
	\subfigure[$k=4$]{
			\includegraphics[width=\thisfigwidth]{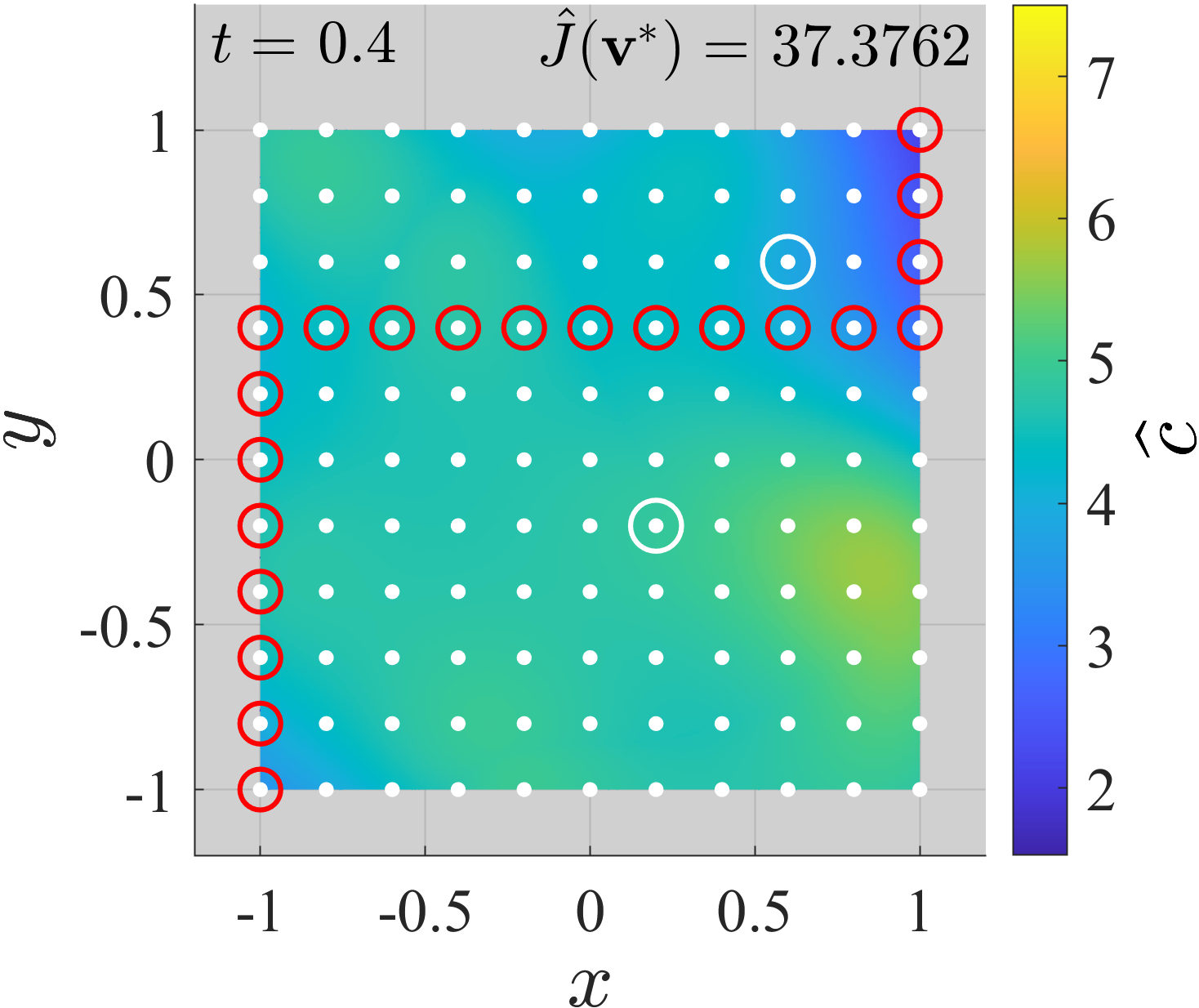}
		}
	\subfigure[$k=10$]{
			\includegraphics[width=\thisfigwidth]{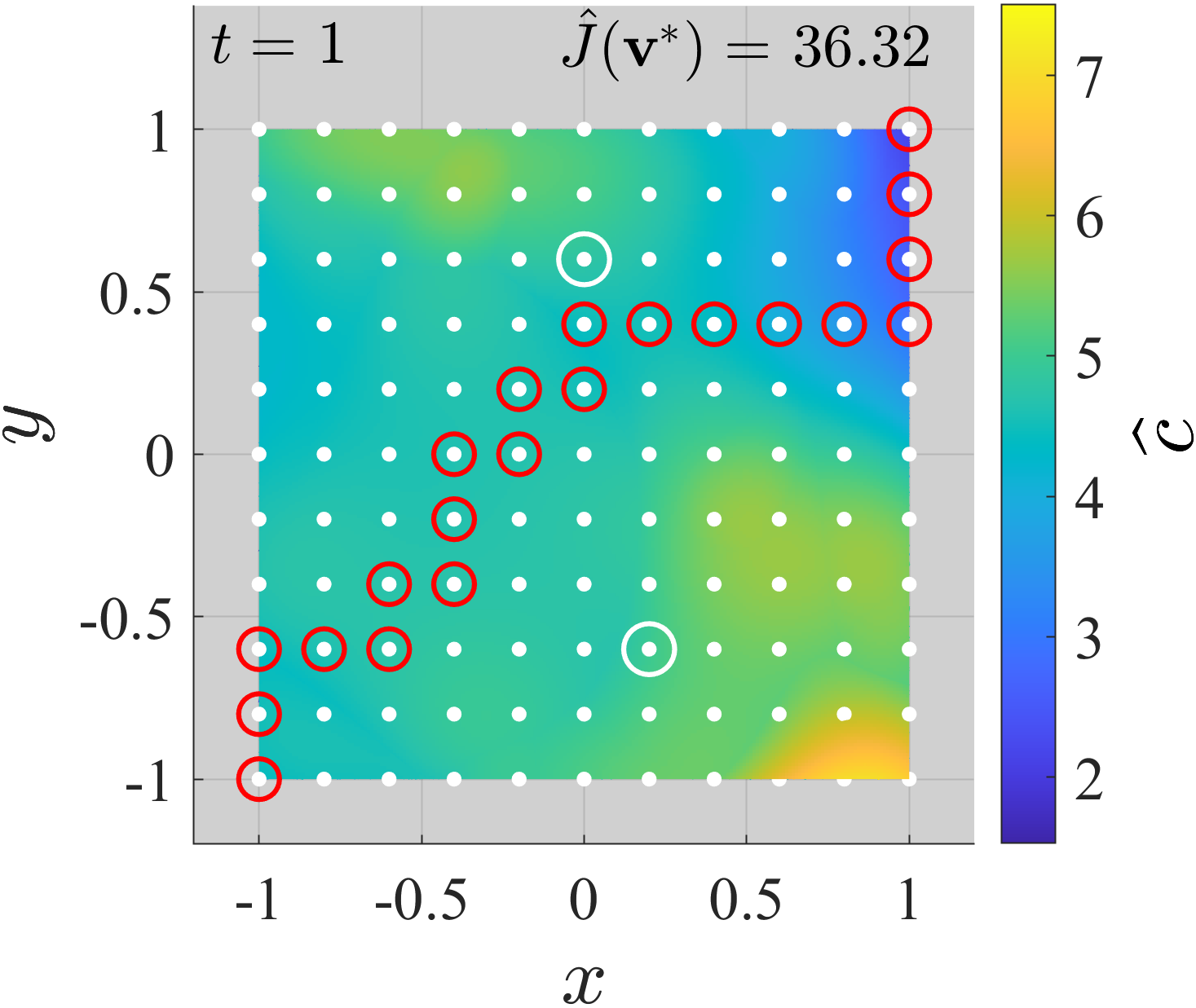}
		}
	\subfigure[$k=15$]{
			\includegraphics[width=\thisfigwidth]{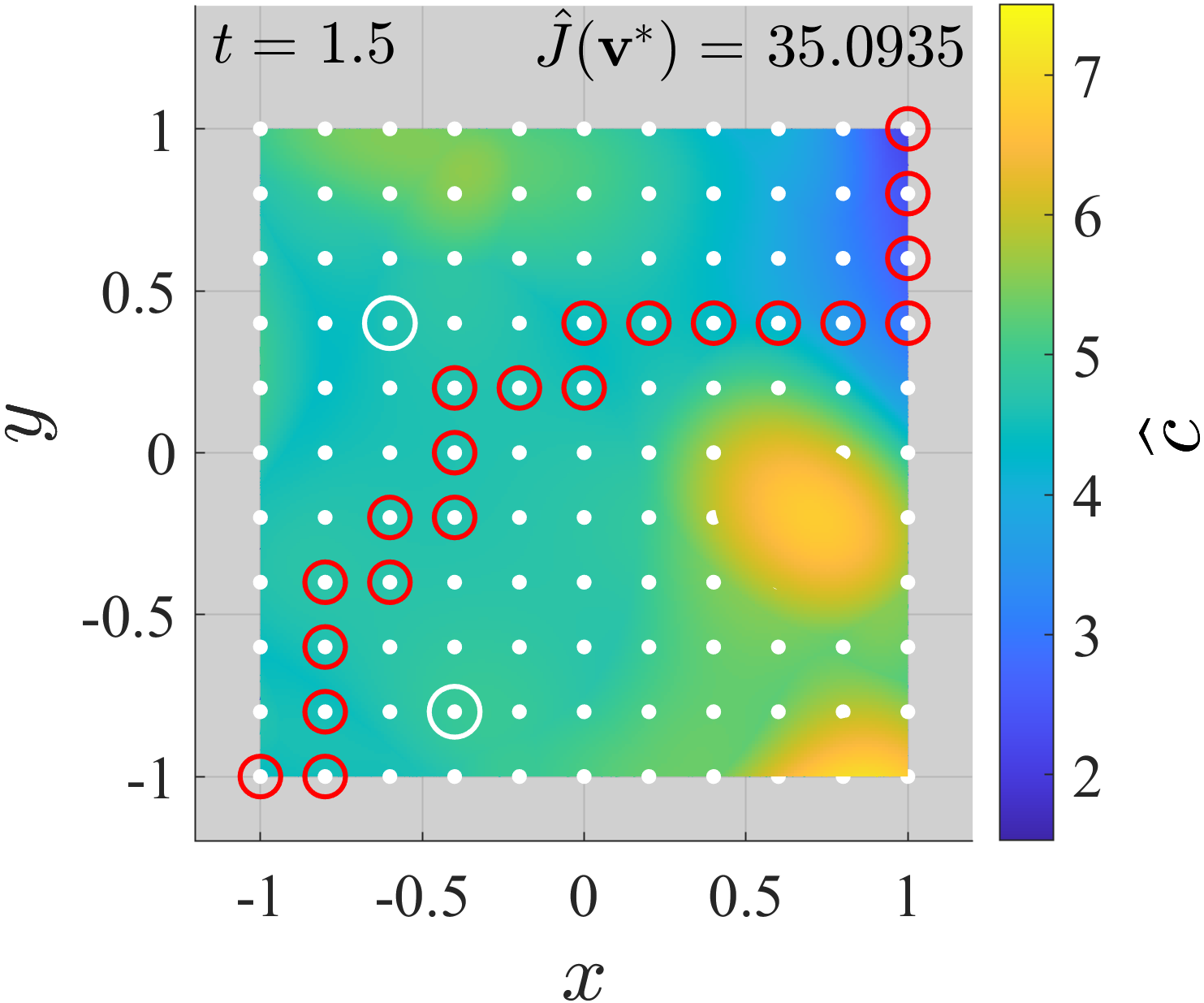}
		}
	\subfigure[$k=19$]{
			\includegraphics[width=\thisfigwidth]{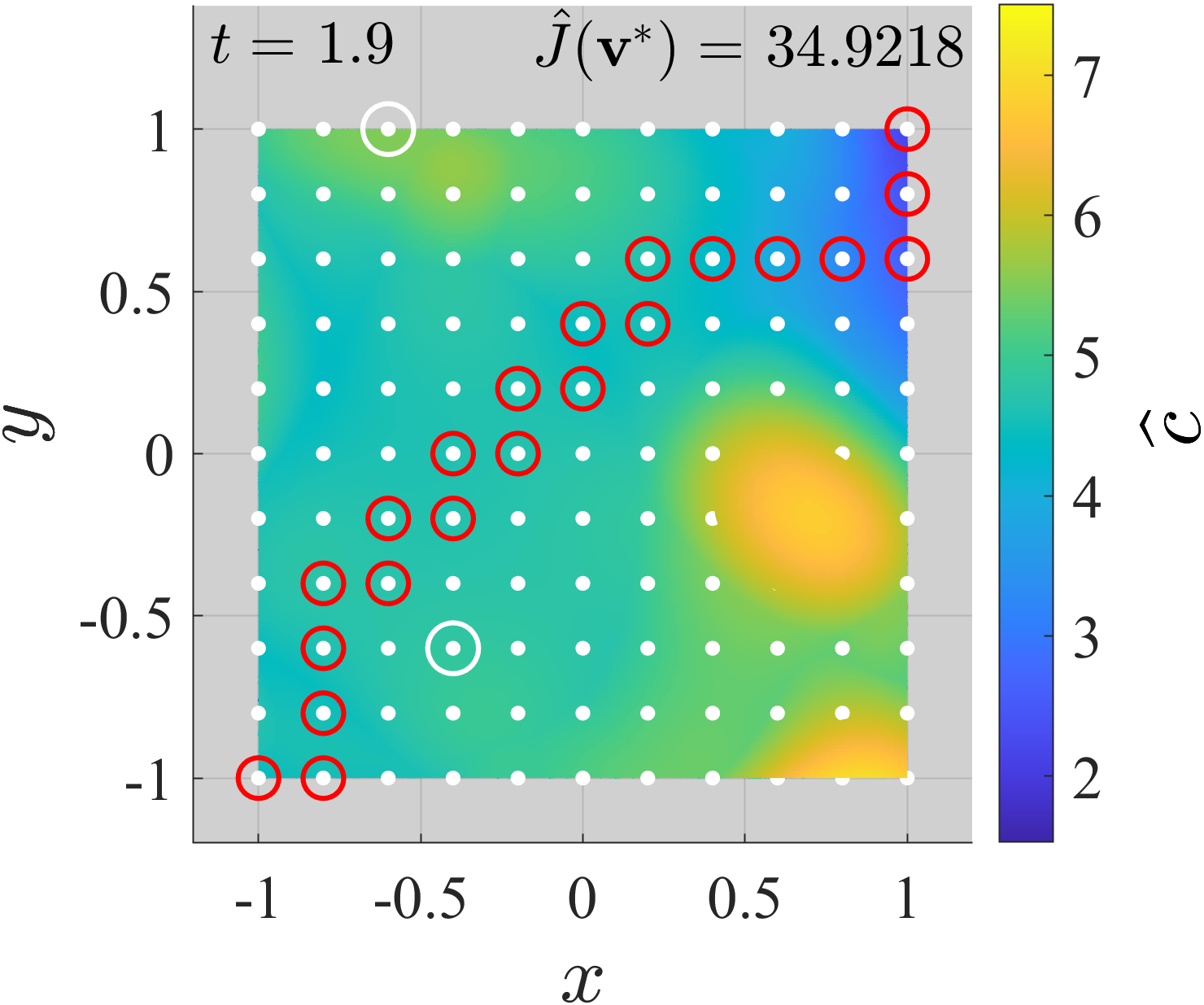}
		}
	\caption{CSCP iterations for 
		non uniformly spaced basis functions, with
		$N_P = 49$, $N_G = 121,$ and $N_S = 2$.}
	\label{fig:CSCP_non_unifrom_field}
\end{figure}

\subsection{Computational Complexity}

The computational complexity of the \algoref{alg-cscp} depends on the complexity
of CRMI optimization in \algline{line-CRMI}, and the number of calls performed
in Dijkstra's algorithm in \algline{line-dijkstra}. CRMI calculation and
maximization involves determinant and inverse computations, thus has a time
complexity of $\mathcal{O}(\nSensor^{3})$. By comparison, the CRMI calculation
for the greedy optimization method has a complexity of $\mathcal{O}(\nSensor)$.
The worst-case time complexity of Dijkstra's algorithm implemented with a
Fibonacci heap is $\mathcal{O}(\nGridPt +\nGridPt\log(\nGridPt))$
\citep{Cooper2019}. The complexity of CRMI optimization depends on the specific
algorithm chosen.
\revnew{CSCP addresses scenarios where sensor resources are limited.
Scaling up should then be considered in the context of the grid size,
which may be achieved by relaxation to a continuous workspace and a suitable
optimization method, e.g., evolutionary global optimization methods, as 
previously mentioned. Further computational efficiency may be achieved
via decentralized estimation, as we had reported earlier in~\cite{cooper2019ICC},
albeit with a na\"ive sensor placement method.}

\section{Conclusions}

\label{sec-conclusions}
In this paper, we discussed a new measure of information gain for optimal sensor
placement to capture coupling of the sensor placement problem with a p\ppl\
problem. This measure, which we call the context-relevant mutual information
(CRMI), addresses the reduction in uncertainty in the path cost, rather than the
environment state. We presented a coupled sensor placement and p\ppl\ algorithm
that iteratively places sensors based on CRMI maximization, updates the
environment threat estimate, and then plans paths with minimum expected cost.
Crucially, we showed CRMI to be a submodular function, due to which we can apply
greedy optimization to arrive at near-optimal results while maintaining
computational efficiency. We performed a comparative study between CSCP-CRMI and
a decoupled CSCP-SMI method. The CSCP-SMI method places sensors by maximizing
the standard (path-independent) mutual information of the measurments and threat
state. We showed via numerical simulation examples that the CSCP-CRMI algorithm
converges in less than half as many iterations compared to CSCP-SMI algorithm,
which indicates a significant reduction in the number of sensor observations
needed to find near-optimal paths. We also introduced and analyzed a modified
cost function that addresses costs associated with distances traveled by sensors
(i.e., sensor reconfiguration cost) after each iteration of placement.
\revnew{For future work, we will consider other real-world sensor constraints
	such as communication limitations, energy consumption, and sensor reliability
	issues.}

\section*{Acknowledgments}
This work is funded in part by NSF 
Dynamics, Control, and Systems Diagnostics program grant \#2126818.

\bibliographystyle{aiaa}
\bibliography{References}

\appendix
\section*{Appendix}

\subsection{Technical Proofs}

\begin{proof}[Proof of \prpf{prop-submodular}]
	%
	%
	%
	By the definition of path relevant set $\mathcal{K}$, for the path 
	$\gridPath^{*}_k=\{v_{0},v_{1},\ldots,v_{L}\}$, the values of 
	$\phi_m(\xState_{\gridPath_{\ell}}) 
	\approx 0$ for each $\ell=(0,1,\dots, L)$ and $m\notin\mathcal{K}$. Therefore, for all basis 
	functions $m\notin\mathcal{K}$,  $\boldsymbol\Phi(\xState_{\gridPath_{\ell}}) = 
	\boldsymbol{0}$, 
	and by \eqnnt{eq-covar2}, $\PEECovar_{{J \zMeas}_{k|k-1}}\approx 0$. By \eqnnt{eq-crmi}, 
	$I({J}_{k};\zMeas_{k}(\qConfig))\approx 0$ at locations in the workspace where the basis 
	functions 
	$\phi_m$ do not overlap with the path $\gridPath$.

	Per the proposed sensor reconfiguration policy, $\qConfig^{*}_{k} =\max_{\qConfig} 
	I({J}_{k};\zMeas_{k}(\qConfig))$, sensors are necessarily placed at locations within the 
	regions of 
	significant support $\mathcal{R}_m\msup{sup}$ defined by $\phi_m$, for each $m\in\mathcal{K},$
	i.e., within the basis support regions of the basis functions contained within a path relevant 
	set. 
	Given the path cost $J(\gridPath)$ is known, the sensor measurements $\zMeas(\qConfig)$ within 
	the 
	path-relevant set $\mathcal{K}$ fully capture the information about the $J(\gridPath)$.
	This shows that the sensor measurements are conditionally independent 
	given the path cost. i.e., $p(\zMeas_{1}, \zMeas_{2}, \ldots \zMeas_{N_{s}} \mid J) 
	= p(z_{1}|J)p(z_{2}|J) \ldots p(\zMeas_{N_{s}} \mid J).$ Next, we show that this conditional
	independence implies submodularity.
	
	Consider two subsets of the sensor configurations, $\mathcal {A}$ and $\mathcal {B}$, 
	such that $\mathcal {A} \subseteq \mathcal {B}\subseteq \qConfig$,
	where $\qConfig \subset \intset{\nGridPt}.$ By~\eqnnt{eqn-entropy-gain}, we write
	$I({J};\mathcal {A}) = \entropy(\mathcal {A}) - \entropy(\mathcal {A}\mid J)$ and 
	$I({J};\mathcal {B}) = \entropy(\mathcal {B}) - \entropy(\mathcal {B}\mid J)$.
	For any $x \in \qConfig \backslash \mathcal {B}$, the marginal CRMI gain 
	due to a sensor placed at $x$ to in addition to those in sets 
	$\mathcal {A}$ and $\mathcal {B}$ is
	\begin{alignat*}{3}
		\Delta_{\mathcal {A}}(x)&= I({J};\mathcal {A}\cup {x}) - I({J};\mathcal {A})
		&&= \entropy(\mathcal {A}\cup {x}) - \entropy(\mathcal {A}\cup {x}\mid J) - \entropy(\mathcal {A}) + \entropy(\mathcal {A}\mid J)	\\
		\Delta_{\mathcal {B}}(x)&= I({J};\mathcal {B}\cup {x}) - I({J};\mathcal {B})
		&&= \entropy(\mathcal {B}\cup {x}) - \entropy(\mathcal {B}\cup {x}\mid J) - \entropy(\mathcal {B}) + \entropy(\mathcal {B}\mid J)
	\end{alignat*}
	By the chain rule of conditional entropy, $\entropy(\mathcal {A}\cup {x}\mid J) = \entropy(\mathcal {A}\mid J) + \entropy(x\mid \mathcal {A},J)$, and $\entropy(\mathcal{B} \cup {x}\mid J) = \entropy(\mathcal {B}\mid J) + \entropy(x\mid \mathcal {B},J)$. Also, the conditional independence of $\zMeas(\qConfig)$ given $J$ implies 
	$\entropy(x\mid \mathcal {A},J) = \entropy(x\mid \mathcal {B},J) = \entropy(x\mid J).$ It follows that
	\begin{align*}
		\Delta_{\mathcal {A}}(x) &= \entropy(\mathcal {A}\cup {x}) -\entropy(x\mid J) - \entropy(\mathcal {A}), &
		\Delta_{\mathcal {B}}(x) &= \entropy(\mathcal {B}\cup {x}) -\entropy(x\mid J) - \entropy(\mathcal {B})
	\end{align*}
	By the chain rule again, 
	$\entropy(\mathcal {A}\cup {x}) = \entropy(\mathcal {A})+ \entropy(x\mid \mathcal {A})$ 
	and $\entropy(\mathcal {B}\cup {x}) = \entropy(\mathcal {B})+ \entropy(x\mid \mathcal {B})$ from the previous
	expressions, we obtain $\Delta_{\mathcal {A}}(x) = \entropy(x\mid \mathcal {A}) - \entropy(x\mid J)$, and 
	$\Delta_{\mathcal {B}}(x) = \entropy(x\mid \mathcal {B}) - \entropy(x\mid J)$.
	Therefore difference in marginal gains is $\Delta_{\mathcal {A}}(x) - \Delta_{\mathcal {B}}(x) = \entropy(x\mid \mathcal {A}) - \entropy(x\mid \mathcal {B})$, which must be nonnegative because 
	$\mathcal {A} \subseteq \mathcal {B}.$ Consequently, $I({J};\mathcal {A}\cup {x}) - 
	I({J};\mathcal {A}) \geq I({J};\mathcal {B}\cup {x}) - I({J};\mathcal {B}) $, which satisfies 
	the 
	submodularity criterion.
\end{proof}

\begin{proof}[Proof of \prpf{cor-submodularity}]
	The sensor reconfiguration cost function can be represented as a set function 
	\begin{align*}
		f(\mathcal{S}) \defeq \alpha_1 - \alpha_2 \min_{j
			\in \intset{\nSensor}, \ell \in \mathcal{S}} 
			\left\{ \| q_\ell - q_{j}^{k*} \| \right\},
	\end{align*}
	where $\mathcal{S} \subseteq \intset{\nSensor}$ is 
	a subset of sensor indices. Consider two subsets $\mathcal{A}$ and $\mathcal{B}$, 
	such that $\mathcal{A} \subseteq \mathcal{B} 
	\subseteq \intset{\nSensor}$ and $x\in\intset{\nSensor}\backslash \mathcal{B} $. 
	In the rest of this proof, the symbol $j$ is an index over $\intset{\nSensor},$
	i.e., $j \in \intset{\nSensor},$ which we avoid writing explicitly for notational
	convenience. Without loss of generality, we assume $\alpha_2 = 1.$
	
	To calculate the marginal costs due to the inclusion of the index $x$ in
	either $\mathcal{A}$ or $\mathcal{B},$ we note:
	\begin{align*}
		f(\mathcal{A} \cup x ) &= \alpha_1 -  \min_{\ell \in \{\mathcal{A}\cup
			x\} }  \| q_\ell - q_{j}^{k*} \| \\ %
		f(\mathcal{B} \cup x)&=\alpha_1 - \min_{\ell \in \{\mathcal{B} \cup
			x\} } \| q_\ell - q_{j}^{k*} \| \\ %
		\Rightarrow f(\mathcal{A}\cup x) - f(\mathcal{A}) &= - \min_{\ell \in
			\{\mathcal{A} \cup x \}}  \| q_\ell - q_{j}^{k*} \| + \min_{\ell \in
			\mathcal{A}}  \| q_\ell- q_{j}^{k*} \| \\ %
		\Rightarrow f(\mathcal{B}\cup x) - f(\mathcal{B}) &= - \min_{\ell \in
			\{\mathcal{B} \cup x\} }  \| q_\ell - q_{j}^{k*} \| + \min_{\ell \in
			\mathcal{B}} \| q_\ell- q_{j}^{k*} \|
	\end{align*}
	
	Since $\mathcal{A} \subseteq \mathcal{B}$, $\min_{\ell \in 
	\mathcal{A}} \| q_\ell- q_{j}^{k*} \| 
	\geq \min_{\ell \in \mathcal{B}} \| q_\ell- q_{j}^{k*} \|.$ 
	It follows that, for sufficiently 
	large $\alpha_1,$  $f(\mathcal{A})\leq f(\mathcal{B})$. 
	Then we consider the following three cases:
	\begin{align*}
		\| q_x- q_{j}^{k*} \| \leq \min_{\ell \in \mathcal{B}}
		\| q_\ell- q_{j}^{k*} \| \leq 
		\min_{\ell \in \mathcal{A}} \| q_\ell- q_{j}^{k*} \|,  \tag{i}\\
		\min_{\ell \in \mathcal{B}}  \| q_\ell- q_{j}^{k*} \|  \leq \|
		q_x- q_{j}^{k*} \| \leq \min_{\ell \in \mathcal{A}} \| q_\ell-
		q_{j}^{k*} \|, \tag{ii}\\
		\min_{\ell \in \mathcal{B}} \| q_\ell- q_{j}^{k*} 
		\| \leq \min_{\ell \in \mathcal{A}} \| q_\ell-  q_{j}^{k*} \| 
		\leq \| q_x- q_{j}^{k*} \|. \tag{iii}
	\end{align*}
	
	For case (i) we note: 
	\begin{align*}
		f(\mathcal{A}\cup x) - f(\mathcal{A}) &=  
		\min_{\ell \in \mathcal{A}} \| q_\ell
		- q_{j}^{k*} \| - \| q_x- q_{j}^{k*} \|, \\
		f(\mathcal{B}\cup x) - f(\mathcal{B}) &= 
		\min_{\ell \in \mathcal{B}} \| q_\ell- q_{j}^{k*} \| - 
		\| q_x - q_{j}^{k*} \|
	\end{align*}
	It follows that $f(\mathcal{A}\cup x) - f(\mathcal{A}) > 
	f(\mathcal{B}\cup x) - f(\mathcal{B}) $. For case (ii) we note:
	\begin{align*}
		f(\mathcal{A}\cup x) - f(\mathcal{A}) &=  
		\min_{\ell \in \mathcal{A}} \| q_\ell
		- q_{j}^{k*} \| - \| q_x- q_{j}^{k*} \|, \\
		f(\mathcal{B}\cup x) - f(\mathcal{B}) &= 
		\min_{\ell \in \mathcal{B}} \| q_\ell - q_{j}^{k*} \| 
		 - \min_{\ell \in \mathcal{B}} \| q_\ell- q_{j}^{k*} \| = 0.
	\end{align*}
	Again, it follows that $f(\mathcal{A}\cup x) - f(\mathcal{A}) > 
	f(\mathcal{B}\cup x) - f(\mathcal{B}) $. Finally, for case (iii) we note:
	\begin{align*}
		f(\mathcal{A}\cup x) - f(\mathcal{A}) =
		f(\mathcal{B}\cup x) - f(\mathcal{B}) = 0.
	\end{align*}
	Therefore, the inequality
	$f(\mathcal{A}\cup x) - f(\mathcal{A}) \geq 
	f(\mathcal{B}\cup x) - f(\mathcal{B}) $ is always true
	and $f$ is submodular.
	By \prpf{prop-submodular} and the additive property of submodular 
	functions~\cite{nemhauser1978analysis}, $I\msup{mod}(\qConfig)$ is submodular.
\end{proof}

\subsection{Unscented Kalman Filter for Threat Estimation}
\label{ssec-ukf}

The estimated state parameters and the error covariance of the system 
are calculated by an Unscented Kalman Filter (UKF) \citep{Julier2004}. 
Although the scope of this paper is limited to linear threat field dynamics, 
we would like to ensure generality for nonlinear threat models to be considered 
in the future. For the reader's convenience we provide a brief overview of 
the UKF here and refer to \citep{Julier2004} for further details.

An augmented state variable $\paramVec_{k-1}^{a} \in 
\real[N_{\paramVec}+N_{\boldsymbol{\omega}}+N_{\boldsymbol{\eta}}]$ is defined as
$\paramVec_{k-1}^{a}=[\paramVec^{T}_{k-1} 
\quad\boldsymbol{\omega}^{T}_{k-1} \quad\boldsymbol{\eta}^{T}_{k-1}]^{T}$ 
and $2N+1$ sigma vectors are 
initialized using augmented estimated states 
$\widehat{\paramVec}_{k-1}^{a}$ and error 
covariance $\PEECovar^{a}_{k-1}$ as:
\begin{align*}
	\boldsymbol\sigma^{a}_{k-1} 
	&\defeq\left[\widehat{\paramVec}_{k-1}^{a}\cdots\widehat{\paramVec}_{k-1}^{a}\right] 
	+\left[\textbf{0}\quad\sqrt{(N+\lambda)\PEECovar^{a}_{k-1}}\quad-
	\sqrt{(N+\lambda)\PEECovar^{a}_{k-1}}\right].
\end{align*}
The generated sigma vectors are propagated through the process model as 
$\boldsymbol\sigma^{\paramVec}_{k|k-1} = \textbf{A} \boldsymbol\sigma^{\paramVec}_{k-1} + 
\boldsymbol\sigma^{\boldsymbol\omega}_{k-1}$, where $\boldsymbol\sigma^{\paramVec}_{k-1}$ and 
$\boldsymbol\sigma^{\boldsymbol\omega}_{k-1}$ are the components of $\boldsymbol\sigma^{a}$ that 
correspond 
to the state variables and process noise, respectively. At time instants where measurements are not 
available, the estimated state and error covariance are predicted as
\begin{align*}
	\widehat{\paramVec}_{k}^{-} &= \sum_{i}^{2N}W_{i}^{(m)}\sigma^{\paramVec}_{i,k|k-1}, \\
	{\PEECovar}_{k}^{-} &= 
	\sum_{i}^{2N}W_{i}^{(c)}\left(\boldsymbol\sigma^{\paramVec}_{i,k|k-1}-
	\widehat{\paramVec}_{k}^{-}\right){\left(\boldsymbol\sigma^{\paramVec}_{i,k|k-1}-
		\widehat{\paramVec}_{k}^{-}\right)}^{\intercal},
\end{align*}
\noindent
where, $W_{i}^{(m)}$ and $W_{i}^{(c)}$ are weights associated with sigma vectors. 
After the sensor 
measurements are taken, the measurement model is propagated using $\boldsymbol\gamma_{k|k-1} = 
\HMeas_{k}(\qConfig)\boldsymbol\sigma^{\paramVec}_{k|k-1} + 
\boldsymbol\sigma^{\boldsymbol\eta}_{k-1}$. Next, 
the mean and covariance of the measurement, and the cross-covariance of the state and measurement 
are 
calculated as:
\begin{align*}
	\widehat{\zMeas}_{k}^{-} &= \sum_{i}^{2N}W_{i}^{(m)}\boldsymbol\gamma_{i,k|k-1}, \\
	\PEECovar_{{\zMeas\zMeas}_{k}} &= 
	\sum_{i}^{2N}W_{i}^{(c)}\left(\boldsymbol\gamma_{i,k|k-1}-\widehat{\zMeas}_{k}^{-}\right)
	{\left(\boldsymbol\gamma_{i,k|k-1}-\widehat{\zMeas}_{k}^{-}\right)}^{\intercal}, \\
	\PEECovar_{{\paramVec \zMeas}_{k}} &= 
	\sum_{i}^{2N}W_{i}^{(c)}\left(\boldsymbol\sigma^{\paramVec}_{i,k|k-1}-
	\widehat{\paramVec}_{k}^{-}\right){\left(\boldsymbol\gamma_{i,k|k-1}-
		\widehat{\zMeas}_{k}^{-}\right)}^{\intercal}.
\end{align*}
Finally, the filter gain is calculated as $\UKFGain_k = \PEECovar_{{\paramVec 
		\zMeas}_{k}}{\PEECovar}^{-1}_{{\zMeas\zMeas}_{k}}$ and the estimated states, 
$\widehat{\paramVec}_{k}$ 
and 
covariance, $\PEECovar_{k}$  are updated as,
$\widehat{\paramVec}_{k}=\widehat{\paramVec}_{k}^{-} + 
\UKFGain_k(\zMeas_{k}-\widehat{\zMeas}_{k}^{-})$, and 
$\PEECovar_{k}=\PEECovar_{k}^{-}-\UKFGain_k\PEECovar_{{\zMeas\zMeas}_{k}}\UKFGain_k^{\intercal}$.

\end{document}